\documentclass[superscriptaddress,twocolumn,prx,preprintnumbers,amsmath,amssymb,longbibliography]{revtex4}
\usepackage{graphicx,afterpage}
\usepackage{xcolor}
\usepackage{ulem}
\usepackage[colorlinks=true,plainpages=false,linkcolor=blue,urlcolor=blue,citecolor=blue,pdfpagemode=UseNone,pdfstartview=FitBH]
{hyperref}

\begin{document}
\title{Competition between spin ordering and superconductivity near the pseudogap boundary in ${\mathbf{La}}_{\mathbf{2\ensuremath{-}x}}{\mathbf{Sr}}_{\mathbf{x}}{\mathbf{CuO}}_{\bf{4}}$: insights from NMR}

\author{I. Vinograd}
\email{grvngrd@gmail.com}
\affiliation{CNRS, LNCMI, Univ. Grenoble Alpes, INSA-T, UPS, EMFL, Grenoble, France}
\author{R. Zhou}
\affiliation{CNRS, LNCMI, Univ. Grenoble Alpes, INSA-T, UPS, EMFL, Grenoble, France}
\author{H. Mayaffre}
\affiliation{CNRS, LNCMI, Univ. Grenoble Alpes, INSA-T, UPS, EMFL, Grenoble, France}
\author{S. Kr\"amer}
\affiliation{CNRS, LNCMI, Univ. Grenoble Alpes, INSA-T, UPS, EMFL, Grenoble, France}

\author{S. K. Ramakrishna}
\affiliation{National High Magnetic Field Laboratory, Florida State University, Tallahassee, FL 32310, USA}
\author{A. P. Reyes}
\affiliation{National High Magnetic Field Laboratory, Florida State University, Tallahassee, FL 32310, USA}

\author{T.~Kurosawa}
\affiliation{Department of Physics, Hokkaido University, Sapporo 060-0810, Japan}

\author{N. Momono}
\affiliation{Muroran Institute of Technology, Muroran 050-8585, Japan}

\author{M. Oda}
\affiliation{Department of Physics, Hokkaido University, Sapporo 060-0810, Japan}

\author{S. Komiya}
\affiliation{Central Research Institute of Electric Power Industry, Yokosuka, 240-0196, Japan}

\author{S. Ono}
\affiliation{Central Research Institute of Electric Power Industry, Yokosuka, 240-0196, Japan}

\author{M. Horio}
\affiliation{Department of Physics, University of Z\"{u}rich, CH-8057 Zurich, Switzerland}

\author{J. Chang}
\affiliation{Department of Physics, University of Z\"{u}rich, CH-8057 Zurich, Switzerland}

\author{M.-H. Julien}
\email{marc-henri.julien@lncmi.cnrs.fr}
\affiliation{CNRS, LNCMI, Univ. Grenoble Alpes, INSA-T, UPS, EMFL, Grenoble, France}
\date{\today}

\begin{abstract}

When superconductivity is suppressed by high magnetic fields in ${\mathrm{La}}_{\mathrm{2\ensuremath{-}x}}{\mathrm{Sr}}_{\mathrm{x}}{\mathrm{CuO}}_{4}$, striped antiferromagnetic (AFM) order becomes the magnetic ground state of the entire pseudogap regime, up to its end at the doping $p^*$ [M. Frachet, I. Vinograd \textit{et al.}, Nat. Phys. 16, 1064 (2020)]. Glass-like freezing of this state is detected in $^{139}$La NMR measurements of the spin-lattice relaxation rate $T^{-1}_{1}$. Here, we present a quantitative analysis of $T^{-1}_{1}$ data in the hole-doping range $p=x=0.12-0.171$, based on the Bloembergen-Purcell-Pound (BPP) theory, modified to include statistical distribution of parameters arising from strong spatial inhomogeneity. We observe spin fluctuations to slow down at temperatures $T$ near the onset of static charge order and, overall, the effect of the field $B$ may be seen as equivalent to strengthening stripe order by approaching $p=0.12$ doping. In details however, our analysis reveals significant departure from usual field-induced magnetic transitions. The continuous growth of the amplitude of the fluctuating moment with increasing $B$ suggests a nearly-critical state in the $B\rightarrow 0$ limit, with very weak quasi-static moments possibly confined in small areas like vortex cores. Further, the nucleation of spin order in the vortex cores is shown to account quantitatively for both the value and the $p$ dependence of a field scale characterizing bulk spin freezing. The correlation time of the fluctuating moment appears to depend exponentially on $B/T$ (over the investigated range). This explains the timescale dependence of various experimental manifestations, including why, for transport measurements, the AFM moments may be considered static over a considerable range of $B$ and $T$. These results make the high-field magnetic ground state up to $p^*$ an integral part of the discussion on putative quantum criticality.

\end{abstract}
\maketitle

\section{Introduction}

There are various reasons for which high $T_c$ superconductivity in the cuprates is a hard problem. It is now clear that one of these reasons is that the superconducting state impedes the ordering of spin or charge degrees of freedom. Because of these competing effects, ordered phases may thus remain partially or entirely "hidden", which hampers full understanding of the cuprate electronic properties. Nevertheless, tremendous progress has been accomplished in the last two decades as experiments using magnetic fields to quench superconductivity have played a pivotal role in exposing the spin and/or charge orders that compete with superconductivity~\cite{Lake2002,Hoffman2002,Mitrovic2003,Wu2011}. 

   \begin{figure}[b!]
  \includegraphics[width=8.5cm]{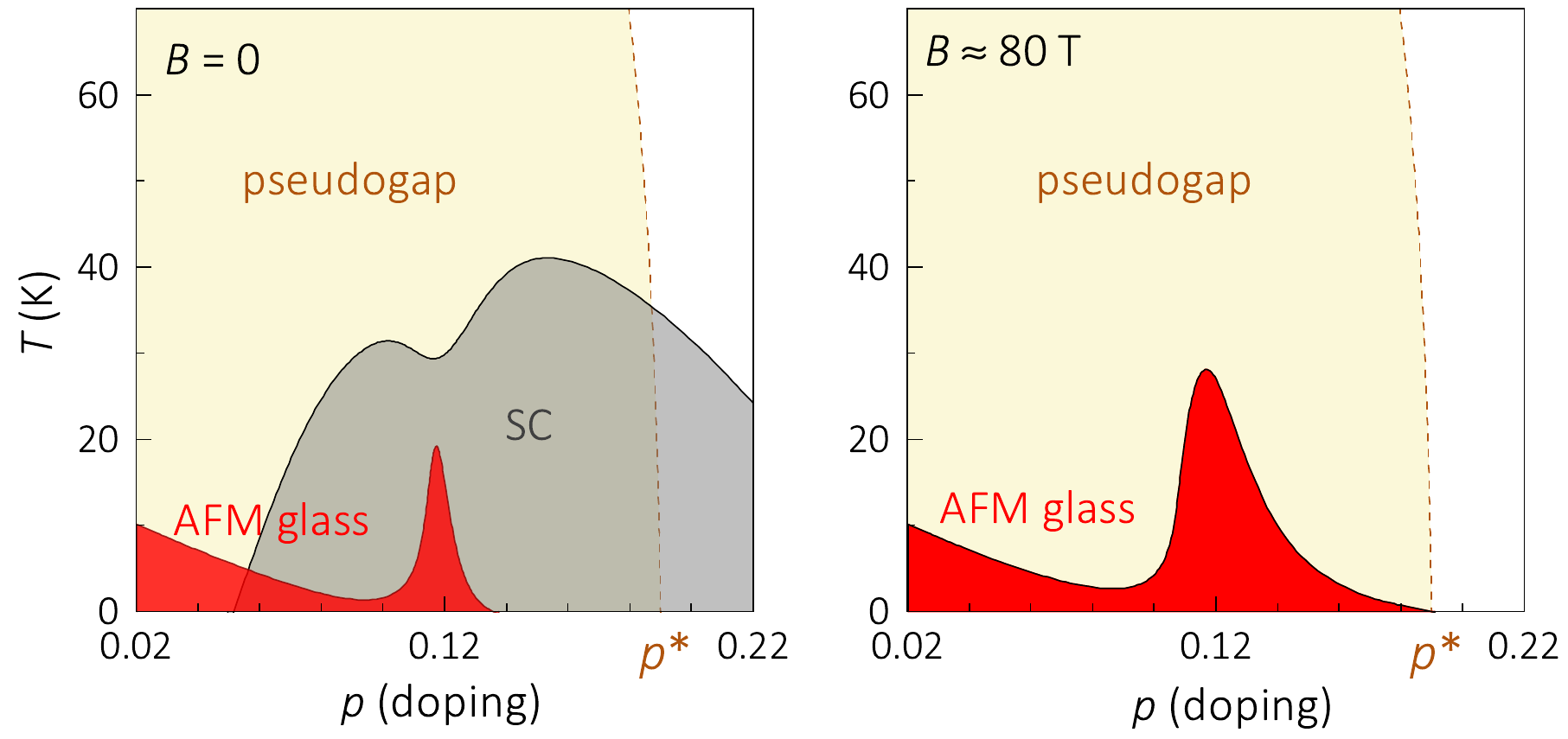}
  \caption{\label{Fig_intro} Schematic phase diagram of LSCO showing that the pseudogap and antiferromagnetic (AFM) glass phases have separate doping endpoints in zero field ($B=0$, left) but coinciding endpoints in high fields (right), according to the NMR and ultrasound results in ref.~\cite{Frachet2020}.}
\end{figure} 
A recent study combining nuclear magnetic resonance (NMR) and sound velocity measurements in ${\mathrm{La}}_{\mathrm{2\ensuremath{-}x}}{\mathrm{Sr}}_{\mathrm{x}}{\mathrm{CuO}}_{4}$ (LSCO)~\citep{Frachet2020} has provided the latest illustration of how high fields can uncover a hidden piece of the cuprate puzzle. This work revealed a connection between magnetism and the pseudogap phase that had been hitherto hidden by superconductivity: when superconductivity is removed by high fields, the striped antiferromagnetic (AFM) glass  (sometimes referred to as spin-glass or spin-stripe phase) persists well above its end-doping in zero field $p_{\rm sg}\simeq 0.135$ (Fig.~\ref{Fig_intro}a), actually up to the endpoint of the pseudogap phase, $p^*\simeq 0.19$ (Fig.~\ref{Fig_intro}b). Note that in ${\mathrm{La}}_{\mathrm{2\ensuremath{-}x}}{\mathrm{Sr}}_{\mathrm{x}}{\mathrm{CuO}}_{4}$ the hole doping $p$ is considered to be equal to the Sr concentration $x$.

At the qualitative level, that the same frozen state of AFM moments~\cite{Julien2003} extends from the doped Mott insulating state at $p\simeq 0.02$ up to $p^*$ underlines the relevance of Mott physics throughout the pseudogap state, even suggesting a possible connection between local-moment magnetism of the doped Mott insulator and the pseudogap state. This work, however, raises a number of important questions. How is the magnetic quantum phase transition connected with the sharp changes in the electronic properties observed across $p^*$ in high-field measurements~\cite{Collignon2017,Michon2019,Fang2022}? How do we describe the zero-field ground state? Is it sharply distinct from the high-field (AFM glass) ground state? Could the slow spin fluctuations have any impact on the transport properties? 

The purpose of this paper is to perform a quantitative analysis of the NMR results of ref.~\cite{Frachet2020} in order to gain insight into these questions. A central issue in the analysis will be the presence of strong spatial inhomogeneity of various origins, as discussed below. 

The paper is organized as follows: section II describes the fitting model, section III discusses the results and their interpretation. The readers who are not interested in the details may go directly to section IV, an extended summary of the main points discussed in the previous section. Perspectives are mentioned in section V.

\section{NMR background and model}     

\subsection{Why $T_1$ measurements?}

The most direct information on magnetic order is the magnitude of the ordered moment, which is in principle extracted from the broadening or splitting of the NMR lines. In La-based cuprates (La214 in short), internal fields $\langle h_\perp \rangle$ of $\sim 10 - 40$~mT produced by ordered moments up to 0.3~$\mu_B$ within the CuO$_2$ planes have been successfully detected from measurements in zero external field~\cite{Kitaoka1987,Ohsugi1996} or with the field applied parallel to the planes~\cite{Arsenault2018}.
 
In the specific case of field-induced order, however, measurements of NMR spectra cannot provide information on the ordered moment. Indeed, because the ordered moments lie within the planes, such measurement require in-plane fields ($B\parallel ab$) while the existence of the field-induced order itself requires out-of-plane fields ($B\parallel c$)~\cite{Wu2013,Frachet2020}. As a matter of fact, the $^{139}$La linewidth broadens only slightly upon cooling for $B\parallel c$ and saturates when spins freeze (see Fig.~\ref{linewidth} for an example).  Therefore, magnetic ordering will be detected here through the low-energy spin fluctuations, which are probed by measurements of the spin-lattice relaxation rate $T_1^{-1}$ of $^{139}$La nuclei.

 \subsection{The standard BPP model}
 
Our NMR experiments probe properties of the hyperfine field $\boldsymbol{h} \propto \bar{A}\,  \boldsymbol{S} $ produced at the $^{139}$La nuclear positions by the electronic spins  $\boldsymbol{S}$ in CuO$_2$ planes ($\bar{A}$ is the hyperfine coupling tensor). In their original model~\cite{BPP1948}, Bloembergen, Purcell and Pound assume an auto-correlation function of the fluctuating hyperfine field $\boldsymbol{h}(t)$ that decays exponentially with time ($t$):
\begin{equation}
\label{eq:hperp}
\langle \boldsymbol{h}(t) \boldsymbol{h}(0)\rangle = \langle h^2\rangle e^{-t/\tau_c} ,
\end{equation}
where $\tau_c$ is called the correlation time.

Fourier transformation of this expression and evaluation of the resulting spectral density of fluctuations at the NMR frequency $\omega_{L}$ leads to the following expression of the relaxation rate $T^{-1}_{1}$:
\begin{equation}\label{eq:BPP}
T^{-1}_{1,\, \rm BPP} =  \langle h_{\bot}^2 \rangle \, \gamma_{n}^{2}\frac{2\tau_{c}}{1+(\omega_{L}\tau_{c})^2} .
\end{equation}
with the nuclear gyromagnetic ratio $\gamma_{n}$ 
and the angular Larmor frequency $\omega_{L}=2\pi f$, $f$ being the actual resonance frequency. 

Here, $\langle h_{\bot}^2 \rangle= \langle h_{xx}^2 \rangle + \langle h_{yy}^2 \rangle $ is the time-averaged, squared hyperfine field (also called "fluctuating field") transverse to the direction ($z$) of the applied field ($B$). For a diagonal hyperfine tensor $\bar{A}$, $T_1^{-1}$ is thus related to the correlation function $\langle S_+(t)S_-(0)\rangle$.
 
Eq.~\ref{eq:BPP} has a maximum when $\tau_c=\omega_L^{-1}$, that is, when electronic fluctuations are as slow as the NMR frequency. Therefore, if electronic fluctuations slow down so much upon cooling that they eventually become slower than the NMR frequency, there must be a temperature $T_{\rm peak}$ at which $\omega_{L}=\frac{1}{\tau_{c}}$. Then, $T_1^{-1}$ reaches a maximum value:
\begin{equation}\label{eq:max T1}
(T_1^{-1})_{\rm max} =  \frac{\langle h_{\bot}^2 \rangle\,\gamma_{n}^{2}}{\omega_{L}} \quad .
\end{equation} 
Since $\omega_L \propto B$ in NMR and since $B$ does not affect the physical properties in general, $(T_1^{-1})_{\rm max}$ usually varies as $B^{-1}$. As we shall see, this is no longer true in superconducting cuprates as $\langle h_{\bot}^2 \rangle$ strongly increases with $B$.

\subsection{Basic assumptions}

We follow previous studies of spin-freezing in cuprates~\cite{Suh2000,Curro2000,Simovic2003,Mitrovic2008,Wu2013} that have shown that the peak in $T_1^{-1}$ could be reproduced using the BPP formula (Eq.~\ref{eq:BPP}) in which all of the temperature ($T$) dependence arises from a diverging correlation time $\tau_{c}$ upon cooling:
\begin{equation}\label{eq:tau}
\tau_{c}(T) = \tau_{\infty}\,e^{E_{0}/k_B T}\, .
\end{equation}
The activation energy $E_0$ for spin fluctuations is assumed to be $T$ (but not $B$) independent over the full fitting range (and so is $\langle h_{\bot}^2 \rangle$ in Eq.~\ref{eq:BPP}). Its physical interpretation in terms of spin stiffness will be discussed in section IV. The exponential $\tau_c$ has no onset temperature: Eq.~\ref{eq:tau} describes fluctuations that continuously slow down upon cooling, eventually becoming frozen ({\it i.e.} static) at $T=0$ without any phase transition intervening at finite $T$. In section~\ref{survey}, we also briefly discuss fitting to a Vogel-Fulcher-Tammann dependence:
\begin{equation}
\label{eq:VF}
\tau_c(T) = \tau_{\infty}\,e^{E_{0}/k_B (T-T_{\rm VF})}.
\end{equation}

Experimentally, however, the spin system appears to be frozen as soon as fluctuations become slower than the time scale of the  technique. The larger the time scale, the higher the temperature of the apparent freezing. The BPP peak temperature $T_{\rm peak}$, namely the freezing temperature at the NMR timescale, is readily obtained by inserting Eq.~\ref{eq:tau} into the condition $\omega_{L}=\tau_{c}^{-1}$ and $\omega_L= \gamma_n \, B$ to first approximation (neglecting the Knight shift and the quadrupole shift that are small corrections to $\omega_L$):
\begin{equation}\label{eq:peak temperature}
T_{\rm peak} = \frac{-E_{0}}{\ln(\omega_L \tau_{\infty})}\, \simeq \frac{-E_{0}}{\ln(\gamma_n \, \tau_{\infty} \, B)}\,.
\end{equation}

Notice that we shall evaluate $E_0$ in Kelvin and $\omega = \tau_c^{-1}$ in meV, thus implicitly setting $k_B=1$ and $\hbar=1$.

 \subsection{The spatial-inhomogeneity problem}

Quantitative analysis of the results is a daunting task as multiple levels of spatial inhomogeneity make the problem very intricate and they strongly affect the results for a local probe such as NMR. Moreover, we are not aware of a model that takes into account the full complexity of this problem. In the superconducting mixed state, electronic properties are intrinsically inhomogeneous at relatively short length scales as the competing magnetic order is thought to be enhanced in and around the vortex cores~\cite{Arovas1997,Demler2001,Sachdev2002,Hu2002,Kivelson2002,Franz2002,Zhu2002,Ghosal2002,Andersen2003}. In addition to this, there is evidence that spatial heterogeneity characterizes the spin-freezing process, regardless of the presence or absence of superconductivity~\cite{Julien1999,Suh2000,Curro2000,Teitelbaum2000,Julien2001,Hunt2001,Simovic2003,Mitrovic2008,Baek2015,Baek2017,Arsenault2020,Singer2020,Wu2013}. A third source of inhomogeneity lies in the possible phase separation between magnetic and non-magnetic regions, as suggested by neutron scattering studies~\cite{Wakimoto2007}. In principle, there is also significant spatial inhomogeneity of the hole doping in LSCO~\cite{Singer2002}. Nevertheless, this appears to have little influence on measurements of the spin-lattice relaxation rate $T_1^{-1}$ of $^{139}$La~\cite{Mitrovic2008,Arsenault2020}.

Direct information on inhomogeneity is encoded in the probability density of relaxation rates $T_1^{-1}$. This quantity can in principle be determined without any assumption on the distribution by performing an inverse Laplace transform of the NMR relaxation curve (the time evolution of the nuclear magnetization returning back to its equilibrium value). This technique has been put forward recently in the cuprate context~\cite{Singer2020,Arsenault2020}. However, it requires levels of signal-to-noise ratio that are difficult to achieve in time-constrained experiments at high-field facilities and its application is also not straightforward in general~\cite{Choi2021}. Furthermore, while  inverse Laplace transform gives the full probability density of relaxation rates, its result still needs to be subsequently interpreted within a model in order to relate $T_1$ to physical parameters and to characterize the evolution of these parameters with $B, p$ or $T$. This has not been attempted yet. 

Here we use a simpler approach first developed by Curro~\textit{et al.}~\cite{Curro2000,Suh2000,Curro2004}: the relaxation curves are fitted by stretched exponentials (a standard way to determine a "typical" relaxation rate~\cite{Mitrovic2008,Johnston2006}) and the temperature dependence of the thus-determined $T^{-1}_{1}$ values is then fitted by a modified BPP model in which the characteristic energy $E_0$ is distributed, e.g. by a Gaussian. We generalize this approach by introducing ad-hoc distributions of all relevant fitting parameters to account for the inhomogeneous relaxation mechanism in LSCO. 

\subsection{Statistical distribution of parameters \label{distribution}}

NMR studies of spin-freezing in superconducting cuprates~\citep{Cho1992,Julien1999,Julien2000,Suh2000,Curro2000,Teitelbaum2000,Hunt2001,Julien2001,Mitrovic2008,Wu2013,Baek2015,Baek2017,Arsenault2018,Frachet2020} generally find that the relaxation curves are stretched which indicates that there is a distribution of relaxation rates (in the absence of any significant quadrupolar relaxation mechanism, which is true at low $T$). In this situation, a standard procedure is to fit the relaxation curves with a stretched exponential form, which yields a "typical" relaxation rate $T_1^{-1}$, equivalent to the median of the distribution, and a stretching exponent $\beta$ that quantifies the breadth of the distribution ~\citep{Mitrovic2008,Baek2017,Arsenault2020}. Indeed, as shown by Johnston~\cite{Johnston2006}, for $0.5\leq \beta\leq 1$, $1-\beta$ is proportional to the {\it logarithmic} full width at half maximum (FWHM) of the distribution ($\beta=0.5$ corresponding to a distribution over an order of magnitude and $\beta=1$ to a Dirac function) and the value of $T_1^{-1}$ is within about 10\% of the median of the distribution ({\it i.e.} the value that splits the cumulative probability into equal halves, see Fig.~\ref{distributions} for an example). For $\beta\simeq 0.3$, $T_1^{-1}$ is distributed over two orders of magnitude and the value of $T_1^{-1}$ becomes about twice the median.

In order to fit the $T$ dependence of $T_1$, we convolute the BPP formula (Eqs.~\ref{eq:BPP} and ~\ref{eq:tau}), with positive-definite Gaussian distributions of $E_0$ and $\tau_{\infty}$ (for convenience we define $a=\ln\tau_{\infty}$ and distribute $a$) as follows:

   \begin{figure*}[t!]
\hspace*{-0.3cm}    
  \includegraphics[width=17cm]{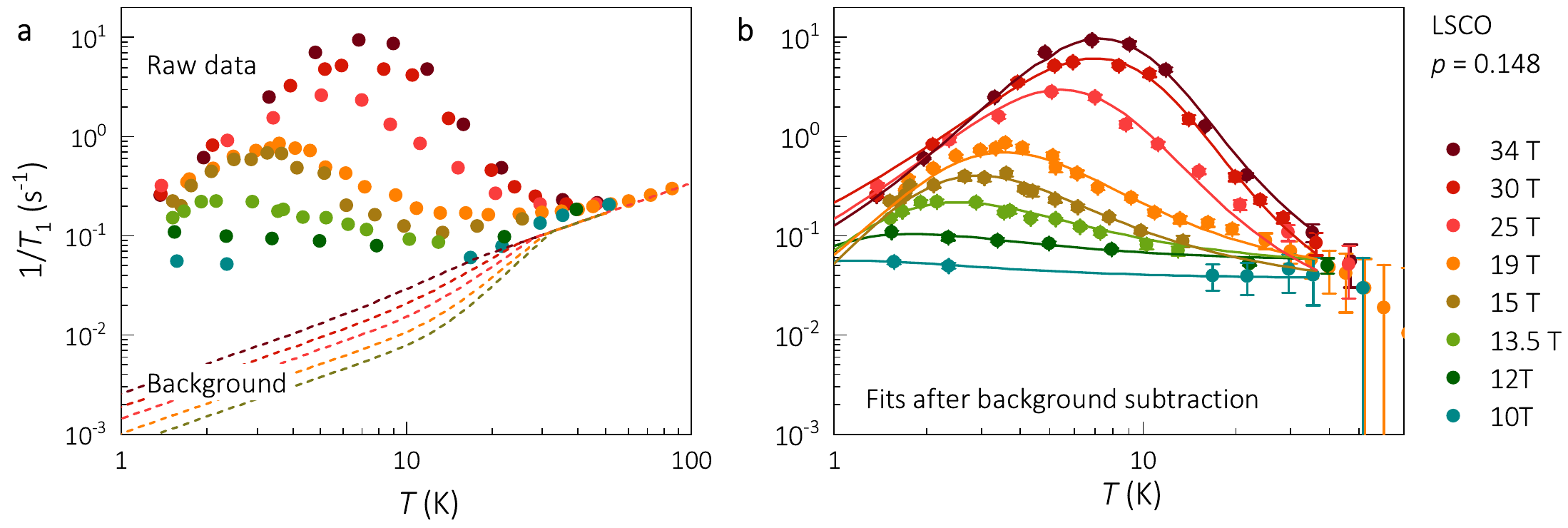}
  \caption{\label{T1} Relaxation data and fits for LSCO $p=0.148$. a) Raw $T^{\,-1}_{1}$ from stretched fits \textit{vs.} $T$ at different fields. For $B=10$~T,  $T^{\,-1}_{1}(T)$ could not be measured at intermediate temperature ($T\sim 7$~K), because the signal becomes very weak due to strong superconductivity which impedes the penetration of radio frequency pulses. However, some signal can be recovered at lower $T$ since the nuclear magnetization is proportional to $1/T$. Lines are calculated background relaxation rates based on linear-in-$T$ $T_1^{-1}$ (constant $(T_{1}T)^{-1}$) above $T_{\rm{c}}$ and exponentially gapped relaxation below $T_{\rm{c}}(B)$ (see Appendix and ref.~\cite{Frachet2020}). b) $T^{-1}_{1,\, \rm subtr.}$ after background subtraction fitted by $T^{-1}_{1,\, \rm BPP\, dist.}(T)$ (continuous lines). The subtraction affects only the high $T$ range. For details about the fits, see text and Fig.~\ref{parameters14p8}.}
\end{figure*} 

\begin{widetext}
\begin{equation}\label{eq:T1BPPfull}
\begin{split}
 T^{-1}_{1,\, \rm BPP\, dist.}(T)&=\frac{1}{\mathcal{N}}\int_{0}^{\infty}\int_{-\infty}^{\infty} T^{-1}_{1,\rm BPP} (E_0,a,T)\cdot e^{-\frac{(E_0-E_{0,c})^2}{2\mathit{\Delta} \!E_0^{2}}} e^{-\frac{(a - a_c)^2}{2\mathit{\Delta} a^{2}}}{\rm d}a \,{\rm d}E_0 \\
& =\frac{1}{\mathcal{N}}\int_{0}^{\infty}\int_{-\infty}^{\infty} \langle h_{\bot}^2 \rangle^{2}\gamma_{n}^{2}\frac{2\tau_{c}(E_0,a)}{1+(\omega_{L}\tau_{c}(E_0,a))^2}\cdot e^{-\frac{(E_0-E_{0,c})^2}{2\mathit{\Delta} \!E_0^{2}}} e^{-\frac{(a - a_c)^2}{2\mathit{\Delta} a^{2}}} {\rm d}a \,{\rm d}E_0\\
& =\frac{1}{\mathcal{N}}\int_{0}^{\infty}\int_{-\infty}^{\infty} \langle h_{\bot}^2 \rangle^{2}\gamma_{n}^{2}\frac{2 e^{a+\frac{E_0}{T}}}{1+(\omega_{L}e^{a+\frac{E_0}{T}})^2}\cdot e^{-\frac{(E_0-E_{0,c})^2}{2\mathit{\Delta} \!E_0^{2}}} e^{-\frac{(a - a_c)^2}{2\mathit{\Delta} a^{2}}} {\rm d}a \,{\rm d}E_0 \, .
\end{split}
\end{equation}
\end{widetext}

Here $\mathcal{N}$ is the normalisation \[\mathcal{N} = \int_{0}^{\infty}\int_{-\infty}^{\infty} e^{-\frac{(E_0-E_{0,c})^2}{2\mathit{\Delta}\!E_0^{2}}}e^{-\frac{(a - a_c)^2}{2\mathit{\Delta} a^{2}}} {\rm d}a \, {\rm d} E_0 \,.\]

Notice that, in the following, on $E_{0,c}$ and $a_c$ we drop the index $c$ (that indicates the center of the distribution) in order to simplify the notations, so the fit parameters $E_0$ and $a$ will actually refer to the center of the distribution of $E_0$ and $a$, respectively.

Introducing the distribution of $\tau_{\infty}$ improves the fit quality significantly at $T$ above the peak (see Appendix, Fig.~\ref{delta_a}). What $\Delta a$ essentially does, and which $\Delta E_0$ does not, is to create a constant $T_1^{-1}$ where $E_0$ is small. The value $a=-31.5$ (with $\tau_{\infty}$ measured in seconds), corresponding to $\tau_{\infty}=0.02$~ps, was found to fit data well for different La-compounds~\citep{Curro2000,Suh2000}. We thus keep this value fixed. Integrating $a$ over the full numerical range is computationally demanding, but limiting the integration range to $a - 3 \Delta a\leq a \leq a + 3 \Delta a$ leads to equivalent results when $\Delta a > 1$ . When $\Delta a \leq 1$, we keep a fixed $\Delta a$-independent integration range of $\pm 3$. 

We point out that distributing $\langle h_{\bot}^2 \rangle$ would have no effect on  $T^{-1}_{1,\, \rm BPP\, dist.}(T)$ because the integral over $\langle h_{\bot}^2 \rangle$ can be factorized out and simply gives $\langle h_{\bot}^2 \rangle=h_{\bot, \rm{mean}}^2$, as long as this distribution is uncorrelated with those of $E_0$ and $\tau_{\infty}$. Nevertheless, this does not imply that the distribution of $T^{-1}_{1}$ would be unaffected if $\langle h_{\bot}^2 \rangle$ were to be distributed. In fact, the width of the $T^{-1}_{1}$ distribution, of which the stretching exponent $\beta$ is a measure, would very well increase (see Appendix).

\begin{figure*}[t!]
\hspace*{-0.3cm}
\includegraphics[width=18cm]{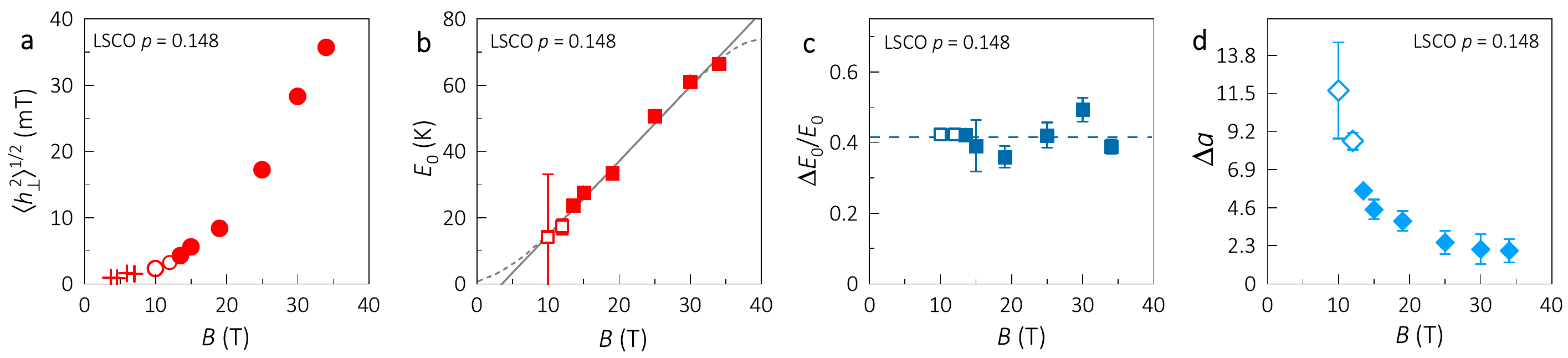}
\caption{\label{parameters14p8} Field and temperature dependence of fittting parameters for LSCO $p=0.148$. (a) Root mean squared fluctuating hyperfine field $\langle h_{\bot}^2 \rangle^{1/2}$, (b) activation energy $E_0$, (c) relative width of the distribution of $E_0$ and (d) width of the distribution of $a=\ln \tau_\infty$, all as a function of magnetic field. For $B\geq 13.5$~T, all fits parameters are free and unconstrained (solid symbols). For $B\leq 12$~T, $T_1^{-1}(T)$ is flat (Fig.~\ref{T1}), so the peak temperature is ill-defined and the constraint $\Delta \!E_0 / E_0 = 0.42$ has been used (open symbols), as determined from a fit of higher-field data to a constant (dashed line in panel c). The crosses in (a) where determined from a single value of $T_1$ at base temperature (see text). The solid line in (b) is a linear fit through the data while the dashes represent another possible dependence, also consistent with the data. The distribution of $\tau_{\infty}$ increases by an order of magnitude as $\Delta a$ increases by a factor of $2.3 \approx \ln10$. At 10~T, $\tau_\infty$ is spread over 5 orders of magnitude. At the quantitative level, $E_0$ appears to depend approximately linearly on $B$ while $\langle h_{\bot}^2 \rangle^{1/2}$ varies more strongly ($\propto B^{5/2}$).}
\end{figure*}

\subsection{Analysis in the absence of a peak in $T_1^{-1}$}

At low fields ($B \lesssim 10$~T), it is impossible to determine the BPP parameters from regular fits as no $T_1^{-1}$ peak is discernible in the data. As explained in Appendix (Fig.~\ref{delta_a}b), a single $T_{1}^{-1}$ value at low $T$ is sufficient to determine $\langle h_{\bot}^2 \rangle$ because $T_1^{-1}$ becomes effectively $T$ independent as $\Delta a$ values become large. In doing so, we of course implicitly assume that freezing described by the BPP model still applies in this field range, which is justified by the fact that the $T_{1}^{-1}$ values are still considerably larger than the estimated background (Fig.~\ref{T1}a).

\subsection{Assumptions, limitations, caveats}

$\bullet$ Following Curro and coworkers~\cite{Curro2000,Suh2000,Curro2004}, the present analysis posits that the width of the peak in $T_1^{-1}(T)$ arises from a distribution of $E_0$ values, or equivalently from a distribution of $T_{\rm peak}$ values (Eq.~\ref{eq:peak temperature}), and that $\tau_c$ varies exponentially with $T$. 

$\bullet$  Our approach assumes unimodal distributions of the parameters. This could be an oversimplification if the distribution is bimodal or multimodal, for example if the sample separates into regions that either undergo spin freezing or do not show it at all. Accurate analysis of $T_1^{-1}$ data in La$_{1.885}$Sr$_{0.115}$CuO$_4$ and La$_{1.875}$Ba$_{0.125}$CuO$_4$ has actually uncovered a bimodal distribution of the relaxation rate. However, both modes show qualitatively similar behavior, namely slowing down of spin fluctuations~\cite{Arsenault2020,Singer2020}, so unimodal distribution is still a reasonable approximation in that case. In fact, while our approach determines properties of the distribution indirectly and is in principle less powerful than the direct use of the Inverse Laplace Transform (ILT), its value lies in the capability to obtain a consistent parametrization of $^{139}$La datasets in LSCO~\cite{Frachet2020,Mitrovic2008,Frachet2021} and to characterize how the parameters evolve as a function of field and doping. 
Furthermore, we show in Appendix an example of a $T_1^{-1}$ distribution that is not unimodal even though the distributions of BPP parameters are unimodal.

$\bullet$  We are fitting $T^{-1}_{1}$ data from stretched fits, that correspond to the median of the $T^{-1}_{1}$ distribution, by an expression $T^{-1}_{1,\, \rm BPP\, dist.}(T)$ (Eq.~\ref{eq:T1BPPfull}) that gives the mean $T^{-1}_{1}$. While this may in principle be incorrect because the mean and the median may be very different, we explain in Appendix that our model is justified by the fact that $T^{-1}_{1,\, \rm mean} \approx T^{-1}_{1,\, \rm median}$ for a realistic distribution of relaxation rates that cannot have a tail extending to infinity. 

\begin{figure*}[t!]
  \includegraphics[width=16cm]{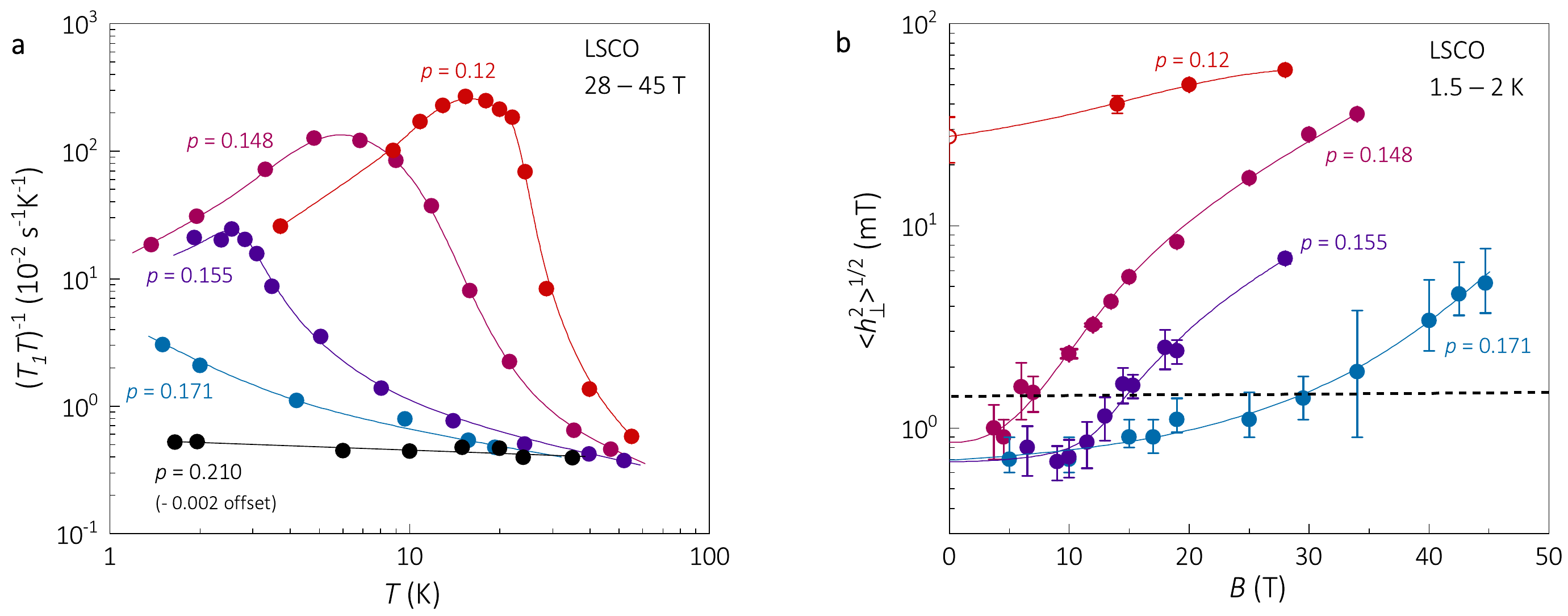}
  \caption{\label{hperp} Doping evolution of magnetism in high fields. a) $T$ dependence of the relaxation rate $T_{1}^{-1}$ divided by $T$, at different hole doping levels. The data for $p=0.21$ has been lowered by 0.002~s$^{-1}$K$^{-1}$ corresponding to the approximated contribution from electric field gradient fluctuations~\cite{Frachet2020}. Field values for each doping are: 28~T ($p=0.12$), 34~T ($p=0.148$), 28~T ($p=0.155$), 44.7~T ($p=0.171$) and 40 and 45~T ($p=0.21$). Complete datasets and fits of doping levels other than $p=0.148$ are presented in Appendix, Fig.~\ref{allfits}). b) Root mean squared fluctuating hyperfine field $\langle h_\bot \rangle^{1/2}$ determined either from from fits to $T^{-1}_{1,\, \rm BPP\, dist.}(T)$ for fields at which a peak is visible in $T_{1}^{-1}(T)$ (\textit{i.e.} $B\geq 12$~T for $p=0.148$, $B\geq 19$~T for $p=0.155$ and all fields for $p=0.12$) or from a single $T_{1}^{-1}(B)$ value at low $T$ otherwise (see text). In this latter case, temperatures are in the range $1.5 - 2$~K, depending on sample. For $p=0.12$, the value of $\langle h_\bot^2 \rangle^{1/2}$ at $B=0$ (open symbol) is taken from a fit of $T_{1}^{-1}$ peak measured at 20~T with the field oriented in the $\rm{CuO}_2$ plane, \textit{i.e.} a situation for which the field weakly affects superconductivity. The horizontal dashed line marks the 1.5~mT value used as a criterion to define the onset of quasi-static spin fluctuations (see text).}

\end{figure*}  

$\bullet$ We fix $\tau_\infty=0.02$~ps for all dopings but add a distribution of $\tau_\infty$. Good BPP fits would be possible without this distribution, provided $\tau_\infty$ is a free, {\it i.e.} field and material dependent parameter~\cite{Hammerath2013}. However, $\tau_{\infty}$ is the correlation time at $T\gg T_{\rm peak}$, so it is unphysical to assume that it is strongly field dependent. $\tau_\infty$ is strongly cross-correlated with other fitting parameters such as $\langle h_{\bot}^2 \rangle$ and $E_0$, so fixing $\tau_\infty$ helps to identify the intrinsic $B$ dependence of these parameters across the investigated hole doping range. In principle, a field dependence of the distribution $\Delta a$ ($a= \ln \tau_\infty$) is also unphysical but we consider that this betrays an actual, but nonessential, shortcoming of the model (see discussion in section IV). 

$\bullet$ No correlation between the distributions of $E_0$, $\tau_{\infty}$ and $h_{\bot}$ is introduced.

$\bullet$ Part of the nuclear relaxation is due to processes other than the spin-freezing: there are other contributions to the spin fluctuation spectrum as well as electric-field gradient fluctuations~\cite{Frachet2020}. Since we are only interested in fluctuations from the local moments, these contributions should thus be subtracted out before fitting with the BPP model. We take a semi-phenomenological approach that assumes uncorrelated relaxation mechanisms, namely a background $T_1^{-1}$ on top of which the enhanced relaxation due to glassy magnetic ordering develops. As explained in ref.~\cite{Frachet2020} (supplementary information), this background is taken to be $T$ and $B$ independent above $T_c$ (in agreement with available data below $\sim$100~K) and with the $B$ and $T$ dependence expected for a superconductor below $T_c$. We thus fit the $T^{\,-1}_{1}$ from stretched fits after subtracting the background. It is important to remark that the field dependence of the assumed background makes a negligible difference for fields where a clear peak in the relaxation rate is visible, but in the following we subtract the background at all fields for consistency. 

$\bullet$ We reiterate that our analysis is an ad-hoc parametrization of the data that aims at effectively accounting for spatial inhomogeneity in a tractable model but that does not pretend to be a realistic representation of the actual physical parameters. Our goal with this "first-order approach" (a zeroth-order approach would assume no distribution at all) is to give an idea of the physical quantities that determine $T_1$, to gain physical insight into the glassy freezing and its competition with superconductivity as well as to motivate more sophisticated approaches.

\section{Results and interpretations} 

\subsection{Organization of the discussion}
Having defined the distributed BPP model we can now show that it captures the full $T$ and $B$ dependence of $T^{-1}_{1}$ (see Appendix for details about samples and experimental methods). We shall first provide a brief survey of the fitting results and then proceed with a more detailed discussion. 
The discussion will revolve around three main physical parameters that characterize glassy antiferromagnetism: the fluctuating field $\langle h_{\bot}^2 \rangle$ and the activation energy $E_0$ (both of which are $B$, but not $T$, dependent in our model) as well the correlation time $\tau_c$ that depends exponentially on $E_0$ (thus on $B$) and on $T$ (Eq.~\ref{eq:tau}). 

Since we are presenting data as a function of temperature ($T$), field ($B$) and hole doping ($p$), we have organized the discussion in the following way: we first discuss aspects that are more related to the $T$ dependence and then we discuss aspects that are more related to the $B$ dependence. Since our most comprehensive set of data is for the crystal with $p=0.148$ doping, we systematically describe results for this sample first, and then for other doping levels when relevant.

\subsection{Brief survey of the fit results \label{survey}}
Fig.~\ref{T1} shows $T^{\,-1}_{1}$ before and after the background subtraction, together with the fits described above for La$_{1.852}$Sr$_{0.148}$CuO$_4$. Fig.~\ref{parameters14p8} shows the field dependence of the fitting parameters. The results of the fits rationalize three characteristic behaviors of the data:

$\bullet$ The decreasing amplitude of the peak in $T^{\,-1}_{1}(T)$ upon decreasing field is rooted in the strong decrease of the fluctuating field $\langle h_{\bot}^2 \rangle$ (Fig.~\ref{parameters14p8}a). 

$\bullet$  The shift of the peak towards low temperatures upon decreasing field arises from both the term $-1/\ln(B)$ in Eq.~\ref{eq:peak temperature} (a property of the BPP model, irrespective of the exact physics at play) but also from the field dependence of $E_0$ (Fig.~\ref{parameters14p8}b) that is specific of LSCO.

$\bullet$ The broadening of the peak upon decreasing field is accounted for by a rapid increase in the width of the distribution of $\tau_\infty$ (Fig.~\ref{parameters14p8}d) whereas the distribution of $E_0$ relative to $E_0$ is constant vs. field (Fig.~\ref{parameters14p8}c). Recall that $a=\ln \tau_\infty$ is a logarithmic quantity, so an order of magnitude change in $\Delta\tau_\infty$ increases or decreases $\Delta a$ by $\ln10^{\pm 1}= \pm 2.3$. At 10~T, $\tau_\infty$ is thus distributed over five orders of magnitude.

\begin{figure}[t!]
\hspace*{-0.3cm}
\includegraphics[width=7cm]{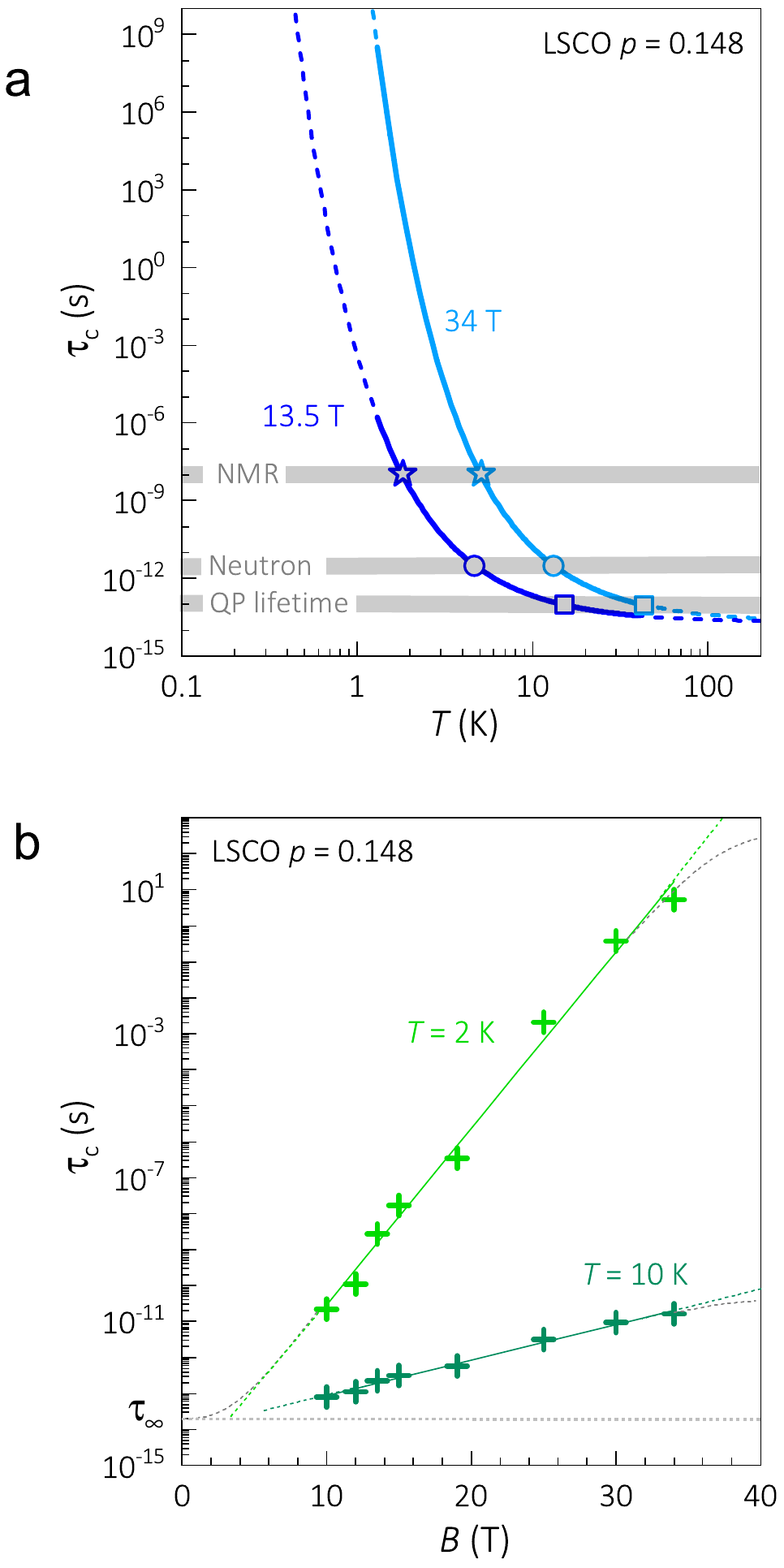}
\caption{\label{tau_c} (a) Correlation time $\tau_c$ {\it vs.} temperature in the $T$ range where the BPP fits are actually performed (continuous traces) and extrapolated over the whole $T$ range (dashes). The typical time scales of a $T_1$ experiment in NMR (for a $\sim$100~MHz resonance frequency) and of elastic neutron scattering (for an an energy resolution of $\sim$1~meV) are indicated. The quasiparticle (QP) lifetime of 0.1~ps is taken from ref.~\cite{Fang2022}. (b) $\tau_c$ {\it vs.} magnetic field at $T=2$~K. Lines are linear fits to the data. Dashes represent linear and nonlinear extrapolation of the data at the low- and high-field ends. The horizontal dotted line marks the value $\tau_\infty=0.02$~ps~(Eq.~\ref{eq:tau}) used in our analysis. In both (a) and (b), $\tau_c$ values are deduced from the $E_0$ values in Fig.~\ref{parameters14p8}b and Eq.~\ref{eq:tau}. }
\end{figure}

We point out that Vogel-Fulcher-Tammann fits (Eq.~\ref{eq:VF}) did not bring significant improvement: for fixed $a = \ln\tau_\infty$, the introduction of a distribution of $T_{\rm VF}$ values was completely unable to achieve the fitting goodness that $\Delta a$ does and $T_{\rm VF}$ values were negligibly small.

In order to visualize the weakening of slow spin fluctuations upon increasing hole doping, Fig.~\ref{hperp}a shows the $(T_1T)^{-1}$ data vs. $T$ at the highest-measured field for different doping levels (dividing $T_1^{-1}$ by $T$ helps to visualize the difference with the nearly-constant behavior of the $p=0.21$ sample). The steep peak in $T_1^{-1}$ {\it vs.}~$T$ for $p=0.12$ shifts to lower $T$, decreases in amplitude and broadens upon increasing $p$ and then entirely disappears for $p=0.21 > p^*$. The weakening of quasi-static magnetism upon decreasing $B$ or increasing $p$ is also reflected in the $B$ and $p$ dependence of $\langle h_{\bot}^2 \rangle^{1/2}$ shown in Fig.~\ref{hperp} b.

\subsection{Probe-frequency dependence}
The activated behavior of $\tau_c$ for $p=0.148$, as deduced from Eq.~\ref{eq:tau} and the fitted $E_0$ values (Fig.~\ref{parameters14p8}b), is shown in Fig.~\ref{tau_c}a at two different fields. $\tau_c$ is seen to cross the typical time scale of various experimental techniques at different temperatures. This means that the freezing temperatures $T_g$ depends on the time scale of the measurement. 

In neutron scattering (NS), the timescale is defined by the energy resolution, which is rather coarse in standard experiments ($\sim$meV). This results in a quasi-elastic, rather than purely elastic, signal over a substantial range of $T$ above $T_g^{\rm NMR}$~\cite{Haug2010}. In the data at 34~T for example, $T_g^{\rm NS} \sim$13~K whereas $T_g^{\rm NMR}=4$~K.

For $p=0.12$, $T_g^{\rm NMR}=10$~K while a similar analysis (not shown) gives $T_g\simeq23$~K at the neutron timescale. This is in reasonable agreement with the experimental value $T_g^{\rm NS}\simeq30$~K~\cite{Lake2002,Chang2008,Romer2013} that (fortuitously) coincides with the zero-field $T_c$. The slight difference between our model's prediction and the neutron onset temperature may result from $\tau_c(T)$ not exactly diverging exponentially and/or from a distribution of $T_{\rm peak}$ values (Eq.~\ref{eq:peak temperature}): indeed, the distribution does not contribute to the value predicted from the median $T_1^{-1}$ but it tends to increase the temperature onset of the signal detected in neutron scattering (or muon spin rotation) experiments. 

The probe-frequency dependence of the freezing temperature arises from the divergence of $\tau_c$ being slower than the critical slowing down at a second-order phase transition. Such a gradual slowing down is typical of glassy systems and glassiness may indeed be expected from quenched disorder induced by the dopant atoms and by lattice inhomogeneity~\cite{Bozin1999,Horibe2000,Pelc2021}. Disorder is likely to induce frustration in the spin system, especially as long as AFM order is intertwined with charge-stripe order, the latter being more sensitive to disorder than the former, as witnessed by its shorter correlation length. Furthermore, characteristic properties of spin glasses (irreversibility, scaling behavior and remanent magnetization) are observed in non-superconducting LSCO at low doping $p\simeq 0.04 - 0.05$~\cite{Chou1995,Wakimoto2000}.

\subsection{Physical interpretation of $E_0$}
An alternative explanation for the relatively slow divergence of $\tau_c$ is that it is an intrinsic consequence of the quasi two-dimensional, nearly isotropic (Heisenberg) nature of the spin system that orders only at $T=0$~\cite{Haug2010}. While this latter view appears to be at odds with three sets of experimental observations (traditional glass-like properties at low doping~\cite{Chou1995,Wakimoto2000}, evidence that disorder plays a role~\cite{Sasagawa2002,Mendels1994} and the three-dimensionality of spin correlations for $p\simeq0.12$~\cite{Lake2005,Romer2015}), it is interesting to note the similarity with the renormalized classical regime of the 2D Heisenberg AFM already pointed out in several NMR works~\cite{Cho1992,Julien1999,Julien2000,Curro2000,Teitelbaum2000,Julien2001,Hunt2001}: in the limit $T>T_{\rm peak}$, $\omega_L \tau_c \ll 1$ and thus Eq.~\ref{eq:BPP} reduces to $T_1^{-1} \propto \tau_c =\tau_\infty \exp{\frac{E_0}{k_BT}}$, which ressembles the expression derived by Chakravarty and Orbach~\citep{Chakravarty1990} for the renormalized classical regime:
\begin{equation}\label{eq:Renormalised classical}
\frac{1}{T_1} \propto \xi(T) \propto \exp(\frac{2\pi \rho_s}{T})\, ,
\end{equation}
where the spin stiffness $\rho_s$ is proportional to the nearest-neighbor AFM coupling $J$. Eq.~\ref{eq:Renormalised classical} is found to describe the undoped parent compound LCO very accurately~\citep{Birgeneau1995}. Even if the correlation length in optimally doped LSCO is probably limited at low temperature by quenched disorder and by competing effects from superconductivity, it is plausible that the dynamics of freezing moments retain characteristics of the renormalized classical regime, consistent with the idea that spin fluctuations are "nearly singular" above $T_c$~\cite{Aeppli1997}. The activation energy $E_0$ characterizing the low-frequency dynamics (Eq.~\ref{eq:tau}) may thus be interpreted as a measure of the spin stiffness.

\subsection{Slow spin fluctuations {\it vs.} CDW}

 \begin{figure}[t!]
\hspace*{-0.3cm}    
  \includegraphics[width=7.5cm]{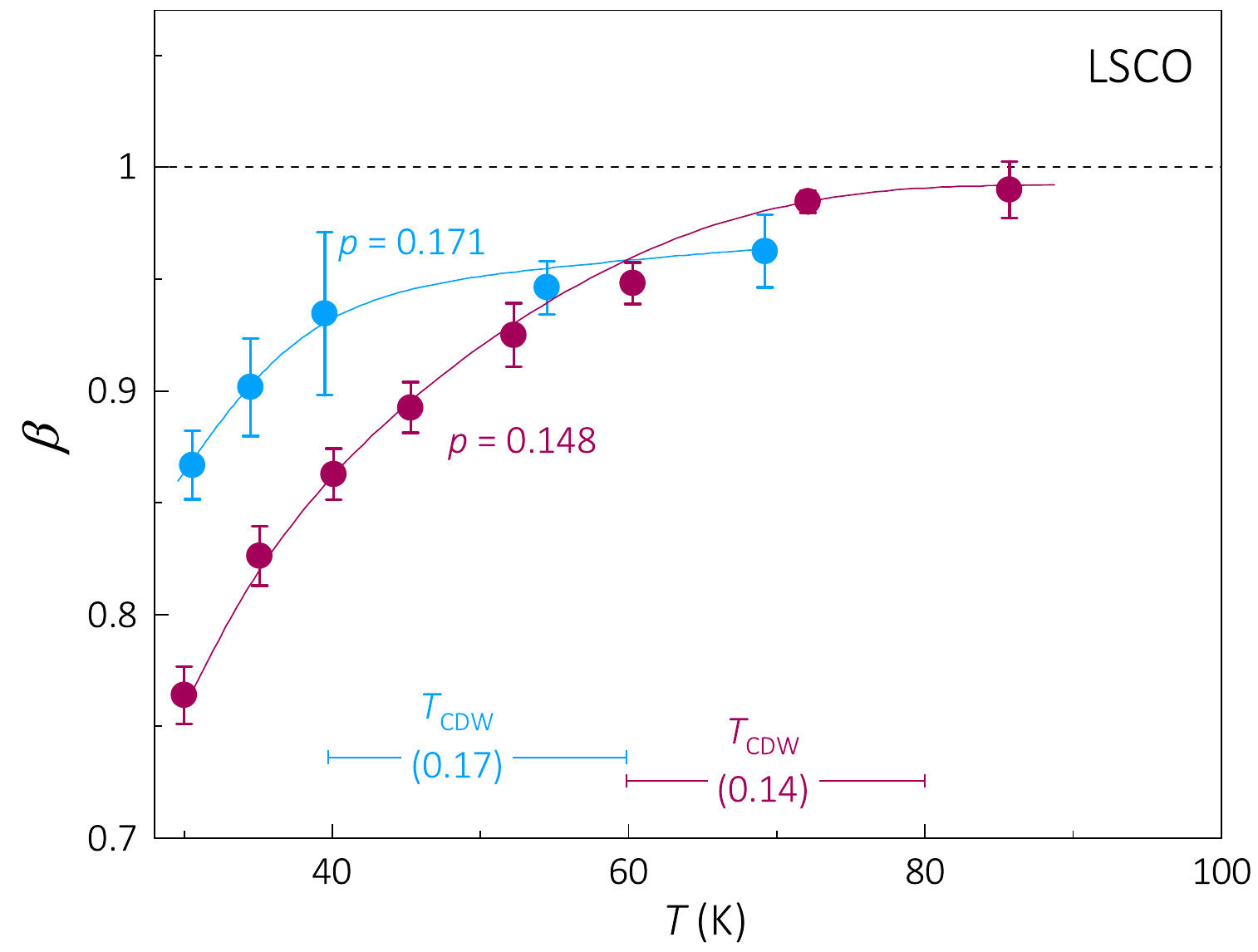}
  \caption{\label{beta} $T$ dependence of the stretching exponent $\beta$ above the zero-field $T_c$. The departure from $\beta=1$ signals spatial heterogeneity of $T_1$ values (due to slowing down of spin dynamics developing inhomogeneously across the sample) and appears to be mostly correlated with the presence of static CDW correlations detected in x-ray scattering. $T_{\rm CDW}$ data are from refs.~\cite{Wen2019,Miao2021}. For $p=0.171$, $\beta$ decreases again above 80~K (and thus does not reach 1) because of quadrupole fluctuations related to the structural transition at 141~K~\cite{Frachet2020}.}    
\end{figure}
 \begin{figure}[t!]
\hspace*{-0.3cm}    
  \includegraphics[width=7cm]{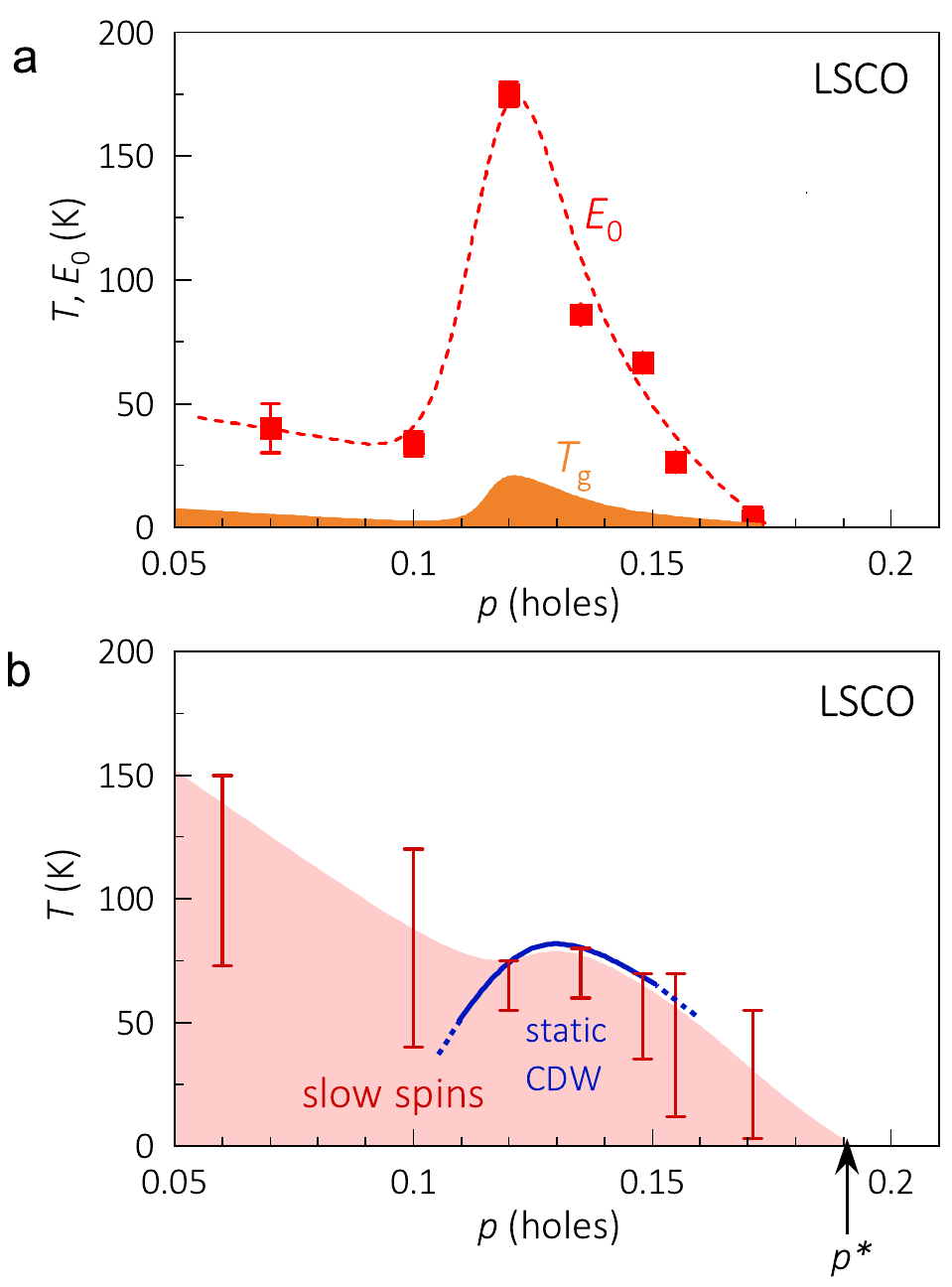}
  \caption{\label{phasediag} (a) Doping ($p$) dependence of the spin stiffness $E_0$ that tracks the doping dependence of the freezing temperature at the NMR timescale $T_g$ ($E_0$ and $T_g$ values are from our highest-field data, see text). (b) Doping dependence of the onset temperature of slow spin fluctuations (from our highest-field data) represented as vertical bars: the upper value corresponds to the deviation of $\beta$ from 1 ({\it i.e.} apparition of the first regions with slow fluctuations in the sample) and the lower value to the temperature at which $T_1^{-1}$ shows a minimum (regions with slow fluctuations become large enough to affect the median $T_1^{-1}$). The CDW phase is drawn according to x-ray results in refs.~\cite{Croft2014,Wen2019,Miao2021,vonArx2022}. The dots indicate that the boundaries of the phase are uncertain in zero field (there are issues related to whether the x-ray signal is elastic or inelastic) and even more so in high fields (no experiment yet). NMR data for $p=0.06$, 0.07 and 0.10 are from refs.~\cite{Julien1999}, \cite{Baek2012} and \cite{Julien2001}, respectively. }   
\end{figure}

Eq.~\ref{eq:tau} defines an exponential slowing down that has no onset temperature. Slow fluctuations may thus, in principle, be present up to arbitrarily high temperatures. However, in La$_{1.875}$Ba$_{0.125}$CuO$_4$ that has the longest-range CDW order of all cuprates, the upturn of (the median) $T_1^{-1}$ of $^{139}$La turns out to be very sharp and exactly coinciding with the CDW transition temperature $T_{\rm CDW}\simeq54$~K~\cite{Baek2015,Singer2020}. The stretching exponent $\beta$ (that relates to spatial inhomogeneity as explained in section~\ref{distribution}) also appears to deviate from 1 at $T_{\rm CDW}$~\cite{Baek2015} (in a recent report on  the same material, $\beta$ deviates from 1 below a slightly higher temperature of $\sim$80~K~\cite{Singer2020}, which might arise from slightly different doping or homogeneity of the sample). Clearly, it is principally the CDW transition that abruptly triggers the slow spin fluctuations in La$_{1.875}$Ba$_{0.125}$CuO$_4$ (notice that the concomitance of the structural transition might also play a role in the abruptness). If present above $T_{\rm CDW}$, slow spin fluctuations must have an amplitude lower than the NMR detection threshold (determined by the "background" $T_1^{-1}$ due to other components of the spin fluctuation spectrum and to electric-field-gradient fluctuations). This includes the possibility that slow fluctuations are present in a volume fraction of the sample that is too small to be detected.  

Consistent with earlier works~\cite{Julien2001,Curro2000,Hunt2001}, the link between slow spin fluctuations and charge order is visible in our data for LSCO $p=0.12$. Nonetheless, due to shorter-range charge order in LSCO, things are expectedly a bit less clear-cut than in La$_{1.875}$Ba$_{0.125}$CuO$_4$: $\beta$ still deviates from 1 at the CDW onset $T_{\rm CDW}\simeq 75$~K ({\it i.e.} spins already start to slow down in parts of the sample at that temperature) but the median $T_1^{-1}$ increases less steeply than for La$_{2-x}$Ba$_{x}$CuO$_4$ and only below $\sim$55~K~\cite{Mitrovic2008}. That the median $T_1^{-1}$ does not increase immediately at $T_{\rm CDW}$ indicates that the transition is more gradual because of stronger spatial inhomogeneity in LSCO. This is consistent with conclusions reached in ref.~\cite{Singer2020}.

For $0.14 \lesssim p\lesssim 0.17$, the upturn in $T_1^{-1}(T)$ occurs entirely below the zero-field $T_c$. Its onset is thus field dependent due to the competing effect of superconductivity. However, $\beta$ deviates from 1 already above $T_c$, actually $\sim 50 - 70$~K (Fig.~\ref{beta}), consistent with the onset temperature of static CDW order (see Fig.~\ref{phasediag}b). Therefore, the quasi-static spin fluctuations are still intertwined with CDW correlations in this doping range but the weakening of both orders and the greater strength of superconductivity now tip the balance in favor of superconductivity:  spin freezing is hampered by superconductivity, which results in a nonmagnetic ground state in zero field (Fig.~\ref{Fig_intro}a).

The correlation between CDW order and quasi-static spin fluctuations is also manifested in the doping dependence of the spin stiffness $E_0$ from our highest field data: like the NMR freezing temperature $T_g$, $E_0$ is strongly enhanced around $p=0.12$ (Fig.~\ref{phasediag}a), that is, where charge-density wave (CDW) order is  the strongest. 

It is important to point out that the $T=0$ phase boundaries for static CDW order are not settled yet. On the high doping side in particular, a recent report~\cite{Miao2021} of CDW order in zero field at $p=0.21$ seemingly points to a  disconnection from quasi-static spin fluctuations that are absent at this doping, at any field~\cite{Frachet2020}. However, the CDW signal might be dynamic rather than static~\cite{Miao2021,vonArx2022}. Therefore, where truly static CDW order exactly ends in the phase diagram is at present unclear in zero field, let alone in high fields where no experiments have been performed yet. Quite evidently, it is of paramount interest to investigate the field dependence of the static CDW in the vicinity of $p^*$ and to determine the doping endpoint of static CDW order, both in zero and high field. Is there any connection with $p^*$ and with our results concerning spin order?

Disconnection between spin freezing and CDW may be more evident on the low doping side: for $p\leq 0.10$, CDW order, if any, is weak but the onset of slow fluctuations tends to occur at increasingly higher $T$ as $p$ decreases and the slowing down now takes place over a much larger $T$ range (see refs.~\cite{Julien1999,Baek2017,Hunt1999} and Fig.~\ref{phasediag}). Glassy spin freezing is visibly less and less CDW driven as $p$ is decreased towards the non-superconducting boundary at $p\simeq 0.055$.Fang2022

The AFM spin-glass arises at low doping from strong two-dimensional AFM correlations in the presence of disorder. {\it Per se}, its strength would monotonically decrease upon increasing doping but, in La214 cuprates, it is actually reinforced over a range of doping around $p=0.12$ because it intertwines with charge-stripe order. This is unlike YBa$_2$Cu$_3$O$_y$ for which static magnetism does not extend beyond $p \sim 0.08$ where charge order starts to emerge~\citep{Coneri2010,Wu2011}.

\subsection{Overall field dependence}
The increase of the mean squared fluctuating field $\langle h_{\bot}^2 \rangle$ upon increasing $B$ (Fig.~\ref{hperp}b) is reminiscent the increase of the ordered moment measured in neutron scattering~\cite{Chang2008}. The concomitant increase of the activation energy $E_0$ (Fig.~\ref{parameters14p8}b) signifies that the spin stiffness increases in high fields, which results in a steeper divergence of $\tau_c(T)$ (Fig.~\ref{tau_c}a). A consequence of the approximately linear increase of $E_0(B)$ in the investigated range is that $\tau_c$ grows exponentially with $B$ (Fig.~\ref{tau_c}b): at 2 K, $\tau_c$ increases by eleven orders of magnitude between 10~T and 34~T. Therefore, the correlation time of spin fluctuations grows exponentially upon both decreasing $T$ and increasing $B$: $\tau_c\propto\exp(B/T)$ in the explored range of $B$ and $T$.

As already mentioned, we also see a considerable increase of the distribution $\Delta a$ at low fields (Fig.~\ref{parameters14p8}d). At the qualitative level, this parallels the growing spatial inhomogeneity at low fields indicated by the values of the stretching exponent $\beta$ at low $T$: for $p=0.148$, $\beta\simeq0.5$ at $\sim$30~T and $\beta\simeq0.3$ at $\sim$4~T~\cite{Frachet2020}. The distribution of $\Delta a$, however, seems more extreme than the $T_1$ distribution quantified by $\beta$. It is then possible that the increase in $\Delta a$ does not only reflect a growing spatial inhomogeneity in the superconducting state but also a deviation from the assumed exponential $T$ dependence of $\tau_c$ that becomes more significant at low fields as superconductivity progressively quenches spin freezing.

In summary, the field dependence suggests that superconductivity "dissolves" magnetic ordering by reducing the moment amplitude and by undermining both the stiffness of the spin system and the strong divergence of the correlation time. 

\begin{figure}[t!]
\hspace*{-0.3cm}
\includegraphics[width=8.5cm]{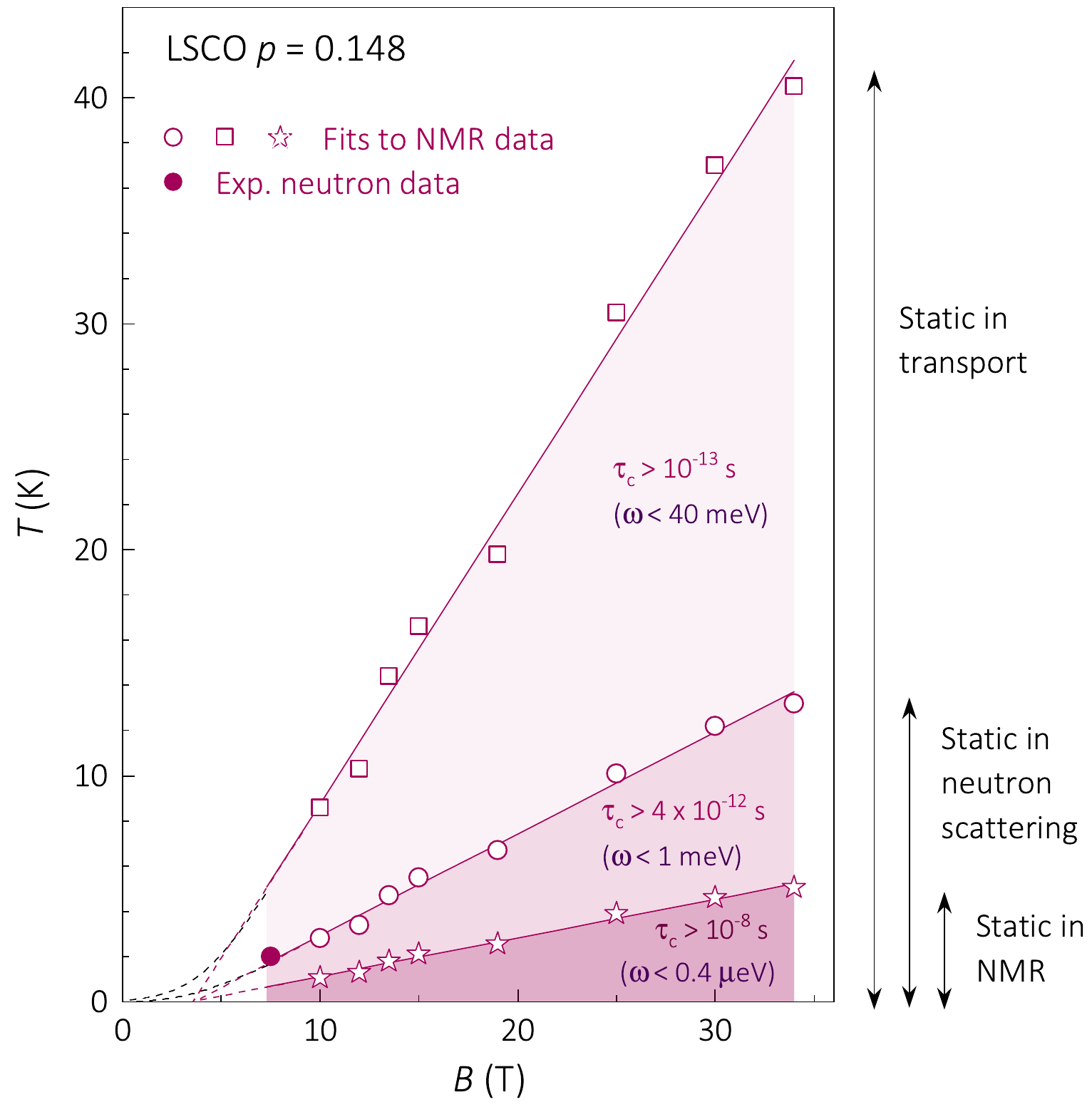}
\caption{\label{B-T} Regions of the field $B - T$ phase diagram where spin fluctuations are frozen at the timescale of transport (10$^{-13}$~s), neutron scattering ($4\times10^{-12}$~s) and NMR (10$^{-8}$~s) measurements (the corresponding frequency values $\omega=\tau_c^{-1}$ are also indicated). The open symbols correspond to the intersections between experimental $\tau_c$ values determined from BPP fits and the horizontal grey bars in Fig.~\ref{tau_c}a (see the symbol correspondance). Solid lines are linear fits to the data. Filled point: experimental onset (7.5~T at 2~K) of quasi-elastic intensity in neutron scattering~\cite{Chang2008}, that nicely matches the low-field extrapolation of the estimate based on NMR fits. Linear extrapolation of the upper line ($\tau_c=10^{-13}$~s) up to 55~T yields a temperature of 70~K, approximately matching the onset of the resistivity upturn at this field~\cite{BourgeoisHope2019}. Magenta and black dashes represent linear and nonlinear extrapolation of the data at low fields, with and without an onset field.}
\end{figure}

 \begin{figure*}[t!]
\hspace*{-0.3cm}    
  \includegraphics[width=12cm]{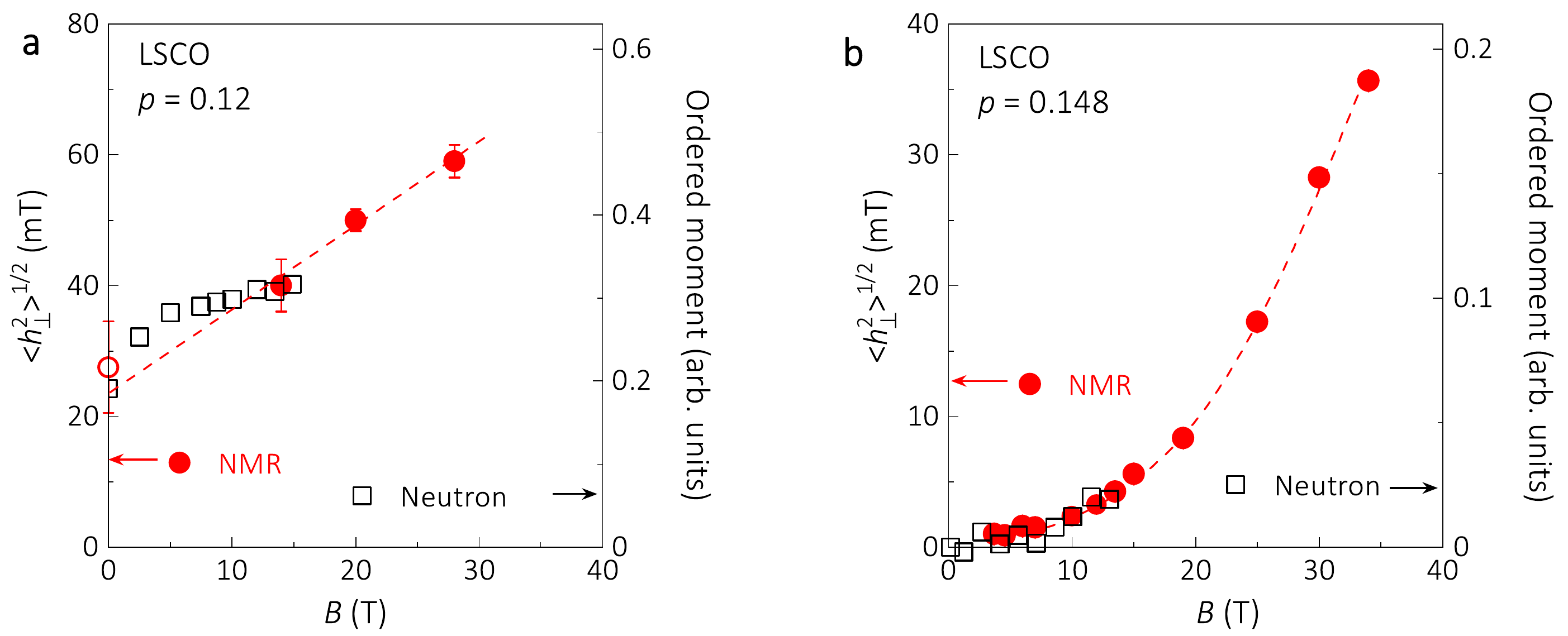}
  \caption{\label{scaling} Comparison of the field dependence of the fluctuating hyperfine field $\langle h_{\bot}^2 \rangle^{1/2}$ measured in NMR and of the ordered moment measured in elastic neutron scattering for $p=x=0.12$ (a) and $p=x=0.148$ (b). Neutron data are from ref.~\cite{Chang2008}. The $p=0.148$ crystal studied in NMR has a precisely-determined doping~\cite{Frachet2020} and is from the same batch as the neutron crystal initially labelled $p=0.145$. Notice that a slight vertical offset of neutron data in panel (b) (implying a tiny, rather than strictly null, ordered moment at low fields) would improve the scaling with $\langle h_{\bot}^2 \rangle^{1/2}$ below 7~T. This is not impossible given the $\mu$SR detection threshold of 0.005~$\mu_B$ in ref.~\cite{Chang2008}.}    
\end{figure*}

\subsection{High-field puzzle}
A most striking evidence that the field-dependent freezing arises primarily from the quenching of superconductivity is the absence of spin freezing when the field is applied parallel to the CuO$_2$ planes, which affects superconductivity much less than perpendicular fields~\cite{Frachet2020}. Within a scenario of competition between superconductivity and spin order, one would also expect the results in perpendicular fields to become field independent above the upper critical field $B_{c2}$, as indeed observed for the CDW state in YBa$_2$Cu$_3$O$_y$~\cite{Wu2013,Zhou2017,Kacmarcik2018}.  

Our data in LSCO, however, does not show any saturation as a function of $B$  (Figs.~\ref{parameters14p8}a,b and \ref{hperp}b). This implies either that superconductivity is still present in some form and competing with spin ordering at these fields or that the field has an additional, direct effect on the moments, irrespective of the presence of superconductivity (see ref.~\cite{Katanin2011} for one such theoretical proposal). It is impossible to  answer this question here since our highest field values (30 - 45~T) are not significantly larger than estimates of $B_{c2}$ based on an extrapolation of the vortex melting line at $T=0$ (see~\cite{Frachet2020,Girod2021} and refs. therein). Still, the absence of saturation up to at least 28~T for $p=0.12$ with $B_{c2}\simeq20$~T is puzzling and we notice that there is also no saturation in the sound velocity results up to at least 80~T at any doping level below $p^*$~\cite{Frachet2020,Frachet2021}. 

This lack of saturation in the field dependence contrasts with neutron scattering measurements: for $p=0.12$ (Fig.~\ref{scaling}a), $\langle h_{\bot}^2 \rangle^{1/2}$ varies much more with $B$ than the ordered moment does. This is also possibly true for $p=0.148$ (Fig.~\ref{scaling}b): for this doping, the NMR and neutron datasets are not inconsistent within error bars but the strong field dependence of $\langle h_{\bot}^2 \rangle^{1/2}$ up to at least 34~T is inconsistent with the presumed saturation of the ordered moment above $\sim$12~T~\cite{Chang2008}. Neutron data at higher fields would definitely be helpful here. Had the field dependence of $\langle h_{\bot}^2 \rangle^{1/2}$ clearly matched that of the ordered moment, we could have scaled the two quantities to predict the ordered moment in putative neutron high field experiments close to $p^*$. In the present situation, we refrain from doing so.

\subsection{Defining a magnetic-field scale}
Inasmuch as spin fluctuations continuously slow down upon decreasing $T$ and freeze at a temperature $T_g$ that depends on the frequency of the experimental probe, one may expect similar behavior as a function of field, namely a freezing field that depends on the experimental time scale. This is indeed the case for LSCO $p=0.148$: from the $\tau_c(B,T)$ results in Fig.~\ref{tau_c}a (and data at other fields, not shown in this figure), we deduce the $\tau_c=4 \times 10^{-12}$~s boundary in $(B,T)$ space, {\it i.e.} where fluctuations become as slow as $\tau_c^{-1}=\omega=1$~meV and thus contribute an elastic signal in neutron scattering (ignoring inhomogeneity aspects that complicate the problem). As Fig.~\ref{B-T} shows, linear extrapolation of this data below 10~T matches the neutron scattering finding that elastic scattering at 2~K onsets at $\sim$7.5~T~\cite{Chang2008}. In other words, for this $p=0.148$ doping, both the transition field at the neutron timescale and the NMR data are consistently accounted for by the same correlation time of spin fluctuations $\tau_c\propto\exp(B/T)$.

At $T=2$~K, the fluctuations reach the NMR frequency at a higher field of $\sim$14~T (Fig.~\ref{B-T}) for $p=0.148$, which corresponds to the field at which a peak becomes discernible in the data of Fig.~\ref{T1}. For other doping levels, however, the same amount of information is unfortunately not available, especially as a field of 45~T was insufficient to produce a measurable peak in $T_1^{-1}$ for the $p=0.171$ sample. Therefore, we have no information on $E_0$ and $\tau_c$ for this doping. The only available data, $\langle h_{\bot}^2 \rangle$, shows a smooth $B$ dependence from which no characteristic field scale naturally emerges.

Therefore, in order to determine a field scale for any doping level, we define the onset field $B_{\rm slow}$ as the field above which an "ordered moment" should be detected in neutron scattering (quotes are because of the quasi-elastic, rather than purely elastic, nature of the signal as discussed above). We use as a criterion the value of $\langle h_{\bot}^2 \rangle$ at the transition field of 7~T as determined from neutron scattering for $p=0.148$~\cite{Chang2008} (horizontal line $\langle h_{\bot}^2 \rangle^{1/2}=1.5$~mT at 7~T in Fig.~\ref{hperp}b). There is thus no independently-measured value of $B_{\rm slow}$ from NMR at $p=0.148$. A merit of this definition is that it agrees with the onset of slow fluctuations detected in sound velocity~\cite{Frachet2020} (notice that a criterion based on the value of $T_1^{-1}$, rather than on $\langle h_{\bot}^2 \rangle$, was used in ref.~\cite{Frachet2020} and gives similar values of $B_{\rm slow}$). As Fig.~\ref{xi} shows, $B_{\rm slow}$ increases with $p$ up to $p^*\simeq0.19$, meaning that, as doping increases, larger fields are necessary to induce spin freezing. We now discuss the interpretation of this observation.

\subsection{Evidence for competing order in vortex cores}
There is a fundamental difference between decreasing temperature and increasing field: in the latter case, the AFM glass emerges from the electronically inhomogeneous vortex state. This situation, in which the competing order is enhanced in and around the vortex cores, is analogous to the emergence of long range CDW order upon increasing field in YBa$_2$Cu$_3$O$_y$~\cite{Wu2013b}. To first approximation, the field at which a bulk frozen state is reached corresponds to the critical vortex density for which the halos of charge and/or spin order start to overlap, just as the superconducting upper critical field $B_{c2}$ corresponds to overlap of vortex cores. By analogy, $B_{\rm slow}$ should then scale with the inverse of the halo radius squared~\cite{Wu2013b}: 
\begin{equation}
\label{eq:Bslow}
B_{\rm slow} \, \simeq \, \frac{\Phi_0}{2\pi\xi_{\rm AF}^2}
\end{equation}
where the halo radius is taken to be set by the AFM correlation length $\xi_{\rm AF}$.  In this picture, the characteristic field scale associated with AFM ordering increases with increasing doping (as indeed observed) because $\xi_{\rm AF}$ decreases as one goes away from the N\'eel phase.

According to Eq.~\ref{eq:Bslow}, $B_{\rm slow}$ values of 2.3  and 7.5~T (taken from neutron scattering results in two different samples with $p\simeq0.145$~\cite{Khaykovich2005,Chang2008}) correspond to $\xi_{\rm AF}$ values of 111~\AA~and 69~\AA. The experimental values, measured in fields of 13--14~T for these samples are: $\xi_{\rm AF}\geq~120$~\AA~\cite{Khaykovich2005} and $\xi_{\rm AF}=75$~\AA~\cite{Chang2008}. The agreement is remarkable, especially as the length scale for the AFM halos to overlap should be proportional but not necessarily equal to $\xi_{\rm AF}$ (it depends on how the staggered magnetization exactly decays with distance from the vortex core). 

\begin{figure}[t!]
\hspace*{-0.3cm}    
  \includegraphics[width=7.5cm]{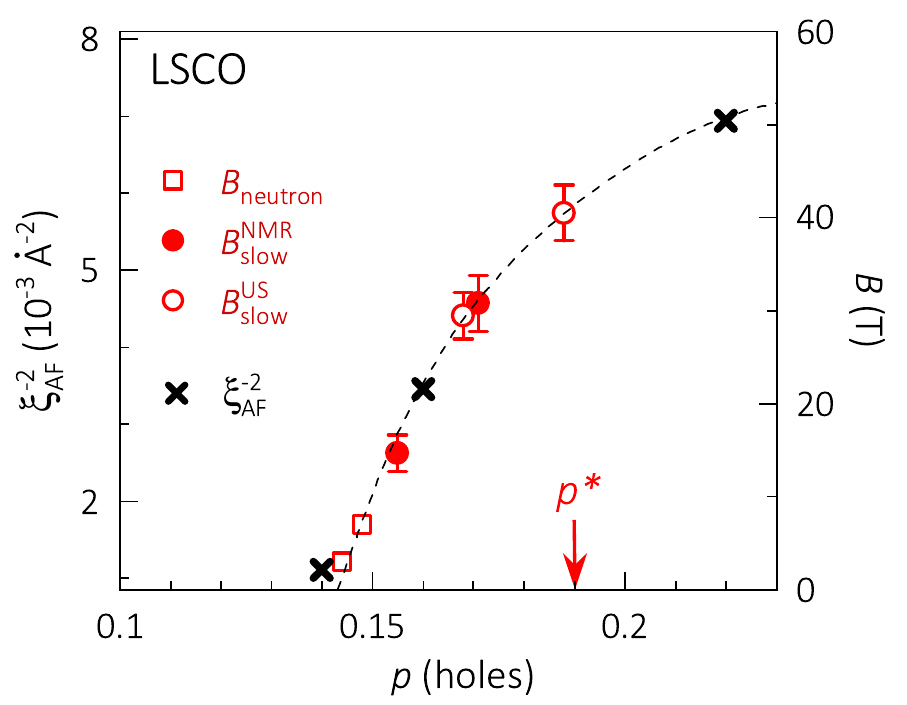}
  \caption{\label{xi} Doping dependence of the characteristic field $B_{\rm slow}$ defined from NMR data (this work) so as to match the transition field $B_{\rm neutron}$ at $p=0.148$~\cite{Chang2008} and compared to $B_{\rm slow}$ from ultrasound measurements (defined as the field above which a softening of the shear elastic constant is observed)~\cite{Frachet2020}. The $B_{\rm neutron}$ value for $p=0.144$ is from ref.~\cite{Khaykovich2005}. $B_{\rm slow}$ appears to vary with $p$ as $\xi_{\rm AF}^{-2}$, with $\xi_{\rm AF}$ the AFM correlation length measured by inelastic neutron scattering at low energy and zero field for $p=0.14$~\cite{Aeppli1997}, $p=0.16$~\cite{Vignolle2007} and $p=0.22$~\cite{Lipscombe2007}. This scaling between $B_{\rm slow}$ and $\xi_{\rm AF}^{-2}$ supports the notion of AFM order emerging from the superconducting vortex cores (see text). The dashes guide the eye through $\xi_{\rm AF}$ data.}    
\end{figure}

In principle, the $\xi_{\rm AF}$ values satisfying Eq.~\ref{eq:Bslow} are those measured in an elastic ($\omega\simeq0$) neutron scattering experiment at $B=B_{\rm slow}$. These are unknown for $p\geq0.15$. Nonetheless, $\xi_{\rm AF}$ values from inelastic neutron scattering studies \cite{Aeppli1997,Vignolle2007,Lipscombe2007} at low energy ($\omega\sim$~few meV) and in zero field are sufficient for the purpose of comparing the doping dependence (notice that, strictly speaking, diffraction peaks being at incommensurate positions, their width defines a spin-stripe, rather than purely AFM, correlation length).  As Fig.~\ref{xi} shows (notice that the $\xi_{\rm AF}^{-2}$ scale does not start from zero, reflecting the fact that the measured $\xi_{\rm AF}$ is not infinite even when magnetic order is present in zero field), $B_{\rm slow}$ values at different doping levels are found to scale with $\xi_{\rm AF}^{-2}$, just as expected from Eq.~\ref{eq:Bslow}. This thus provides quantitative support to the picture of a competing order emerging from the vortex cores. 

In passing, we note that, based on the scaling between $B_{\rm slow}$ and $\xi_{\rm AF}^{-2}$, one would extrapolate $B_{\rm slow}\simeq50$~T for $p=0.22>p^*$ (Fig.~\ref{xi}). Since slow spin fluctuations have not been seen in ultrasound measurements performed up to 90~T for $p=0.21$~\cite{Frachet2020}, spin freezing is surely absent at any field at this doping and by extension above $p^*$.

\subsection{Gap filling {\it vs.} gap closing}

At doping levels $p\geq 0.15$ for which magnetic order is absent in zero field, a spin gap $\Delta_{\rm spin}$ is observed in neutron scattering studies (\cite{Tranquada2020} and references therein). If the excitations at $\Delta_{\rm spin}$ have a triplet character, increasing the field should linearly decrease the energy of the lower branch of the $S=1$ state which then ultimately crosses the singlet ground state at a critical field $B_c=\Delta_{\rm spin}/g\mu_B$, thus giving rise to a magnetic ground state as observed in low-dimensional quantum magnets~\cite{Chaboussant1998,Giamarchi2008}. The field scale for magnetic order in that case is thus set by the spin gap $\Delta_{\rm spin}$.

Here in LSCO $p=0.148$, $\Delta_{\rm spin}\simeq 4$ to 9~meV (depending on the definition used for the spin gap~\cite{Chang2009,Li2018}) whereas the Zeeman energy associated with the field-induced freezing is $g\mu_B\,B\sim$~1~meV for a typical field scale of $\sim$10~T. Therefore, the field-induced order arises before $\Delta_{\rm spin}$ is closed.

Instead, the slow fluctuations appear as in-gap states~\cite{Chang2009}, which is consistent with the phase coexistence expected for a competing order emerging from the vortex cores. Further, the similarity of field-induced and Zn-induced magnetism~\cite{Panagopoulos2003} is also in favor of in-gap states, since these latter are well documented from neutron scattering studies of Zn-doped LSCO~\cite{Kimura2003}. This picture is likely to remain true for $p>0.148$ since $\Delta_{\rm spin}$ is weakly $p$ independent in the doping range $p\simeq0.15 - 0.20$~\cite{Li2018} whereas $B_{\rm slow}$ has a strong $p$ dependence (Fig.~\ref{xi}).

\subsection{Relevance to transport measurements}

Most discussions on "quantum criticality" in cuprates (in the loose sense of ground-state properties changing across $p^*$) have so far overlooked the possible effect of the magnetic field. Our results, however, reveal that there is some difference in magnetic properties between superconducting and nonsuperconducting ground states. As pointed out in ref.~\cite{Frachet2020}, the field-dependent nature of magnetism up to $p^*$ may be relevant for the interpretation of high-field experiments in La214, and particularly for the question of quantum criticality inferred from specific heat measurements~\cite{Michon2019}.  In the present section, we would like to discuss a different but complementary aspect, which is related to experimental time scales.

Looking again at the $T$ dependence of the correlation time $\tau_c$ for $p=0.148$ (Fig.~\ref{tau_c}), one sees that the AFM moments can be considered static on the scale of the quasiparticle lifetime of $\sim0.1$~ps (from ref.~\cite{Fang2022} in Nd-LSCO) below a temperature as high as $\sim 40$~K at 34~T, even though the moments become frozen at the NMR timescale only at $\sim5$~K. Fig.~\ref{B-T} provides an  overview of how fluctuations evolve in the $B - T$ plane. The $T$ range of slow fluctuations visibly expands in high fields.

The same analysis shows that the 0.1~ps timescale is reached far above $T_c$, at $\sim80$~K, for $p=0.12$ but only at $\sim$10 -- 20 K (at 30 Tesla) for $p = 0.155$.

We thus see that, up to $p^*$, the slow spin fluctuations are likely to impact transport properties measured in high fields, not only at the lowest temperatures but also significantly above the temperature of freezing at the NMR timescale. This can even be above $T_c$ when stripe order is already well developed in zero-field, namely close to $p=0.12$ in LSCO and presumably up to $p\sim 0.2$ in LTT (low temperature tetragonal) variants of La214. Therefore, the presence of quasi-static AFM fluctuations is potentially relevant to important optical~\cite{Post2021} and transport~\citep{Boebinger1996,Collignon2017,BourgeoisHope2019,Fang2022} experiments performed in a similar range of $T$, $B$ and $p$ in La214. 

In particular, a correlation between resistivity upturn and magnetic freezing has been recently discussed in the context of field-induced spin order in LSCO $p=0.143$~\citep{BourgeoisHope2019}, thereby extending previous evidence of such correlation at lower doping and zero field~\cite{Harshman1988,Hayden1991,Julien1999,Hunt2001,Panagopoulos2005}. In our $p=0.148$ sample (from the same batch as the $p=0.143$ sample of ref.~\citep{BourgeoisHope2019}), the temperature at which fluctuations reach the 0.1~ps timescale increases linearly with $B$ (squares in Fig.~\ref{B-T}). Extrapolating this linear dependence up to 55~T, we find that the 0.1~ps timescale should be reached at $\sim$70~K, which is about the onset of resistivity upturn at this field~\cite{BourgeoisHope2019}. Of course, such correlation is only approximate: the onset temperature of the upturn is not very precisely defined, the quasiparticle lifetime of 0.1~ps that was determined for Nd-LSCO $p=0.20$ may be different in LSCO $p=0.148$ and the linear extrapolation at high fields may not be justified. Still, that the numbers match shows that our analysis of the NMR data is at least semi-quantitatively consistent with a link between resistivity upturns and quasi-static spin fluctuations (we note that a somewhat weaker field dependence of $E_0$ above 34~T, as suggested by ultrasound measurements~\cite{Frachet-thesis}, leads to an extrapolated temperature of $\sim$55~K at 55~T for the 0.1~ps timescale, which rather corresponds to the resistivity minimum). 

Correlation, however, does not imply causation. Therefore, while the coincidence of $p^*$ with a transition in the high-field magnetic ground state is unlikely to be accidental, we are not claiming that the transformation of the Fermi surface below $p^*$~\citep{Collignon2017,Fang2022} necessarily results from a reconstruction by frozen antiferromagnetism in La214. The credibility of this scenario depends, for instance, on whether the AFM correlation length $\xi_{\rm AF}$ is large enough in high fields up to $p^*$, which is unknown (we nonetheless note that $\xi_{\rm AF}\geq120$~\AA~at 14.5~T for $x=0.144$ in ref.~\cite{Khaykovich2005}). Furthermore, it is possible that the Fermi surface of the pseudogap state ({\it i.e.} in zero field and up to $T\sim T^*$) consists of pockets that are insensitive to a reconstruction by the low-$T$ AFM order observed here. Our point here is simply that, in a striped cuprate such as Nd-LSCO with $p\simeq 0.2$, spin fluctuations should be sufficiently slow at $T\sim T_c$ ({\it i.e.} at temperatures relevant to most transport experiments) for a scenario of reconstruction by quasi-static AFM order to be considered. Our results actually beg the question: why would this state not reconstruct the Fermi surface?

\subsection{Nature of the low-field ground state}

We now come back to the discussion of the field dependence at low fields, which is directly connected to the question of the nature of the ground state in zero or low field. 

As already alluded to in the above,  the values of $T_1^{-1}$ at low $B$ and $T$ remain well above the values expected if the field had only closed the superconducting gap (dashed lines in Fig.~\ref{T1}a for $p=0.148$), even though the peak in $T_1^{-1}(T)$ has disappeared below 10~T.  Within our analysis, this is reflected in the finite and smoothly $B$ dependent values of $\langle h_{\bot}^2 \rangle$ (Fig.~\ref{hperp}b). 

In principle, vestiges of frozen magnetism may arise from the least-doped regions of the sample~\cite{Singer2002}. However, as already noted above, $^{139}$La NMR measurements do not seem to be affected by this type of spatial inhomogeneity in general~\cite{Mitrovic2008,Arsenault2020}, so it is unlikely that this explains the enhanced $T_1^{-1}$ at low fields.

Actually, we have seen from the consistency between neutron scattering and NMR data (Fig.~\ref{B-T}) that the frozen state arises gradually upon increasing the field, in the same way as it arises from a gradual slowing down upon cooling. As Fig.~\ref{B-T} also shows, the $\tau_c=10^{-13}$~s, $\tau_c=4 \times 10^{-12}$~s and $\tau_c=10^{-8}$~s lines all extrapolate linearly to a field as small as 3.6~T at $T=0$ (the common field value regardless of the $\tau_c$ value being a consequence of the linear dependence of $E_0$ on $B$). This already points to quasi-static magnetism very close to the zero-field ground state. Furthermore, it is possible that $\tau_c$ no longer depends linearly on $B$ at low $B$ values: Fig.~\ref{parameters14p8}b shows a possible S-shape dependence - saturating on both high and low field ends - that is as consistent with the data as the purely linear dependence). This thus suggests that there may be no sharp difference between the low-field and high-field ground states: quasi-static moments are probably present at $B\ll B_{\rm slow}$ but of very weak magnitude and/or occupying a relatively small volume fraction of the sample.

It is then possible that the ground state in zero field can be considered as nearly critical~\cite{Chubukov1994} between $p_{\rm sg}\simeq 0.135$ (the end of the frozen ground state in zero-field) and $p=p^*$ (the end of the frozen ground state in high fields): the system "knows" that it is destined to become critical once superconductivity is locally suppressed in the vortex cores. This is different from a situation in which superconductivity simply shifts the end doping of the magnetic phase from $p^*$ down to $p_{\rm sg}$ with no change in the zero-field ground state across $p^*$. 

In this regard, it is interesting to note that theory~\cite{Kivelson2002} predicts the possibility that the order appears at arbitrarily low fields around (and along) each single 3D vortex, before crossing over to a more spatially-homogenous order at higher fields (corresponding to the above-described overlap of AFM halos in each plane). Evidence of such vortex-core magnetism has been claimed from muon spin rotation measurements in fields as low as 0.5~T~\cite{Sonier2007}.
 
 \subsection{Broad picture}
Our work sheds light on how superconductivity coexists and competes with spin order, a question of obvious topical interest in the cuprates~\cite{Fradkin2015,Yamase2016,Cui2020}. In combination with the ultrasound results~\cite{Frachet2020} that are complementary to the NMR results, this work also sheds new light on the magnetism of ${\mathrm{La}}_{\mathrm{2\ensuremath{-}x}}{\mathrm{Sr}}_{\mathrm{x}}{\mathrm{CuO}}_{4}$ by revealing that an AFM glass constitutes the non-superconducting ground state from the weakly-doped insulator at $p = 0.02$ all the way up to $p^* = 0.19$. As argued in ref.~\cite{Frachet2020}, this suggests that the same local-moment antiferromagnetism as found in the doped Mott insulator survives throughout the pseudogap regime. This observation adds to previous experimental evidence that local-moment magnetism prevails in a large part of the cuprate phase diagram~\cite{Tranquada2013}, thus making the doped Mott insulator the most natural starting point for describing the pseudogap state.

The most striking aspect of the NMR and ultrasound results is that the competition between superconductivity and spin order is confined to the pseudogap state: as shown in ref.~\cite{Frachet2020}, no signature of competition is observed for $p=0.21 > p^*$. It may be worth recalling here that short-range AFM correlations have been found to survive above $p^*$ from NMR~\cite{Ohsugi1994}, neutron scattering~\cite{Wakimoto2007,Lipscombe2007} and resonant x-ray scattering~\cite{Dean2013,Wakimoto2015,Monney2016} experiments in LSCO so our results are not explained by the disappearance of AFM correlations. What disappears at $p^*$ is the ability of the moments to freeze out and to compete with superconductivity. 

These results establish an unexpected connection between the pseudogap phase, defined by a relatively high temperature scale $T^*$, and a much lower temperature phenomenon, spin freezing, that may be taken as a signature of the doped Mott insulator. It is even possible that the transition at $p^*$ is precisely associated with the loss of Mott physics. This would be in line with theoretical works using dynamical-mean-field-theory~\cite{Sordi2011} and with a recent pump-probe study of Bi2201~\cite{Peli2017}. Consistent with this view, a resonant x-ray scattering study of Bi2201 and Tl2201 observed a change in the nature of (high-energy) magnetism around $p^*$~\cite{Minola2017}.  An alternative, albeit not necessarily incompatible explanation views the transition at $p^*$ as a percolation phenomenon in a phase-separated system with coexisting magnetic and nonmagnetic patches whose relative areas change with doping~\cite{Tranquada2020}.

The unique ability to sustain local-moment magnetism would then generically distinguish the pseudogap phase from the correlated metal at $p > p^*$. In the specific case of La214 cuprates, this would ultimately result in an ordered magnetic ground state up to $p^*$, provided competing effects from superconductivity have been removed. However, we stress that this is not necessarily so in other cuprates: while the frozen AFM state is a symptom of  local-moment magnetism, not all cuprates are symptomatic from this standpoint. The intertwined spin and charge stripes are known to be specificity of the La214 system while a cuprate like Bi2201 does not seem to feature any resurgence of magnetism in high fields~\cite{Kawasaki2010}. 

Finally, we underline the relevance of our results to recent theoretical work exploring the nature of the doping-driven quantum phase transition from a metallic spin-glass with small Fermi surface to a Fermi liquid with large Fermi surface in the $t-J$ model with random all-to-all hopping and exchange interactions 
(refs.~\cite{Joshi2020,Shackleton2021} and references therein). In this respect, the connection, if any, between the small Fermi surface observed  below $p^*$ in several cuprate families~\cite{Fang2022,Badoux2016,Collignon2017,Putzke2021} and magnetic properties (such as the local-moment nature of the Cu $3d^9$ states or the incipient ordering of the AFM moments) is a fascinating issue.

\section{Summary}

 \begin{figure}[t!]
\hspace*{-0.3cm}    
  \includegraphics[width=9.6cm]{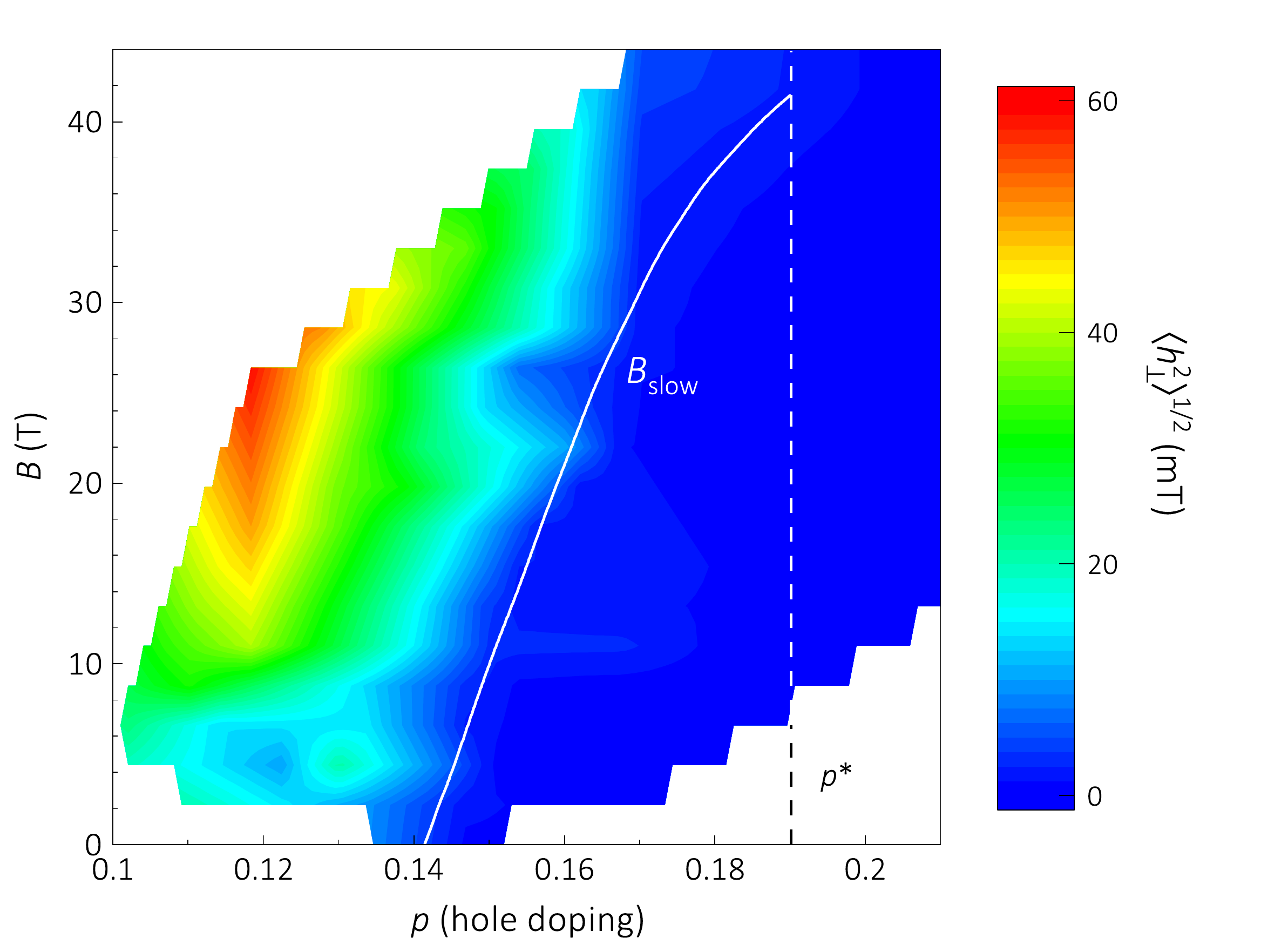}
  \caption{\label{3D} False-color representation of the doping ($p$) and field ($B$) dependence of the mean-squared fluctuating hyperfine field $\langle h_{\bot}^2 \rangle^{1/2}$ from data in Fig.~\ref{hperp}b. $\langle h_{\bot}^2 \rangle$ represents the amplitude of the fluctuating moments. The dashed line ($B_{\rm slow}$) is from Fig.~\ref{xi}. The vertical line shows the end doping of the pseudogap phase at $p^*=0.19$ in LSCO. }    
\end{figure}

In the doping range $p=x=0.14-0.18$, superconducting ${\mathrm{La}}_{\mathrm{2\ensuremath{-}x}}{\mathrm{Sr}}_{\mathrm{x}}{\mathrm{CuO}}_{4}$ has a nonmagnetic ground state protected by a spin-gap. However, the application of magnetic fields $B$ perpendicular to CuO$_2$ planes promotes a ground state with frozen antiferromagnetic moments, thus providing one of the clearest cases of competition between superconductivity and electronic ordering. 

The slow fluctuations of the moments are detected in NMR relaxation rate measurements. In this work, we have described a semi-quantitative BPP model for the relaxation rate of inhomogeneously-freezing electronic spins. The model assumes that the characteristic timescale of the field-induced fluctuations becomes exponentially large upon cooling down to $T=0$. There are also two $T$ independent parameters: the fluctuation amplitude $\langle h_{\bot}^2 \rangle$ and the spin stiffness $E_0$, in addition to distributions of the parameters. We have used this model to fit the relaxation data~\cite{Frachet2020} as a function of $T$, for different values of $B$ and $p$. Our main findings may be summarized as follows:

\begin{itemize}

\item A correlation time that depends exponentially on $B/T$ accounts for the differences in the freezing temperature $T_g$ and in the onset field $B_{\rm slow}$ between NMR and neutron scattering. 

\item Above $T_g$, the spin fluctuations are already slow enough to potentially impact on transport properties. Whether this leads to a reconstruction of the Fermi surface or not is unsettled. Nonetheless, that quasi-static spins are present up to $p^*$ and not beyond makes magnetism obviously relevant to the question of quantum criticality at $p^*$ and more generally to the interpretation of those experiments that find a sharp change in the electronic properties across $p^*$, especially if these experiments are performed in high fields. 

\item Both the fluctuation moment $\langle h_{\bot}^2 \rangle$ and the spin stiffness $E_0$ are monotonously enhanced upon increasing the magnetic field. The relative strength of quasi-static magnetism, as quantified by the fluctuating hyperfine field $\langle h_{\bot}^2 \rangle$ (proportional to the mean-squared or fluctuating moment), is represented as a function of field and doping in Fig.~\ref{3D}. 

\item We defined a characteristic field scale $B_{\rm slow}$ that marks the onset of slow spin fluctuations in NMR and ultrasound measurements as well as the onset of quasi-static scattering in neutron experiments. We find that $B_{\rm slow}$ increases with doping  in a way that is quantitatively consistent with the notion that bulk freezing occurs when the vortex density is large enough that halos of local AFM order around each vortex core start to overlap. 

\item It is possible that the difference between the high-field and low-field ground states is more quantitative than qualitative. Indeed, while the emergence of quasi-static magnetism upon increasing field is reflected in the strong increase of both $\langle h_{\bot}^2 \rangle$ and $E_0$, no sharp transition is observed as a function of field. The data suggests that there may be quasi-static moments of extremely weak amplitude already below $B_{\rm slow}$. These could correspond to quasi-ordering within a single vortex, which would be consistent with the idea of field-induced low-energy fluctuations filling the spin-gap. 

\item Spin order and charge-stripe order are known to be intertwined in superconducting LSCO but, because the $^{139}$La probe lacks sensitivity to CDW order in CuO$_2$ planes, we could investigate the relationship between the two orders only indirectly. We observe that the onset of slow spin fluctuations is essentially triggered by CDW order over the doping range $p=0.12-0.18$. Moreover, both $E_0$ and $T_g$ are sharply enhanced at $p=0.12$ where CDW order is the strongest and they both decrease upon increasing $p$ up to $p^*$. It is thus possible that the primary competitor of superconductivity up to $p^*$ is charge-stripe order. However, an entirely open and fundamental question is whether spin order would remain in the high-field ground state up to $p^*$ if CDW order could be suppressed by any means. Results in Zn-doped LSCO suggest that the answer to this question might be in the affirmative~\cite{Panagopoulos2003,Hirota2003,Hucker2011} but better characterization of the CDW in Zn-doped samples is needed for a firmer answer. 

\item The more stable is the stripe phase (as a function of either doping or field), the stiffer is the spin system and the sharper the transition as a function of $T$. In La$_{1.875}$Ba$_{0.125}$CuO$_4$ that shows the sharpest magnetic transition of all striped cuprates, the difference in $T_g$ values between neutron scattering and NMR does not exceed a few Kelvin~\cite{Baek2015,Singer2020}. 

\item There is a puzzling absence of saturation in the high-field data that is worthy of further investigation. Possible interpretations include the possibility that AFM order competes with superconducting fluctuations well above the bulk $B_{c2}$ determined by specific heat~\cite{Girod2021}, as well as a direct effect of the field on low-energy spin fluctuations ({\it i.e.} irrespective of the presence of superconductivity).
 
\end{itemize}

\section{Perspectives}

\begin{itemize}

\item The described analysis could allow to identify signatures of spin freezing in cuprates where the relaxation rate is enhanced but a clear peak in $T^{-1}_{1}(T)$ is not visible. 

\item Our results provide motivation and guidance for further high-field experiments in LSCO in the vicinity of $p^*$. Neutron scattering in high (most likely pulsed) fields should observe the AFM glass above $B_{\rm slow}$ and allow a much-needed measurement of the AFM correlation length in high fields. Optical measurements could probe whether the field-induced state is a putative 2D superconductor with interlayer decoupling as observed at much lower field/doping values~\cite{Schafgans2010}. Direct probes of CDW order in high fields across $p^*$ would also be highly desirable. Finally, another natural extension of our work is the NMR study of LTT La214 cuprates across $p^*$. A recent neutron scattering study of Nd-doped LSCO finds spin-stripe order above $p^*$~\cite{Ma2021}, seemingly contradicting our NMR results in LSCO. However, as noted by the authors, timescale issues might need to be considered here. Furthermore, the Nd moment might complicate the analysis in this system, thus calling for further investigations in La$_{2-x}$Ba$_{x}$CuO$_4$ or Eu-doped LSCO.

\item Spatial inhomogeneity, due to both disorder and the vortex state, is clearly a key issue. Progress in our understanding will require more sophisticated analysis of the NMR data as well as quantitative predictions from theoretical models that account realistically for this inhomogeneity. 

\item It is possible that quantum simulations using cold atoms~\cite{Mazurenko2017,Koepsell2021} provide in a near future an  appropriate settings for elucidating the particularly intricate interplay between spin, charge and superconducting orders, with or without quenched disorder.

\end{itemize}

\section*{Acknowledgements}

We thank David LeBoeuf, Mehdi Frachet, Adam Dioguardi, Mladen Horvati{\'c}, Jeff Sonier, John Tranquada and Shangfei Wu for useful discussions. 

Work in Grenoble was supported by the Laboratoire d'Excellence LANEF (ANR-10-LABX-51-01) and by the French Agence Nationale de la Recherche (ANR) under reference ANR-19-CE30-0019 (Neptun). Part of this work was performed at the LNCMI, a member of the European Magnetic Field Laboratory (EMFL). A portion of this work was performed at the National High Magnetic Field Laboratory, which is supported by the US National Science Foundation Cooperative Agreement no. DMR-1644779 and the State of Florida. Work in Z\"{u}rich was supported by the Swiss National Science Foundation.

\appendix

\section{Additional datasets}

\begin{figure}[h!]
\hspace*{-0.3cm}    
  \includegraphics[width=7.7cm]{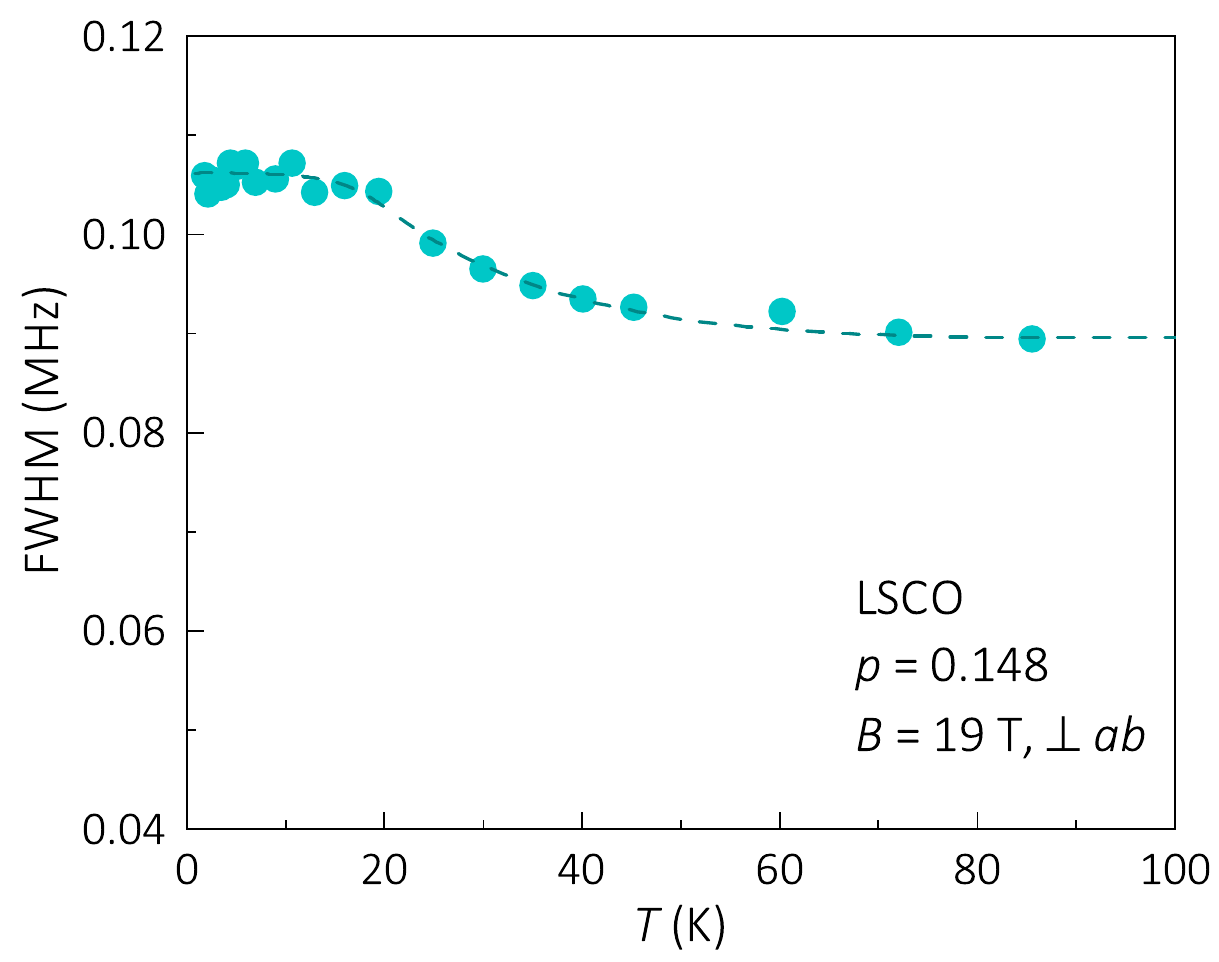}
  \caption{\label{linewidth} Width at half maximum of the $^{139}$La central line for LSCO $p=0.148$. The field is applied perpendicular to the CuO$_2$ planes. The width saturate below $\sim$20~K, the temperature at which $T_1^{-1}$ shows a maximum (Fig.~\ref{T1}).}  
  \end{figure}

Fig.~\ref{linewidth} presents the $T$ dependence of the $^{139}$La linewidth.

\begin{figure*}
\hspace*{-0.3cm}    
  \includegraphics[width=17cm]{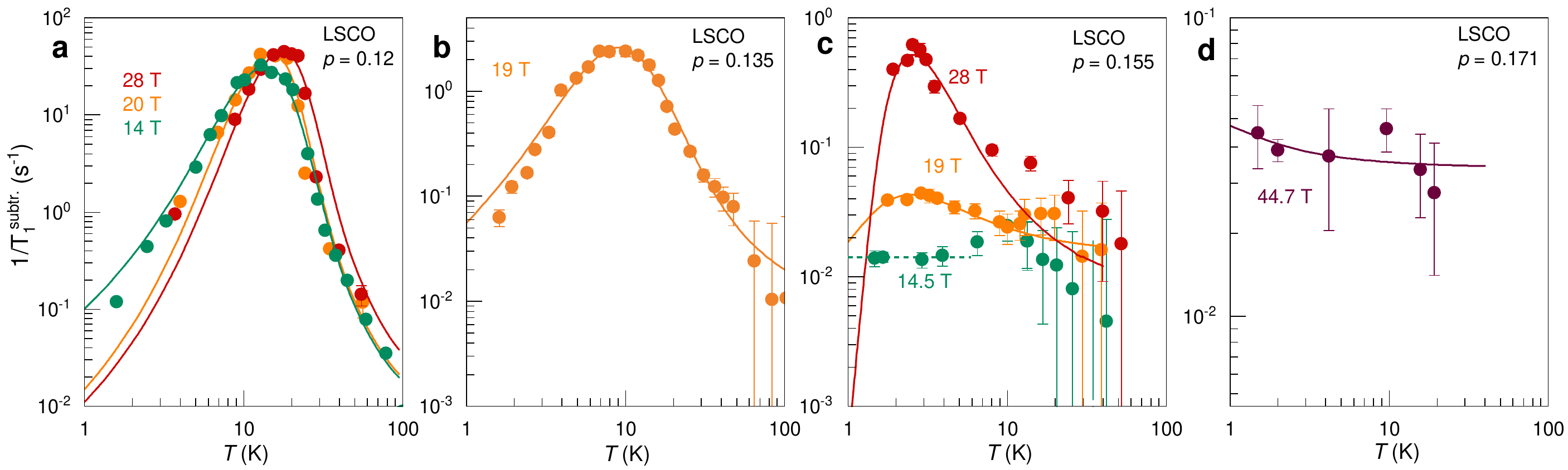}
  \caption{\label{allfits} $T_1^{-1}$ data (symbols) from which a superconducting background has been subtracted as described in ref.~\cite{Frachet2020} for different doping levels and the corresponding fits (continuous lines) with the distributed BPP model described in the text. }
\end{figure*}

Fig.~\ref{allfits} presents additional (background subtracted) $T_1$ datasets and the corresponding fits to the distributed BPP model.

\section{Experimental methods}

The samples were grown by travelling solvent floating zone and the  sample information is summarized in table \ref{tab:sample info}. Where available, the hole doping $p$ is estimated from the High Temperature Tetragonal (HTT) to Low Temperature Orthorhombic (LTO) structural phase transition at $T_s$.

\begin{table}[h!]
\caption{\label{tab:sample info} Sample information}
\begin{ruledtabular}
\begin{tabular}{llll}
\textrm{hole doping $p$}&
\textrm{$T_{\textrm c}$ (K)}&
\multicolumn{1}{c}{\textrm{$T_{\textrm s}$ (K)}}&
\textrm{Ref.}\\
\colrule
 $0.122\pm0.002$ &  $29.3\pm1.5$  &   $252 \pm3$ & \cite{Frachet2021}  \\    
 $0.135\pm0.002$ &  $35.5\pm1.5$ & $225\pm3$  \\  
 $0.148\pm0.001$ &  $36.2\pm1$    & $194\pm2$ & \cite{Frachet2020,Chang2008}\\  
 $0.155\pm0.001$ & $38.1\pm1$     & $177.5\pm2$ & \cite{Frachet2020}\\
 $0.171\pm0.002$ & $37.5\pm1$     & $141.5\pm3$ & \cite{Frachet2020}\\
 $0.210\pm0.005$ &  $25.6\pm1$    & $6\pm10$ & \cite{Frachet2020}\\ 
\end{tabular}
\end{ruledtabular}
\end{table}

We used home-built heterodyne NMR probes and spectrometers, superconducting magnets for fields up to 20~T, the LNCMI M10 resistive magnet for fields up to 30~T and the NHMFL hybrid magnet for fields up to 45~T. Fields were applied along the $c$-axis, \textit{i.e.} perpendicular to the $\rm{CuO}_2$ planes.

The relaxation rate $T_1^{-1}$ was measured on the central transition of $^{139}$La (nuclear spin $I=7/2$) without any contamination from satellite transitions as the quadrupole splitting of $\sim$6~MHz greatly exceeds the excitation width $\Delta f \simeq 50$~kHz. The recoveries were defined by the time dependence of the nuclear magnetization $M(t)$ following a saturating pulse yielding $M(t=0)\simeq 0$. $T_1$ values were determined by fitting these recoveries to a stretched version of the theoretical law for magnetic relaxation between $m_I=\pm 1/2$ levels of a nuclear spin 7/2~\citep{Mitrovic2008,Arsenault2020,Frachet2020}.
\begin{widetext}
\begin{equation}
\begin{split}
M(t) = M_{t\rightarrow\,\infty}-\left(M_{t\rightarrow\,\infty}-M_{t=0}\right) \left(0.714\,e^{-\left(\frac{28\,t}{T_1}\right)^\beta}+0.206\,e^{-\left(\frac{15\,t}{T_1}\right)^\beta}+0.068\,e^{-\left(\frac{6\,t}{T_1}\right)^\beta}+0.012\,e^{-\left(\frac{t}{T_1}\right)^\beta}\right) ,
\end{split}
\end{equation}
\end{widetext}

\begin{figure*}[t!]
\hspace*{-0.3cm}    
  \includegraphics[width=15cm]{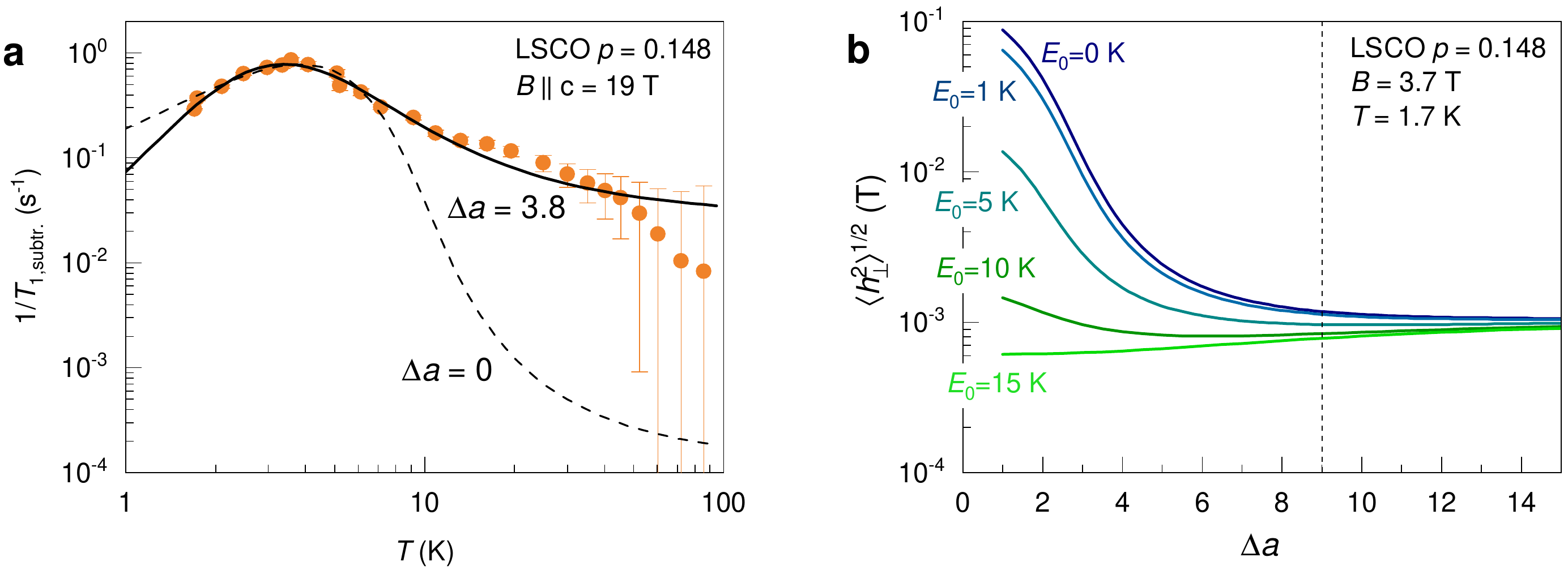}
  \caption{\label{delta_a} (a) Relaxation rate after background subtraction fitted by $T^{-1}_{1,\, \rm BPP\, dist.}(T)$ without distribution of $a=\ln\tau_{\infty}$ (dashed line) and with $\Delta a = 3.8 $ (where $\tau_{\infty}$ is in s$^{-1}$, black line). Notice that the distribution of $a$ is assumed to be uncorrelated from that of $E_0$ (see Eq.~\ref{eq:T1BPPfull}). (b) Simulated $\Delta a$ dependence of $\langle h_{\bot}^2 \rangle^{1/2}$ based on Eq.~\ref{eq:T1BPPfull} for an experimentally determined value $T^{\,-1}_{1}=0.018$~s$^{-1}$ at $T=1.7$~K and $B= 3.7$~T. For small $\Delta a$, $E_0$ affects $\langle h_{\bot}^2 \rangle$ strongly but above $\Delta a \sim 9$, the $\langle h_{\bot}^2 \rangle$ value is essentially independent of $E_0$. This allows to determine $\langle h_{\bot}^2 \rangle$ from a single $T^{\,-1}_{1}$ value at low $T$.}
\end{figure*}
\section{$\langle h_{\bot}^2 \rangle$ in the limit of large $\Delta a$}

The fact that for large $\Delta a$ and small $E_0$ the BPP peak becomes essentially $T$ independent implies that measuring the relaxation rate at a single temperature, e.g. $T=1.7$~K is sufficient to determine the fluctuating field $\langle h_{\bot}^2 \rangle$. It is the non-zero $\langle h_{\bot}^2 \rangle$ which leads to relaxation rates $(T_{1}T)^{-1}$ which are enhanced with respect to the normal state above $T_{\rm{c}}$ and the upturn in $T_{1}^{-1}(B)$ at the lowest fields. Fig.~\ref{delta_a}b illustrates that $\langle h_{\bot}^2 \rangle$ no longer depends on $E_0$ for $\Delta a \gtrsim 9$. According to Fig.~\ref{parameters14p8}d, for the $p=0.148$ sample this limit is reached for $B \lesssim 12$~T. For samples of higher doping it follows that even for fields somewhat larger than $B_{\rm{slow}}$ this limit is still valid.   

\section{Mean and median values of probability distribution functions}

\begin{figure*}[t!]
\hspace*{-0.3cm}    
  \includegraphics[width=17cm]{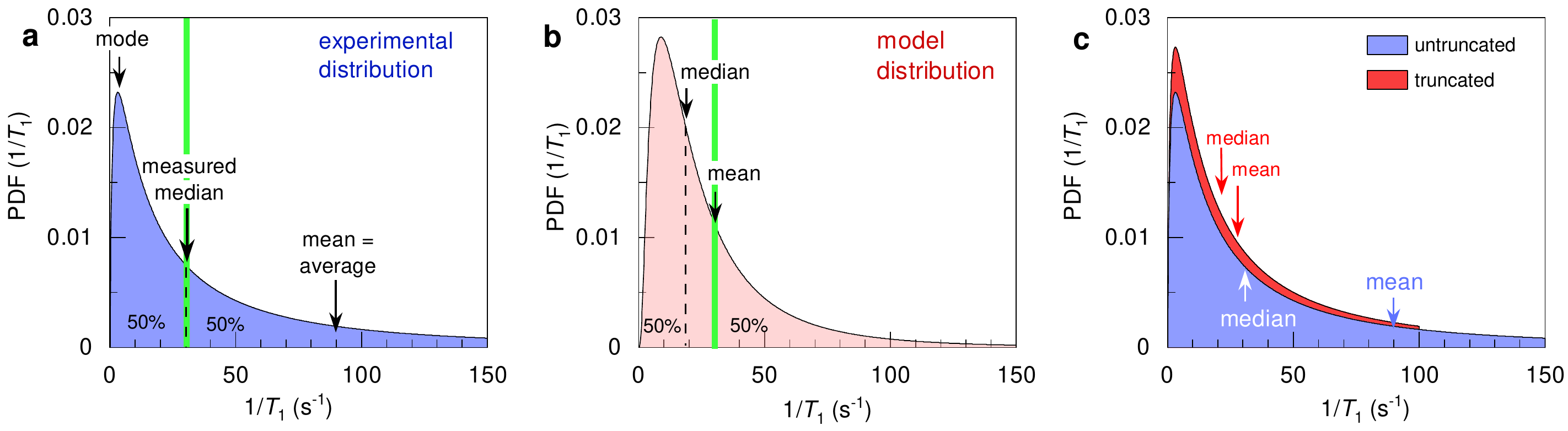}
  \caption{\label{distributions} Situation where the model distribution of relaxation rates (panel b) has a smaller asymmetry than the experimental distribution (panel a) such that $\frac{1}{T_1}_{\rm mean,\, model}(T) = \frac{1}{T_1}_{\rm median,\, experimental}(T)$ (thick vertical line). The model distribution is "wrong" since its median relaxation rate differs from the experimental median. (c) Regular (untruncated) and truncated log-normal probability distribution functions of the relaxation rate $\frac{1}{T_1}$. The regular log-normal distribution has a median value of 30~s$^{-1}$ and a much larger mean. For the truncated distribution ($1/T_1$(max)~=~100~s$^{-1}$) the mean decreases strongly, so one finds $\rm{median_{trunc}} \sim \rm{mean_{trunc}}$. After truncation the distribution is normalized again, so the integrated area is conserved. }
\end{figure*}

As mentioned in the main text, we are fitting the $T^{-1}_{1}$ values that correspond to the {\it median} of the experimental $T^{-1}_{1}$ distribution as shown in Fig.~\ref{distributions}a (because these values are obtained from stretched fits of the recoveries~\cite{Johnston2006}) by an expression $T^{-1}_{1,\, \rm BPP\, dist.}(T)$ (Eq.~\ref{eq:T1BPPfull}) that calculates the {\it mean} $T^{-1}_{1}$ of a model distribution, depicted in Fig.~\ref{distributions}b.

$T^{-1}_{1,\, \rm BPP\, dist.}(T)$ (Eq.~\ref{eq:T1BPPfull}) consists of a convolution of a function $f(x)$ with a probability density function PDF$(x-x_0)$ (in our case we convolute $T^{-1}_{1,\rm BPP}$ with a Gaussian where $x-x_0=E_0-E_{0,c}$) which by construction gives the {\it mean} value, not the median, of that function with respect to the probability density function (PDF):
\begin{equation*}
f_{\rm mean} = \int f(x) \cdot P\!D\!F(x-x_0) \, \rm d \textit{x} \, .
\end{equation*} As mentioned, the mean relaxation rate calculated from the model distribution depends only on the average fluctuating field, so an advantage of the mean $T^{-1}_{1}$ is that we do not need to assume a specific distribution of fluctuating fields. This would be necessary if we needed to calculate the histogram of the model distribution of $T^{-1}_{1}$.

In general, the median and mean of an asymmetric distribution with a long tail can differ significantly. In a realistic situation, however, the mean $T^{-1}_{1}$ does not differ greatly from the median because the distribution of $T^{-1}_{1}$ cannot extend to infinity and must be truncated (Fig.~\ref{distributions}c). Within the BPP model, according to Eq.~\ref{eq:max T1}, an infinite relaxation rate would imply an infinite fluctuating moment.

For the log-normal distribution there are analytical expressions available from Zaninetti~\citep{Zaninetti2017} for the relation of the mean and the median with which one can verify the effect of the truncation: Fig.~\ref{distributions}c displays that the mean of the truncated log-normal distribution, $\rm{mean_{trunc}}$, becomes much more comparable to the distribution median, $\rm{median_{trunc}}$. We thus see that, although the distribution of relaxation rates is asymmetric, the mean and median relaxation rates are similar, which justifies the fitting by Eq. \ref{eq:T1BPPfull}.

If we wanted to correctly model the experimental $T^{-1}_{1}$ by the median of a model distribution, this would require a specific distribution of fluctuating fields to numerically evaluate the histogram of the distributed relaxation rates. While we could make assumptions about the distribution of the fluctuating fields (see for example the inset of Fig.~\ref{PDF_ILT}), we do not have an efficient implementation of the process embedded in an automated fitting routine which finds the best set of model parameters.

\section{Comparison of distributed BPP method with ILT distributions}

\begin{figure*}[t!]
\hspace*{-0.3cm}    
  \includegraphics[width=10cm]{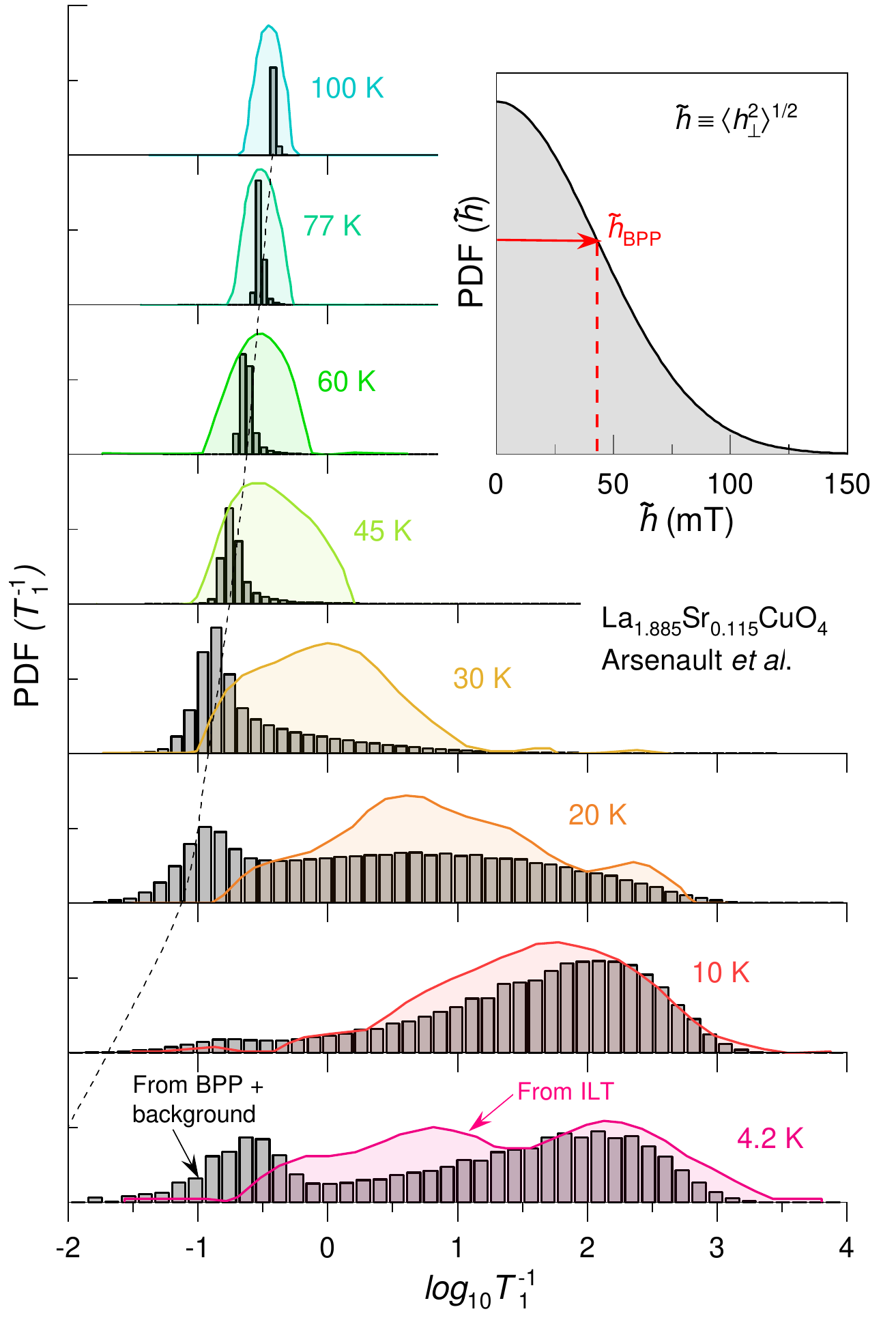}
  \caption{\label{PDF_ILT}Comparison of probability distributions PDF$(T_1^{-1})$ for La$_{1.885}$Sr$_{0.115}$CuO$_4$ at different temperatures based on the Inverse Laplace Transform (ILT) (colored curves, reproduced from Arsenault {\it et al.}~\citep{Arsenault2020}) with calculated distributions based on parameters $E_0=90.5$~K, $\Delta E_0=54.4$~K and $h_\bot=41.3$~mT from a distributed BPP fit of $T_1(T)$ data of Arsenault \textit{et al.} (grey histograms). The dashed trace follows the background relaxation rate discussed in the text (see also Fig.~\ref{T1}) and in ref.~\cite{Frachet2020}. The inset depicts the assumed half-Gaussian distribution of the fluctuating field, $\tilde{h}=\langle h_\bot^2 \rangle ^{1/2}$, whose width $\Delta \tilde{h}$ is taken to be equal to the 41.3~mT found in the BPP fit.}
\end{figure*} 

To assess the quality of the distributed BPP model, we would like to compare the distribution of $T_1^{-1}$ values that corresponds to the fitted BPP parameters  with the distribution obtained from the inverse Laplace transform (ILT). For this purpose, we extract the  parameters ($\langle h_{\bot}^2 \rangle$, $E_0$, $\Delta E_0$) from a distributed BPP fit of the $T_1^{-1}~vs.~T$ data of Arsenault \textit{et al.}~\citep{Arsenault2020} in La$_{1.885}$Sr$_{0.115}$CuO$_4$ and we calculate the corresponding $T_1^{-1}$ distribution, which we then compare with the calculated ILT of Arsenault \textit{et al.}

In our model, the maximum possible $T_1^{-1}$ value is $(T_1^{-1})_{\rm max} =  \langle h_{\bot}^2 \rangle\,\gamma_{n}^{2}/\omega_{L}$. (Eq.~\ref{eq:max T1}). As we did not introduce any distribution of $\langle h_{\bot}^2 \rangle$, $T_1^{-1}$ must reach this exact value at some temperature for  every nucleus. This means that we have introduced an artificial cutoff in the distribution (no higher value is allowed). Consequently, the probability density function (PDF) of $T_1^{-1}$ calculated from the BPP parameters diverges at $(T_1^{-1})_{\rm max}$ for temperatures close to the peak. Since this is unphysical, we introduce a distribution of $\langle h_{\bot}^2 \rangle$ values that smoothes this singularity. Defining $\tilde{h}=\langle h_{\bot}^2 \rangle ^{1/2}$ to simplify notations, we model $\tilde{h}$ by a half Gaussian (see inset of Fig.~\ref{PDF_ILT}) with a width $\Delta \tilde{h}$. For this particular distribution the mean value of $\tilde{h}^2$ is equal to $\Delta \tilde{h}^2$: $\frac{\sqrt{2}}{\sqrt{\pi} \Delta \tilde{h}}\int_0^\infty \tilde{h}^2\exp(-\frac{\tilde{h}^2}{2\Delta \tilde{h}^2})\mathrm{d} \tilde{h}= \Delta \tilde{h}^2$, so this distribution is consistent with the fluctuating fields determined from BPP fits provided that we set $\Delta \tilde{h}=\tilde{h}_{\rm{BPP}}$. Such a distribution, having its most probable value at $\tilde{h}=0$, is consistent with the distribution of static local fields $\langle h_{\bot} \rangle$ inferred by Hunt \textit{et al.} from $^{63}$Cu nuclear quadrupole resonance lineshapes in La$_{2-x}$Ba$_{x}$CuO$_4$~\citep{Hunt2001}.

There is a similar issue on the low $T_1^{-1}$ side of the PDF. While we have modeled the BPP relaxation after subtracting a background $T_1^{-1}$ (see Fig.~\ref{T1} and ref.~\citep{Frachet2020}), the ILT is performed on recoveries that contain all contributions to the relaxation. These include the background $T_1^{-1}$ that actually determines the smallest possible value of $T^{-1}_{1}$ at $T \ll T_{\rm{peak}}$. Indeed, based solely on the BPP mechanism, $T_1^{-1}$ would drop to arbitrarily low values as $T\rightarrow 0$. We thus need to reestablish the subtracted background "by hand" in order to avoid unphysically small values when we reverse-engineer the BPP parameters to get PDF$(T^{-1}_{1})$. Again, if all nuclei had the same $T_1^{-1}$ background, there would be a singularity in PDF$(T^{-1}_{1})$, so we distribute this background relaxation rate by a Gaussian, whose width increases as $1/T$ with cooling. The $1/T$ dependence mimics the increased width of PDF$(T^{-1}_{1})$ from high to low temperature and is chosen for the sake of simplicity (a more realistic choice would probably require introducing an onset temperature near $T_{\rm{CDW}}=80$~K, as discussed in the main text).

The calculation of PDF$(T^{-1}_{1})$ histograms shown in Fig.~\ref{PDF_ILT} takes place in two steps. First, we calculate $T^{-1}$ for each point ($E_0$, $\tilde{h}$, $T^{-1}_{1,\mathrm{backgr.}}$) in parameter space and its probability $P(E_0, \tilde{h}, T^{-1}_{1,\mathrm{backgr.}})$ considering the (half-)Gaussian distributions with widths $\Delta E_0$, $\Delta \tilde{h}$ and $\Delta T^{-1}_{1,\mathrm{backgr.}}$. Second, as we want to compare to ILT distributions that are plotted on logarithmic scale, the probabilities are sorted and binned together in 50 bins from $\log_{10}T^{-1}_{1,\mathrm{min}}$ to $\log_{10}T^{-1}_{1,\mathrm{max}}$ and each binned probability is weighted with its corresponding $T^{-1}_1$. We restrict the comparison to temperatures below 100~K, for which the data is correctly described by the BPP peak and/or the approximated background $T^{-1}_{1,\mathrm{backgr.}}\propto T$.

The agreement between the calculated PDF$(T^{-1}_{1})$ and the ILT is not perfect but some features of the ILT distributions are reproduced (Fig.~\ref{PDF_ILT}): i) the peak position initially shifts to lower values with cooling below 100~K, tracking $T^{-1}_{1,\mathrm{backgr.}}\propto T$. ii) the distribution eventually broadens, leading to the shoulder at intermediate temperatures. Asymmetric histograms, however, systematically underestimate the median value compared to the ILT distributions. iii) At base temperature 4.2~K, the two distributions are similar: two peaks appear, reminiscent of the bimodal ILT distribution, although the peak positions do not match. 
In the calculated histogram, the peak on the left hand side is rooted in $T^{-1}_{1,\mathrm{backgr.}}$ and is not centered at the dashed line but is shifted due to the weighting by $T^{-1}_{1}$ in combination with a large $\Delta T^{-1}_{1,\mathrm{backgr.}}$. Possibly, the left peak of the ILT distribution corresponds to frozen spins that relax at a faster background relaxation rate than assumed in our model.

We are of course aware that our distributed BPP analysis is based on modeling the data with over-simplified parametrization and various assumptions but it is a bold attempt to understand a very complex problem within a physically reasonable model. For instance, it might be argued that assuming additive contributions (the BPP freezing and the residual "background" relaxation) is itself suggestive of some form of phase separation. Nevertheless, any nucleus must always be affected by residual relaxation, for instance, relaxation by paramagnetic impurities at low $T$. Therefore, despite the shortcomings of the analysis, the findings suggest that a unimodal distribution of the BPP parameters can reasonably describe the inhomogeneous relaxation process and that the bimodal nature of the $T_1^{-1}$ distribution found in the ILT analysis could arise from residual relaxation processes yielding a lower bound to $T_1^{-1}$ values at low $T$, rather than from spatial differentiation~\cite{Arsenault2020}. More work is certainly needed to clarify the issues related to inhomogeneous nuclear relaxation.


\begin{thebibliography}{113}
\expandafter\ifx\csname natexlab\endcsname\relax\def\natexlab#1{#1}\fi
\expandafter\ifx\csname bibnamefont\endcsname\relax
  \def\bibnamefont#1{#1}\fi
\expandafter\ifx\csname bibfnamefont\endcsname\relax
  \def\bibfnamefont#1{#1}\fi
\expandafter\ifx\csname citenamefont\endcsname\relax
  \def\citenamefont#1{#1}\fi
\expandafter\ifx\csname url\endcsname\relax
  \def\url#1{\texttt{#1}}\fi
\expandafter\ifx\csname urlprefix\endcsname\relax\def\urlprefix{URL }\fi
\providecommand{\bibinfo}[2]{#2}
\providecommand{\eprint}[2][]{\url{#2}}

\bibitem[{\citenamefont{Lake et~al.}(2002)\citenamefont{Lake, R{\o}nnow,
  Christensen, Aeppli, Lefmann, McMorrow, Vorderwisch, Smeibidl, Mangkorntong,
  Sasagawa et~al.}}]{Lake2002}
\bibinfo{author}{\bibfnamefont{B.}~\bibnamefont{Lake}},
  \bibinfo{author}{\bibfnamefont{H.~M.} \bibnamefont{R{\o}nnow}},
  \bibinfo{author}{\bibfnamefont{N.~B.} \bibnamefont{Christensen}},
  \bibinfo{author}{\bibfnamefont{G.}~\bibnamefont{Aeppli}},
  \bibinfo{author}{\bibfnamefont{K.}~\bibnamefont{Lefmann}},
  \bibinfo{author}{\bibfnamefont{D.~F.} \bibnamefont{McMorrow}},
  \bibinfo{author}{\bibfnamefont{P.}~\bibnamefont{Vorderwisch}},
  \bibinfo{author}{\bibfnamefont{P.}~\bibnamefont{Smeibidl}},
  \bibinfo{author}{\bibfnamefont{N.}~\bibnamefont{Mangkorntong}},
  \bibinfo{author}{\bibfnamefont{T.}~\bibnamefont{Sasagawa}},
  \bibnamefont{et~al.}, \bibinfo{journal}{Nature}
  \textbf{\bibinfo{volume}{415}}, \bibinfo{pages}{299} (\bibinfo{year}{2002}),
  \urlprefix\url{https://doi.org/10.1038/415299a}.

\bibitem[{\citenamefont{Hoffman et~al.}(2002)\citenamefont{Hoffman, Hudson,
  Lang, Madhavan, Eisaki, Uchida, and Davis}}]{Hoffman2002}
\bibinfo{author}{\bibfnamefont{J.~E.} \bibnamefont{Hoffman}},
  \bibinfo{author}{\bibfnamefont{E.~W.} \bibnamefont{Hudson}},
  \bibinfo{author}{\bibfnamefont{K.~M.} \bibnamefont{Lang}},
  \bibinfo{author}{\bibfnamefont{V.}~\bibnamefont{Madhavan}},
  \bibinfo{author}{\bibfnamefont{H.}~\bibnamefont{Eisaki}},
  \bibinfo{author}{\bibfnamefont{S.}~\bibnamefont{Uchida}}, \bibnamefont{and}
  \bibinfo{author}{\bibfnamefont{J.~C.} \bibnamefont{Davis}},
  \bibinfo{journal}{Science} \textbf{\bibinfo{volume}{295}},
  \bibinfo{pages}{466} (\bibinfo{year}{2002}), ISSN \bibinfo{issn}{0036-8075},
  \urlprefix\url{https://science.sciencemag.org/content/295/5554/466}.

\bibitem[{\citenamefont{Mitrovi\ifmmode~\acute{c}\else \'{c}\fi{}
  et~al.}(2003)\citenamefont{Mitrovi\ifmmode~\acute{c}\else \'{c}\fi{},
  Sigmund, Halperin, Reyes, Kuhns, and Moulton}}]{Mitrovic2003}
\bibinfo{author}{\bibfnamefont{V.~F.}
  \bibnamefont{Mitrovi\ifmmode~\acute{c}\else \'{c}\fi{}}},
  \bibinfo{author}{\bibfnamefont{E.~E.} \bibnamefont{Sigmund}},
  \bibinfo{author}{\bibfnamefont{W.~P.} \bibnamefont{Halperin}},
  \bibinfo{author}{\bibfnamefont{A.~P.} \bibnamefont{Reyes}},
  \bibinfo{author}{\bibfnamefont{P.}~\bibnamefont{Kuhns}}, \bibnamefont{and}
  \bibinfo{author}{\bibfnamefont{W.~G.} \bibnamefont{Moulton}},
  \bibinfo{journal}{Phys. Rev. B} \textbf{\bibinfo{volume}{67}},
  \bibinfo{pages}{220503} (\bibinfo{year}{2003}),
  \urlprefix\url{https://link.aps.org/doi/10.1103/PhysRevB.67.220503}.

\bibitem[{\citenamefont{Wu et~al.}(2011)\citenamefont{Wu, Mayaffre, Kr{\"a}mer,
  Horvati{\'c}, Berthier, Hardy, Liang, Bonn, and Julien}}]{Wu2011}
\bibinfo{author}{\bibfnamefont{T.}~\bibnamefont{Wu}},
  \bibinfo{author}{\bibfnamefont{H.}~\bibnamefont{Mayaffre}},
  \bibinfo{author}{\bibfnamefont{S.}~\bibnamefont{Kr{\"a}mer}},
  \bibinfo{author}{\bibfnamefont{M.}~\bibnamefont{Horvati{\'c}}},
  \bibinfo{author}{\bibfnamefont{C.}~\bibnamefont{Berthier}},
  \bibinfo{author}{\bibfnamefont{W.~N.} \bibnamefont{Hardy}},
  \bibinfo{author}{\bibfnamefont{R.}~\bibnamefont{Liang}},
  \bibinfo{author}{\bibfnamefont{D.~A.} \bibnamefont{Bonn}}, \bibnamefont{and}
  \bibinfo{author}{\bibfnamefont{M.-H.} \bibnamefont{Julien}},
  \bibinfo{journal}{Nature} \textbf{\bibinfo{volume}{477}},
  \bibinfo{pages}{191} (\bibinfo{year}{2011}),
  \urlprefix\url{https://doi.org/10.1038/nature10345}.

\bibitem[{\citenamefont{Frachet et~al.}(2020)\citenamefont{Frachet, Vinograd,
  Zhou, Benhabib, Wu, Mayaffre, Kr{\"a}mer, Ramakrishna, Reyes, Debray
  et~al.}}]{Frachet2020}
\bibinfo{author}{\bibfnamefont{M.}~\bibnamefont{Frachet}},
  \bibinfo{author}{\bibfnamefont{I.}~\bibnamefont{Vinograd}},
  \bibinfo{author}{\bibfnamefont{R.}~\bibnamefont{Zhou}},
  \bibinfo{author}{\bibfnamefont{S.}~\bibnamefont{Benhabib}},
  \bibinfo{author}{\bibfnamefont{S.}~\bibnamefont{Wu}},
  \bibinfo{author}{\bibfnamefont{H.}~\bibnamefont{Mayaffre}},
  \bibinfo{author}{\bibfnamefont{S.}~\bibnamefont{Kr{\"a}mer}},
  \bibinfo{author}{\bibfnamefont{S.~K.} \bibnamefont{Ramakrishna}},
  \bibinfo{author}{\bibfnamefont{A.~P.} \bibnamefont{Reyes}},
  \bibinfo{author}{\bibfnamefont{J.}~\bibnamefont{Debray}},
  \bibnamefont{et~al.}, \bibinfo{journal}{Nature Physics}
  \textbf{\bibinfo{volume}{16}}, \bibinfo{pages}{1064} (\bibinfo{year}{2020}),
  \urlprefix\url{https://doi.org/10.1038/s41567-020-0950-5}.

\bibitem[{\citenamefont{Julien}(2003)}]{Julien2003}
\bibinfo{author}{\bibfnamefont{M.-H.} \bibnamefont{Julien}},
  \bibinfo{journal}{Physica B: Condensed Matter}
  \textbf{\bibinfo{volume}{329-333}}, \bibinfo{pages}{693}
  (\bibinfo{year}{2003}), ISSN \bibinfo{issn}{0921-4526},
  \bibinfo{note}{proceedings of the 23rd International Conference on Low
  Temperature Physics},
  \urlprefix\url{https://www.sciencedirect.com/science/article/pii/S092145260201997X}.

\bibitem[{\citenamefont{Collignon et~al.}(2017)\citenamefont{Collignon, Badoux,
  Afshar, Michon, Lalibert\'e, Cyr-Choini\`ere, Zhou, Licciardello, Wiedmann,
  Doiron-Leyraud et~al.}}]{Collignon2017}
\bibinfo{author}{\bibfnamefont{C.}~\bibnamefont{Collignon}},
  \bibinfo{author}{\bibfnamefont{S.}~\bibnamefont{Badoux}},
  \bibinfo{author}{\bibfnamefont{S.~A.~A.} \bibnamefont{Afshar}},
  \bibinfo{author}{\bibfnamefont{B.}~\bibnamefont{Michon}},
  \bibinfo{author}{\bibfnamefont{F.}~\bibnamefont{Lalibert\'e}},
  \bibinfo{author}{\bibfnamefont{O.}~\bibnamefont{Cyr-Choini\`ere}},
  \bibinfo{author}{\bibfnamefont{J.-S.} \bibnamefont{Zhou}},
  \bibinfo{author}{\bibfnamefont{S.}~\bibnamefont{Licciardello}},
  \bibinfo{author}{\bibfnamefont{S.}~\bibnamefont{Wiedmann}},
  \bibinfo{author}{\bibfnamefont{N.}~\bibnamefont{Doiron-Leyraud}},
  \bibnamefont{et~al.}, \bibinfo{journal}{Phys. Rev. B}
  \textbf{\bibinfo{volume}{95}}, \bibinfo{pages}{224517}
  (\bibinfo{year}{2017}),
  \urlprefix\url{https://link.aps.org/doi/10.1103/PhysRevB.95.224517}.

\bibitem[{\citenamefont{Michon et~al.}(2019)\citenamefont{Michon, Girod,
  Badoux, Ka{\v c}mar{\v c}{\'\i}k, Ma, Dragomir, Dabkowska, Gaulin, Zhou, Pyon
  et~al.}}]{Michon2019}
\bibinfo{author}{\bibfnamefont{B.}~\bibnamefont{Michon}},
  \bibinfo{author}{\bibfnamefont{C.}~\bibnamefont{Girod}},
  \bibinfo{author}{\bibfnamefont{S.}~\bibnamefont{Badoux}},
  \bibinfo{author}{\bibfnamefont{J.}~\bibnamefont{Ka{\v c}mar{\v c}{\'\i}k}},
  \bibinfo{author}{\bibfnamefont{Q.}~\bibnamefont{Ma}},
  \bibinfo{author}{\bibfnamefont{M.}~\bibnamefont{Dragomir}},
  \bibinfo{author}{\bibfnamefont{H.~A.} \bibnamefont{Dabkowska}},
  \bibinfo{author}{\bibfnamefont{B.~D.} \bibnamefont{Gaulin}},
  \bibinfo{author}{\bibfnamefont{J.~S.} \bibnamefont{Zhou}},
  \bibinfo{author}{\bibfnamefont{S.}~\bibnamefont{Pyon}}, \bibnamefont{et~al.},
  \bibinfo{journal}{Nature} \textbf{\bibinfo{volume}{567}},
  \bibinfo{pages}{218} (\bibinfo{year}{2019}),
  \urlprefix\url{https://doi.org/10.1038/s41586-019-0932-x}.

\bibitem[{\citenamefont{Fang et~al.}(2022)\citenamefont{Fang, Grissonnanche,
  Legros, Verret, Lalibert{\'e}, Collignon, Ataei, Dion, Zhou, Graf
  et~al.}}]{Fang2022}
\bibinfo{author}{\bibfnamefont{Y.}~\bibnamefont{Fang}},
  \bibinfo{author}{\bibfnamefont{G.}~\bibnamefont{Grissonnanche}},
  \bibinfo{author}{\bibfnamefont{A.}~\bibnamefont{Legros}},
  \bibinfo{author}{\bibfnamefont{S.}~\bibnamefont{Verret}},
  \bibinfo{author}{\bibfnamefont{F.}~\bibnamefont{Lalibert{\'e}}},
  \bibinfo{author}{\bibfnamefont{C.}~\bibnamefont{Collignon}},
  \bibinfo{author}{\bibfnamefont{A.}~\bibnamefont{Ataei}},
  \bibinfo{author}{\bibfnamefont{M.}~\bibnamefont{Dion}},
  \bibinfo{author}{\bibfnamefont{J.}~\bibnamefont{Zhou}},
  \bibinfo{author}{\bibfnamefont{D.}~\bibnamefont{Graf}}, \bibnamefont{et~al.},
  \bibinfo{journal}{Nature Physics} \textbf{\bibinfo{volume}{18}},
  \bibinfo{pages}{558} (\bibinfo{year}{2022}),
  \urlprefix\url{https://doi.org/10.1038/s41567-022-01514-1}.

\bibitem[{\citenamefont{Kitaoka et~al.}(1987)\citenamefont{Kitaoka, Hiramatsu,
  Ishida, Kohara, and Asayama}}]{Kitaoka1987}
\bibinfo{author}{\bibfnamefont{Y.}~\bibnamefont{Kitaoka}},
  \bibinfo{author}{\bibfnamefont{S.}~\bibnamefont{Hiramatsu}},
  \bibinfo{author}{\bibfnamefont{K.}~\bibnamefont{Ishida}},
  \bibinfo{author}{\bibfnamefont{T.}~\bibnamefont{Kohara}}, \bibnamefont{and}
  \bibinfo{author}{\bibfnamefont{K.}~\bibnamefont{Asayama}},
  \bibinfo{journal}{Journal of the Physical Society of Japan}
  \textbf{\bibinfo{volume}{56}}, \bibinfo{pages}{3024} (\bibinfo{year}{1987}),
  \eprint{https://doi.org/10.1143/JPSJ.56.3024},
  \urlprefix\url{https://doi.org/10.1143/JPSJ.56.3024}.

\bibitem[{\citenamefont{Ohsugi}(1996)}]{Ohsugi1996}
\bibinfo{author}{\bibfnamefont{S.}~\bibnamefont{Ohsugi}},
  \bibinfo{journal}{Czechoslovak Journal of Physics}
  \textbf{\bibinfo{volume}{46}}, \bibinfo{pages}{2669} (\bibinfo{year}{1996}),
  \urlprefix\url{https://doi.org/10.1007/BF02570321}.

\bibitem[{\citenamefont{Arsenault et~al.}(2018)\citenamefont{Arsenault,
  Takahashi, Imai, He, Lee, and Fujita}}]{Arsenault2018}
\bibinfo{author}{\bibfnamefont{A.}~\bibnamefont{Arsenault}},
  \bibinfo{author}{\bibfnamefont{S.~K.} \bibnamefont{Takahashi}},
  \bibinfo{author}{\bibfnamefont{T.}~\bibnamefont{Imai}},
  \bibinfo{author}{\bibfnamefont{W.}~\bibnamefont{He}},
  \bibinfo{author}{\bibfnamefont{Y.~S.} \bibnamefont{Lee}}, \bibnamefont{and}
  \bibinfo{author}{\bibfnamefont{M.}~\bibnamefont{Fujita}},
  \bibinfo{journal}{Phys. Rev. B} \textbf{\bibinfo{volume}{97}},
  \bibinfo{pages}{064511} (\bibinfo{year}{2018}),
  \urlprefix\url{https://link.aps.org/doi/10.1103/PhysRevB.97.064511}.

\bibitem[{\citenamefont{Wu et~al.}(2013{\natexlab{a}})\citenamefont{Wu,
  Mayaffre, Kr\"amer, Horvati\ifmmode~\acute{c}\else \'{c}\fi{}, Berthier, Lin,
  Haug, Loew, Hinkov, Keimer et~al.}}]{Wu2013}
\bibinfo{author}{\bibfnamefont{T.}~\bibnamefont{Wu}},
  \bibinfo{author}{\bibfnamefont{H.}~\bibnamefont{Mayaffre}},
  \bibinfo{author}{\bibfnamefont{S.}~\bibnamefont{Kr\"amer}},
  \bibinfo{author}{\bibfnamefont{M.}~\bibnamefont{Horvati\ifmmode~\acute{c}\else
  \'{c}\fi{}}}, \bibinfo{author}{\bibfnamefont{C.}~\bibnamefont{Berthier}},
  \bibinfo{author}{\bibfnamefont{C.~T.} \bibnamefont{Lin}},
  \bibinfo{author}{\bibfnamefont{D.}~\bibnamefont{Haug}},
  \bibinfo{author}{\bibfnamefont{T.}~\bibnamefont{Loew}},
  \bibinfo{author}{\bibfnamefont{V.}~\bibnamefont{Hinkov}},
  \bibinfo{author}{\bibfnamefont{B.}~\bibnamefont{Keimer}},
  \bibnamefont{et~al.}, \bibinfo{journal}{Phys. Rev. B}
  \textbf{\bibinfo{volume}{88}}, \bibinfo{pages}{014511}
  (\bibinfo{year}{2013}{\natexlab{a}}),
  \urlprefix\url{https://link.aps.org/doi/10.1103/PhysRevB.88.014511}.

\bibitem[{\citenamefont{Bloembergen et~al.}(1948)\citenamefont{Bloembergen,
  Purcell, and Pound}}]{BPP1948}
\bibinfo{author}{\bibfnamefont{N.}~\bibnamefont{Bloembergen}},
  \bibinfo{author}{\bibfnamefont{E.~M.} \bibnamefont{Purcell}},
  \bibnamefont{and} \bibinfo{author}{\bibfnamefont{R.~V.} \bibnamefont{Pound}},
  \bibinfo{journal}{Phys. Rev.} \textbf{\bibinfo{volume}{73}},
  \bibinfo{pages}{679} (\bibinfo{year}{1948}),
  \urlprefix\url{https://link.aps.org/doi/10.1103/PhysRev.73.679}.

\bibitem[{\citenamefont{Suh et~al.}(2000)\citenamefont{Suh, Hammel, H\"ucker,
  B\"uchner, Ammerahl, and Revcolevschi}}]{Suh2000}
\bibinfo{author}{\bibfnamefont{B.~J.} \bibnamefont{Suh}},
  \bibinfo{author}{\bibfnamefont{P.~C.} \bibnamefont{Hammel}},
  \bibinfo{author}{\bibfnamefont{M.}~\bibnamefont{H\"ucker}},
  \bibinfo{author}{\bibfnamefont{B.}~\bibnamefont{B\"uchner}},
  \bibinfo{author}{\bibfnamefont{U.}~\bibnamefont{Ammerahl}}, \bibnamefont{and}
  \bibinfo{author}{\bibfnamefont{A.}~\bibnamefont{Revcolevschi}},
  \bibinfo{journal}{Phys. Rev. B} \textbf{\bibinfo{volume}{61}},
  \bibinfo{pages}{R9265} (\bibinfo{year}{2000}),
  \urlprefix\url{https://link.aps.org/doi/10.1103/PhysRevB.61.R9265}.

\bibitem[{\citenamefont{Curro et~al.}(2000)\citenamefont{Curro, Hammel, Suh,
  H\"ucker, B\"uchner, Ammerahl, and Revcolevschi}}]{Curro2000}
\bibinfo{author}{\bibfnamefont{N.~J.} \bibnamefont{Curro}},
  \bibinfo{author}{\bibfnamefont{P.~C.} \bibnamefont{Hammel}},
  \bibinfo{author}{\bibfnamefont{B.~J.} \bibnamefont{Suh}},
  \bibinfo{author}{\bibfnamefont{M.}~\bibnamefont{H\"ucker}},
  \bibinfo{author}{\bibfnamefont{B.}~\bibnamefont{B\"uchner}},
  \bibinfo{author}{\bibfnamefont{U.}~\bibnamefont{Ammerahl}}, \bibnamefont{and}
  \bibinfo{author}{\bibfnamefont{A.}~\bibnamefont{Revcolevschi}},
  \bibinfo{journal}{Phys. Rev. Lett.} \textbf{\bibinfo{volume}{85}},
  \bibinfo{pages}{642} (\bibinfo{year}{2000}),
  \urlprefix\url{https://link.aps.org/doi/10.1103/PhysRevLett.85.642}.

\bibitem[{\citenamefont{Simovi\ifmmode~\check{c}\else \v{c}\fi{}
  et~al.}(2003)\citenamefont{Simovi\ifmmode~\check{c}\else \v{c}\fi{}, Hammel,
  H\"ucker, B\"uchner, and Revcolevschi}}]{Simovic2003}
\bibinfo{author}{\bibfnamefont{B.}~\bibnamefont{Simovi\ifmmode~\check{c}\else
  \v{c}\fi{}}}, \bibinfo{author}{\bibfnamefont{P.~C.} \bibnamefont{Hammel}},
  \bibinfo{author}{\bibfnamefont{M.}~\bibnamefont{H\"ucker}},
  \bibinfo{author}{\bibfnamefont{B.}~\bibnamefont{B\"uchner}},
  \bibnamefont{and}
  \bibinfo{author}{\bibfnamefont{A.}~\bibnamefont{Revcolevschi}},
  \bibinfo{journal}{Phys. Rev. B} \textbf{\bibinfo{volume}{68}},
  \bibinfo{pages}{012415} (\bibinfo{year}{2003}),
  \urlprefix\url{https://link.aps.org/doi/10.1103/PhysRevB.68.012415}.

\bibitem[{\citenamefont{Mitrovi\ifmmode~\acute{c}\else \'{c}\fi{}
  et~al.}(2008)\citenamefont{Mitrovi\ifmmode~\acute{c}\else \'{c}\fi{}, Julien,
  de~Vaulx, Horvati\ifmmode~\acute{c}\else \'{c}\fi{}, Berthier, Suzuki, and
  Yamada}}]{Mitrovic2008}
\bibinfo{author}{\bibfnamefont{V.~F.}
  \bibnamefont{Mitrovi\ifmmode~\acute{c}\else \'{c}\fi{}}},
  \bibinfo{author}{\bibfnamefont{M.-H.} \bibnamefont{Julien}},
  \bibinfo{author}{\bibfnamefont{C.}~\bibnamefont{de~Vaulx}},
  \bibinfo{author}{\bibfnamefont{M.}~\bibnamefont{Horvati\ifmmode~\acute{c}\else
  \'{c}\fi{}}}, \bibinfo{author}{\bibfnamefont{C.}~\bibnamefont{Berthier}},
  \bibinfo{author}{\bibfnamefont{T.}~\bibnamefont{Suzuki}}, \bibnamefont{and}
  \bibinfo{author}{\bibfnamefont{K.}~\bibnamefont{Yamada}},
  \bibinfo{journal}{Phys. Rev. B} \textbf{\bibinfo{volume}{78}},
  \bibinfo{pages}{014504} (\bibinfo{year}{2008}),
  \urlprefix\url{https://link.aps.org/doi/10.1103/PhysRevB.78.014504}.

\bibitem[{\citenamefont{Arovas et~al.}(1997)\citenamefont{Arovas, Berlinsky,
  Kallin, and Zhang}}]{Arovas1997}
\bibinfo{author}{\bibfnamefont{D.~P.} \bibnamefont{Arovas}},
  \bibinfo{author}{\bibfnamefont{A.~J.} \bibnamefont{Berlinsky}},
  \bibinfo{author}{\bibfnamefont{C.}~\bibnamefont{Kallin}}, \bibnamefont{and}
  \bibinfo{author}{\bibfnamefont{S.-C.} \bibnamefont{Zhang}},
  \bibinfo{journal}{Phys. Rev. Lett.} \textbf{\bibinfo{volume}{79}},
  \bibinfo{pages}{2871} (\bibinfo{year}{1997}),
  \urlprefix\url{https://link.aps.org/doi/10.1103/PhysRevLett.79.2871}.

\bibitem[{\citenamefont{Demler et~al.}(2001)\citenamefont{Demler, Sachdev, and
  Zhang}}]{Demler2001}
\bibinfo{author}{\bibfnamefont{E.}~\bibnamefont{Demler}},
  \bibinfo{author}{\bibfnamefont{S.}~\bibnamefont{Sachdev}}, \bibnamefont{and}
  \bibinfo{author}{\bibfnamefont{Y.}~\bibnamefont{Zhang}},
  \bibinfo{journal}{Phys. Rev. Lett.} \textbf{\bibinfo{volume}{87}},
  \bibinfo{pages}{067202} (\bibinfo{year}{2001}),
  \urlprefix\url{https://link.aps.org/doi/10.1103/PhysRevLett.87.067202}.

\bibitem[{\citenamefont{Sachdev and Zhang}(2002)}]{Sachdev2002}
\bibinfo{author}{\bibfnamefont{S.}~\bibnamefont{Sachdev}} \bibnamefont{and}
  \bibinfo{author}{\bibfnamefont{S.-C.} \bibnamefont{Zhang}},
  \bibinfo{journal}{Science} \textbf{\bibinfo{volume}{295}},
  \bibinfo{pages}{452} (\bibinfo{year}{2002}),
  \eprint{https://www.science.org/doi/pdf/10.1126/science.1068718},
  \urlprefix\url{https://www.science.org/doi/abs/10.1126/science.1068718}.

\bibitem[{\citenamefont{Hu and Zhang}(2002)}]{Hu2002}
\bibinfo{author}{\bibfnamefont{J.-P.} \bibnamefont{Hu}} \bibnamefont{and}
  \bibinfo{author}{\bibfnamefont{S.-C.} \bibnamefont{Zhang}},
  \bibinfo{journal}{Journal of Physics and Chemistry of Solids}
  \textbf{\bibinfo{volume}{63}}, \bibinfo{pages}{2277} (\bibinfo{year}{2002}).

\bibitem[{\citenamefont{Kivelson et~al.}(2002)\citenamefont{Kivelson, Lee,
  Fradkin, and Oganesyan}}]{Kivelson2002}
\bibinfo{author}{\bibfnamefont{S.~A.} \bibnamefont{Kivelson}},
  \bibinfo{author}{\bibfnamefont{D.-H.} \bibnamefont{Lee}},
  \bibinfo{author}{\bibfnamefont{E.}~\bibnamefont{Fradkin}}, \bibnamefont{and}
  \bibinfo{author}{\bibfnamefont{V.}~\bibnamefont{Oganesyan}},
  \bibinfo{journal}{Phys. Rev. B} \textbf{\bibinfo{volume}{66}},
  \bibinfo{pages}{144516} (\bibinfo{year}{2002}),
  \urlprefix\url{https://link.aps.org/doi/10.1103/PhysRevB.66.144516}.

\bibitem[{\citenamefont{Franz et~al.}(2002)\citenamefont{Franz, Sheehy, and
  Te\ifmmode \check{s}\else \v{s}\fi{}anovi\ifmmode~\acute{c}\else
  \'{c}\fi{}}}]{Franz2002}
\bibinfo{author}{\bibfnamefont{M.}~\bibnamefont{Franz}},
  \bibinfo{author}{\bibfnamefont{D.~E.} \bibnamefont{Sheehy}},
  \bibnamefont{and} \bibinfo{author}{\bibfnamefont{Z.}~\bibnamefont{Te\ifmmode
  \check{s}\else \v{s}\fi{}anovi\ifmmode~\acute{c}\else \'{c}\fi{}}},
  \bibinfo{journal}{Phys. Rev. Lett.} \textbf{\bibinfo{volume}{88}},
  \bibinfo{pages}{257005} (\bibinfo{year}{2002}),
  \urlprefix\url{https://link.aps.org/doi/10.1103/PhysRevLett.88.257005}.

\bibitem[{\citenamefont{Zhu et~al.}(2002)\citenamefont{Zhu, Martin, and
  Bishop}}]{Zhu2002}
\bibinfo{author}{\bibfnamefont{J.-X.} \bibnamefont{Zhu}},
  \bibinfo{author}{\bibfnamefont{I.}~\bibnamefont{Martin}}, \bibnamefont{and}
  \bibinfo{author}{\bibfnamefont{A.~R.} \bibnamefont{Bishop}},
  \bibinfo{journal}{Phys. Rev. Lett.} \textbf{\bibinfo{volume}{89}},
  \bibinfo{pages}{067003} (\bibinfo{year}{2002}),
  \urlprefix\url{https://link.aps.org/doi/10.1103/PhysRevLett.89.067003}.

\bibitem[{\citenamefont{Ghosal et~al.}(2002)\citenamefont{Ghosal, Kallin, and
  Berlinsky}}]{Ghosal2002}
\bibinfo{author}{\bibfnamefont{A.}~\bibnamefont{Ghosal}},
  \bibinfo{author}{\bibfnamefont{C.}~\bibnamefont{Kallin}}, \bibnamefont{and}
  \bibinfo{author}{\bibfnamefont{A.~J.} \bibnamefont{Berlinsky}},
  \bibinfo{journal}{Phys. Rev. B} \textbf{\bibinfo{volume}{66}},
  \bibinfo{pages}{214502} (\bibinfo{year}{2002}),
  \urlprefix\url{https://link.aps.org/doi/10.1103/PhysRevB.66.214502}.

\bibitem[{\citenamefont{Andersen et~al.}(2003)\citenamefont{Andersen,
  Hedeg\aa{}rd, and Bruus}}]{Andersen2003}
\bibinfo{author}{\bibfnamefont{B.~M.} \bibnamefont{Andersen}},
  \bibinfo{author}{\bibfnamefont{P.}~\bibnamefont{Hedeg\aa{}rd}},
  \bibnamefont{and} \bibinfo{author}{\bibfnamefont{H.}~\bibnamefont{Bruus}},
  \bibinfo{journal}{Phys. Rev. B} \textbf{\bibinfo{volume}{67}},
  \bibinfo{pages}{134528} (\bibinfo{year}{2003}),
  \urlprefix\url{https://link.aps.org/doi/10.1103/PhysRevB.67.134528}.

\bibitem[{\citenamefont{Julien et~al.}(1999)\citenamefont{Julien, Borsa,
  Carretta, Horvati\ifmmode~\acute{c}\else \'{c}\fi{}, Berthier, and
  Lin}}]{Julien1999}
\bibinfo{author}{\bibfnamefont{M.-H.} \bibnamefont{Julien}},
  \bibinfo{author}{\bibfnamefont{F.}~\bibnamefont{Borsa}},
  \bibinfo{author}{\bibfnamefont{P.}~\bibnamefont{Carretta}},
  \bibinfo{author}{\bibfnamefont{M.}~\bibnamefont{Horvati\ifmmode~\acute{c}\else
  \'{c}\fi{}}}, \bibinfo{author}{\bibfnamefont{C.}~\bibnamefont{Berthier}},
  \bibnamefont{and} \bibinfo{author}{\bibfnamefont{C.~T.} \bibnamefont{Lin}},
  \bibinfo{journal}{Phys. Rev. Lett.} \textbf{\bibinfo{volume}{83}},
  \bibinfo{pages}{604} (\bibinfo{year}{1999}),
  \urlprefix\url{https://link.aps.org/doi/10.1103/PhysRevLett.83.604}.

\bibitem[{\citenamefont{Teitel'baum et~al.}(2000)\citenamefont{Teitel'baum,
  Abu-Shiekah, Bakharev, Brom, and Zaanen}}]{Teitelbaum2000}
\bibinfo{author}{\bibfnamefont{G.~B.} \bibnamefont{Teitel'baum}},
  \bibinfo{author}{\bibfnamefont{I.~M.} \bibnamefont{Abu-Shiekah}},
  \bibinfo{author}{\bibfnamefont{O.}~\bibnamefont{Bakharev}},
  \bibinfo{author}{\bibfnamefont{H.~B.} \bibnamefont{Brom}}, \bibnamefont{and}
  \bibinfo{author}{\bibfnamefont{J.}~\bibnamefont{Zaanen}},
  \bibinfo{journal}{Phys. Rev. B} \textbf{\bibinfo{volume}{63}},
  \bibinfo{pages}{020507} (\bibinfo{year}{2000}),
  \urlprefix\url{https://link.aps.org/doi/10.1103/PhysRevB.63.020507}.

\bibitem[{\citenamefont{Julien et~al.}(2001)\citenamefont{Julien, Campana,
  Rigamonti, Carretta, Borsa, Kuhns, Reyes, Moulton,
  Horvati\ifmmode~\acute{c}\else \'{c}\fi{}, Berthier et~al.}}]{Julien2001}
\bibinfo{author}{\bibfnamefont{M.-H.} \bibnamefont{Julien}},
  \bibinfo{author}{\bibfnamefont{A.}~\bibnamefont{Campana}},
  \bibinfo{author}{\bibfnamefont{A.}~\bibnamefont{Rigamonti}},
  \bibinfo{author}{\bibfnamefont{P.}~\bibnamefont{Carretta}},
  \bibinfo{author}{\bibfnamefont{F.}~\bibnamefont{Borsa}},
  \bibinfo{author}{\bibfnamefont{P.}~\bibnamefont{Kuhns}},
  \bibinfo{author}{\bibfnamefont{A.~P.} \bibnamefont{Reyes}},
  \bibinfo{author}{\bibfnamefont{W.~G.} \bibnamefont{Moulton}},
  \bibinfo{author}{\bibfnamefont{M.}~\bibnamefont{Horvati\ifmmode~\acute{c}\else
  \'{c}\fi{}}}, \bibinfo{author}{\bibfnamefont{C.}~\bibnamefont{Berthier}},
  \bibnamefont{et~al.}, \bibinfo{journal}{Phys. Rev. B}
  \textbf{\bibinfo{volume}{63}}, \bibinfo{pages}{144508}
  (\bibinfo{year}{2001}),
  \urlprefix\url{https://link.aps.org/doi/10.1103/PhysRevB.63.144508}.

\bibitem[{\citenamefont{Hunt et~al.}(2001)\citenamefont{Hunt, Singer,
  Cederstr\"om, and Imai}}]{Hunt2001}
\bibinfo{author}{\bibfnamefont{A.~W.} \bibnamefont{Hunt}},
  \bibinfo{author}{\bibfnamefont{P.~M.} \bibnamefont{Singer}},
  \bibinfo{author}{\bibfnamefont{A.~F.} \bibnamefont{Cederstr\"om}},
  \bibnamefont{and} \bibinfo{author}{\bibfnamefont{T.}~\bibnamefont{Imai}},
  \bibinfo{journal}{Phys. Rev. B} \textbf{\bibinfo{volume}{64}},
  \bibinfo{pages}{134525} (\bibinfo{year}{2001}),
  \urlprefix\url{https://link.aps.org/doi/10.1103/PhysRevB.64.134525}.

\bibitem[{\citenamefont{Baek et~al.}(2015)\citenamefont{Baek, Utz, H\"ucker,
  Gu, B\"uchner, and Grafe}}]{Baek2015}
\bibinfo{author}{\bibfnamefont{S.-H.} \bibnamefont{Baek}},
  \bibinfo{author}{\bibfnamefont{Y.}~\bibnamefont{Utz}},
  \bibinfo{author}{\bibfnamefont{M.}~\bibnamefont{H\"ucker}},
  \bibinfo{author}{\bibfnamefont{G.~D.} \bibnamefont{Gu}},
  \bibinfo{author}{\bibfnamefont{B.}~\bibnamefont{B\"uchner}},
  \bibnamefont{and} \bibinfo{author}{\bibfnamefont{H.-J.} \bibnamefont{Grafe}},
  \bibinfo{journal}{Phys. Rev. B} \textbf{\bibinfo{volume}{92}},
  \bibinfo{pages}{155144} (\bibinfo{year}{2015}),
  \urlprefix\url{https://link.aps.org/doi/10.1103/PhysRevB.92.155144}.

\bibitem[{\citenamefont{Baek et~al.}(2017)\citenamefont{Baek, Erb, and
  B\"uchner}}]{Baek2017}
\bibinfo{author}{\bibfnamefont{S.-H.} \bibnamefont{Baek}},
  \bibinfo{author}{\bibfnamefont{A.}~\bibnamefont{Erb}}, \bibnamefont{and}
  \bibinfo{author}{\bibfnamefont{B.}~\bibnamefont{B\"uchner}},
  \bibinfo{journal}{Phys. Rev. B} \textbf{\bibinfo{volume}{96}},
  \bibinfo{pages}{094519} (\bibinfo{year}{2017}),
  \urlprefix\url{https://link.aps.org/doi/10.1103/PhysRevB.96.094519}.

\bibitem[{\citenamefont{Arsenault et~al.}(2020)\citenamefont{Arsenault, Imai,
  Singer, Suzuki, and Fujita}}]{Arsenault2020}
\bibinfo{author}{\bibfnamefont{A.}~\bibnamefont{Arsenault}},
  \bibinfo{author}{\bibfnamefont{T.}~\bibnamefont{Imai}},
  \bibinfo{author}{\bibfnamefont{P.~M.} \bibnamefont{Singer}},
  \bibinfo{author}{\bibfnamefont{K.~M.} \bibnamefont{Suzuki}},
  \bibnamefont{and} \bibinfo{author}{\bibfnamefont{M.}~\bibnamefont{Fujita}},
  \bibinfo{journal}{Phys. Rev. B} \textbf{\bibinfo{volume}{101}},
  \bibinfo{pages}{184505} (\bibinfo{year}{2020}),
  \urlprefix\url{https://link.aps.org/doi/10.1103/PhysRevB.101.184505}.

\bibitem[{\citenamefont{Singer et~al.}(2020)\citenamefont{Singer, Arsenault,
  Imai, and Fujita}}]{Singer2020}
\bibinfo{author}{\bibfnamefont{P.~M.} \bibnamefont{Singer}},
  \bibinfo{author}{\bibfnamefont{A.}~\bibnamefont{Arsenault}},
  \bibinfo{author}{\bibfnamefont{T.}~\bibnamefont{Imai}}, \bibnamefont{and}
  \bibinfo{author}{\bibfnamefont{M.}~\bibnamefont{Fujita}},
  \bibinfo{journal}{Phys. Rev. B} \textbf{\bibinfo{volume}{101}},
  \bibinfo{pages}{174508} (\bibinfo{year}{2020}),
  \urlprefix\url{https://link.aps.org/doi/10.1103/PhysRevB.101.174508}.

\bibitem[{\citenamefont{Wakimoto et~al.}(2007)\citenamefont{Wakimoto, Yamada,
  Tranquada, Frost, Birgeneau, and Zhang}}]{Wakimoto2007}
\bibinfo{author}{\bibfnamefont{S.}~\bibnamefont{Wakimoto}},
  \bibinfo{author}{\bibfnamefont{K.}~\bibnamefont{Yamada}},
  \bibinfo{author}{\bibfnamefont{J.~M.} \bibnamefont{Tranquada}},
  \bibinfo{author}{\bibfnamefont{C.~D.} \bibnamefont{Frost}},
  \bibinfo{author}{\bibfnamefont{R.~J.} \bibnamefont{Birgeneau}},
  \bibnamefont{and} \bibinfo{author}{\bibfnamefont{H.}~\bibnamefont{Zhang}},
  \bibinfo{journal}{Phys. Rev. Lett.} \textbf{\bibinfo{volume}{98}},
  \bibinfo{pages}{247003} (\bibinfo{year}{2007}),
  \urlprefix\url{https://link.aps.org/doi/10.1103/PhysRevLett.98.247003}.

\bibitem[{\citenamefont{Singer et~al.}(2002)\citenamefont{Singer, Hunt, and
  Imai}}]{Singer2002}
\bibinfo{author}{\bibfnamefont{P.~M.} \bibnamefont{Singer}},
  \bibinfo{author}{\bibfnamefont{A.~W.} \bibnamefont{Hunt}}, \bibnamefont{and}
  \bibinfo{author}{\bibfnamefont{T.}~\bibnamefont{Imai}},
  \bibinfo{journal}{Phys. Rev. Lett.} \textbf{\bibinfo{volume}{88}},
  \bibinfo{pages}{047602} (\bibinfo{year}{2002}),
  \urlprefix\url{https://link.aps.org/doi/10.1103/PhysRevLett.88.047602}.

\bibitem[{\citenamefont{Choi et~al.}(2021)\citenamefont{Choi, Vinograd,
  Chaffey, and Curro}}]{Choi2021}
\bibinfo{author}{\bibfnamefont{H.}~\bibnamefont{Choi}},
  \bibinfo{author}{\bibfnamefont{I.}~\bibnamefont{Vinograd}},
  \bibinfo{author}{\bibfnamefont{C.}~\bibnamefont{Chaffey}}, \bibnamefont{and}
  \bibinfo{author}{\bibfnamefont{N.}~\bibnamefont{Curro}},
  \bibinfo{journal}{Journal of Magnetic Resonance}
  \textbf{\bibinfo{volume}{331}}, \bibinfo{pages}{107050}
  (\bibinfo{year}{2021}), ISSN \bibinfo{issn}{1090-7807},
  \urlprefix\url{https://www.sciencedirect.com/science/article/pii/S1090780721001397}.

\bibitem[{\citenamefont{Curro}(2004)}]{Curro2004}
\bibinfo{author}{\bibfnamefont{N.~J.} \bibnamefont{Curro}}, in
  \emph{\bibinfo{booktitle}{Fluctuations and Noise in Materials}}, edited by
  \bibinfo{editor}{\bibfnamefont{D.}~\bibnamefont{Popovic}},
  \bibinfo{editor}{\bibfnamefont{M.~B.} \bibnamefont{Weissman}},
  \bibnamefont{and} \bibinfo{editor}{\bibfnamefont{Z.~A.} \bibnamefont{Racz}},
  \bibinfo{organization}{International Society for Optics and Photonics}
  (\bibinfo{publisher}{SPIE}, \bibinfo{year}{2004}), vol.
  \bibinfo{volume}{5469}, pp. \bibinfo{pages}{114 -- 124},
  \urlprefix\url{https://doi.org/10.1117/12.537625}.

\bibitem[{\citenamefont{Johnston}(2006)}]{Johnston2006}
\bibinfo{author}{\bibfnamefont{D.~C.} \bibnamefont{Johnston}},
  \bibinfo{journal}{Phys. Rev. B} \textbf{\bibinfo{volume}{74}},
  \bibinfo{pages}{184430} (\bibinfo{year}{2006}),
  \urlprefix\url{https://link.aps.org/doi/10.1103/PhysRevB.74.184430}.

\bibitem[{\citenamefont{Cho et~al.}(1992)\citenamefont{Cho, Borsa, Johnston,
  and Torgeson}}]{Cho1992}
\bibinfo{author}{\bibfnamefont{J.~H.} \bibnamefont{Cho}},
  \bibinfo{author}{\bibfnamefont{F.}~\bibnamefont{Borsa}},
  \bibinfo{author}{\bibfnamefont{D.~C.} \bibnamefont{Johnston}},
  \bibnamefont{and} \bibinfo{author}{\bibfnamefont{D.~R.}
  \bibnamefont{Torgeson}}, \bibinfo{journal}{Phys. Rev. B}
  \textbf{\bibinfo{volume}{46}}, \bibinfo{pages}{3179} (\bibinfo{year}{1992}),
  \urlprefix\url{https://link.aps.org/doi/10.1103/PhysRevB.46.3179}.

\bibitem[{\citenamefont{Julien et~al.}(2000)\citenamefont{Julien, Carretta, and
  Borsa}}]{Julien2000}
\bibinfo{author}{\bibfnamefont{M.-H.} \bibnamefont{Julien}},
  \bibinfo{author}{\bibfnamefont{P.}~\bibnamefont{Carretta}}, \bibnamefont{and}
  \bibinfo{author}{\bibfnamefont{F.}~\bibnamefont{Borsa}},
  \bibinfo{journal}{Applied Magnetic Resonance} \textbf{\bibinfo{volume}{19}},
  \bibinfo{pages}{287} (\bibinfo{year}{2000}),
  \urlprefix\url{https://doi.org/10.1007/BF03162370}.

\bibitem[{\citenamefont{Frachet et~al.}(2021)\citenamefont{Frachet, Benhabib,
  Vinograd, Wu, Vignolle, Mayaffre, Kr\"amer, Kurosawa, Momono, Oda
  et~al.}}]{Frachet2021}
\bibinfo{author}{\bibfnamefont{M.}~\bibnamefont{Frachet}},
  \bibinfo{author}{\bibfnamefont{S.}~\bibnamefont{Benhabib}},
  \bibinfo{author}{\bibfnamefont{I.}~\bibnamefont{Vinograd}},
  \bibinfo{author}{\bibfnamefont{S.-F.} \bibnamefont{Wu}},
  \bibinfo{author}{\bibfnamefont{B.}~\bibnamefont{Vignolle}},
  \bibinfo{author}{\bibfnamefont{H.}~\bibnamefont{Mayaffre}},
  \bibinfo{author}{\bibfnamefont{S.}~\bibnamefont{Kr\"amer}},
  \bibinfo{author}{\bibfnamefont{T.}~\bibnamefont{Kurosawa}},
  \bibinfo{author}{\bibfnamefont{N.}~\bibnamefont{Momono}},
  \bibinfo{author}{\bibfnamefont{M.}~\bibnamefont{Oda}}, \bibnamefont{et~al.},
  \bibinfo{journal}{Phys. Rev. B} \textbf{\bibinfo{volume}{103}},
  \bibinfo{pages}{115133} (\bibinfo{year}{2021}),
  \urlprefix\url{https://link.aps.org/doi/10.1103/PhysRevB.103.115133}.

\bibitem[{\citenamefont{Hammerath et~al.}(2013)\citenamefont{Hammerath,
  Gr\"afe, K\"uhne, K\"uhne, Kuhns, Reyes, Lang, Wurmehl, B\"uchner, Carretta
  et~al.}}]{Hammerath2013}
\bibinfo{author}{\bibfnamefont{F.}~\bibnamefont{Hammerath}},
  \bibinfo{author}{\bibfnamefont{U.}~\bibnamefont{Gr\"afe}},
  \bibinfo{author}{\bibfnamefont{T.}~\bibnamefont{K\"uhne}},
  \bibinfo{author}{\bibfnamefont{H.}~\bibnamefont{K\"uhne}},
  \bibinfo{author}{\bibfnamefont{P.~L.} \bibnamefont{Kuhns}},
  \bibinfo{author}{\bibfnamefont{A.~P.} \bibnamefont{Reyes}},
  \bibinfo{author}{\bibfnamefont{G.}~\bibnamefont{Lang}},
  \bibinfo{author}{\bibfnamefont{S.}~\bibnamefont{Wurmehl}},
  \bibinfo{author}{\bibfnamefont{B.}~\bibnamefont{B\"uchner}},
  \bibinfo{author}{\bibfnamefont{P.}~\bibnamefont{Carretta}},
  \bibnamefont{et~al.}, \bibinfo{journal}{Phys. Rev. B}
  \textbf{\bibinfo{volume}{88}}, \bibinfo{pages}{104503}
  (\bibinfo{year}{2013}),
  \urlprefix\url{https://link.aps.org/doi/10.1103/PhysRevB.88.104503}.

\bibitem[{\citenamefont{Haug et~al.}(2010)\citenamefont{Haug, Hinkov, Sidis,
  Bourges, Christensen, Ivanov, Keller, Lin, and Keimer}}]{Haug2010}
\bibinfo{author}{\bibfnamefont{D.}~\bibnamefont{Haug}},
  \bibinfo{author}{\bibfnamefont{V.}~\bibnamefont{Hinkov}},
  \bibinfo{author}{\bibfnamefont{Y.}~\bibnamefont{Sidis}},
  \bibinfo{author}{\bibfnamefont{P.}~\bibnamefont{Bourges}},
  \bibinfo{author}{\bibfnamefont{N.~B.} \bibnamefont{Christensen}},
  \bibinfo{author}{\bibfnamefont{A.}~\bibnamefont{Ivanov}},
  \bibinfo{author}{\bibfnamefont{T.}~\bibnamefont{Keller}},
  \bibinfo{author}{\bibfnamefont{C.~T.} \bibnamefont{Lin}}, \bibnamefont{and}
  \bibinfo{author}{\bibfnamefont{B.}~\bibnamefont{Keimer}},
  \bibinfo{journal}{New Journal of Physics} \textbf{\bibinfo{volume}{12}},
  \bibinfo{pages}{105006} (\bibinfo{year}{2010}),
  \urlprefix\url{https://doi.org/10.1088/1367-2630/12/10/105006}.

\bibitem[{\citenamefont{Chang et~al.}(2008)\citenamefont{Chang, Niedermayer,
  Gilardi, Christensen, R\o{}nnow, McMorrow, Ay, Stahn, Sobolev, Hiess
  et~al.}}]{Chang2008}
\bibinfo{author}{\bibfnamefont{J.}~\bibnamefont{Chang}},
  \bibinfo{author}{\bibfnamefont{C.}~\bibnamefont{Niedermayer}},
  \bibinfo{author}{\bibfnamefont{R.}~\bibnamefont{Gilardi}},
  \bibinfo{author}{\bibfnamefont{N.~B.} \bibnamefont{Christensen}},
  \bibinfo{author}{\bibfnamefont{H.~M.} \bibnamefont{R\o{}nnow}},
  \bibinfo{author}{\bibfnamefont{D.~F.} \bibnamefont{McMorrow}},
  \bibinfo{author}{\bibfnamefont{M.}~\bibnamefont{Ay}},
  \bibinfo{author}{\bibfnamefont{J.}~\bibnamefont{Stahn}},
  \bibinfo{author}{\bibfnamefont{O.}~\bibnamefont{Sobolev}},
  \bibinfo{author}{\bibfnamefont{A.}~\bibnamefont{Hiess}},
  \bibnamefont{et~al.}, \bibinfo{journal}{Phys. Rev. B}
  \textbf{\bibinfo{volume}{78}}, \bibinfo{pages}{104525}
  (\bibinfo{year}{2008}),
  \urlprefix\url{https://link.aps.org/doi/10.1103/PhysRevB.78.104525}.

\bibitem[{\citenamefont{R\o{}mer et~al.}(2013)\citenamefont{R\o{}mer, Chang,
  Christensen, Andersen, Lefmann, M\"ahler, Gavilano, Gilardi, Niedermayer,
  R\o{}nnow et~al.}}]{Romer2013}
\bibinfo{author}{\bibfnamefont{A.~T.} \bibnamefont{R\o{}mer}},
  \bibinfo{author}{\bibfnamefont{J.}~\bibnamefont{Chang}},
  \bibinfo{author}{\bibfnamefont{N.~B.} \bibnamefont{Christensen}},
  \bibinfo{author}{\bibfnamefont{B.~M.} \bibnamefont{Andersen}},
  \bibinfo{author}{\bibfnamefont{K.}~\bibnamefont{Lefmann}},
  \bibinfo{author}{\bibfnamefont{L.}~\bibnamefont{M\"ahler}},
  \bibinfo{author}{\bibfnamefont{J.}~\bibnamefont{Gavilano}},
  \bibinfo{author}{\bibfnamefont{R.}~\bibnamefont{Gilardi}},
  \bibinfo{author}{\bibfnamefont{C.}~\bibnamefont{Niedermayer}},
  \bibinfo{author}{\bibfnamefont{H.~M.} \bibnamefont{R\o{}nnow}},
  \bibnamefont{et~al.}, \bibinfo{journal}{Phys. Rev. B}
  \textbf{\bibinfo{volume}{87}}, \bibinfo{pages}{144513}
  (\bibinfo{year}{2013}),
  \urlprefix\url{https://link.aps.org/doi/10.1103/PhysRevB.87.144513}.

\bibitem[{\citenamefont{Bo\ifmmode~\check{z}\else \v{z}\fi{}in
  et~al.}(1999)\citenamefont{Bo\ifmmode~\check{z}\else \v{z}\fi{}in, Billinge,
  Kwei, and Takagi}}]{Bozin1999}
\bibinfo{author}{\bibfnamefont{E.~S.} \bibnamefont{Bo\ifmmode~\check{z}\else
  \v{z}\fi{}in}}, \bibinfo{author}{\bibfnamefont{S.~J.~L.}
  \bibnamefont{Billinge}}, \bibinfo{author}{\bibfnamefont{G.~H.}
  \bibnamefont{Kwei}}, \bibnamefont{and}
  \bibinfo{author}{\bibfnamefont{H.}~\bibnamefont{Takagi}},
  \bibinfo{journal}{Phys. Rev. B} \textbf{\bibinfo{volume}{59}},
  \bibinfo{pages}{4445} (\bibinfo{year}{1999}),
  \urlprefix\url{https://link.aps.org/doi/10.1103/PhysRevB.59.4445}.

\bibitem[{\citenamefont{Horibe et~al.}(2000)\citenamefont{Horibe, Inoue, and
  Koyama}}]{Horibe2000}
\bibinfo{author}{\bibfnamefont{Y.}~\bibnamefont{Horibe}},
  \bibinfo{author}{\bibfnamefont{Y.}~\bibnamefont{Inoue}}, \bibnamefont{and}
  \bibinfo{author}{\bibfnamefont{Y.}~\bibnamefont{Koyama}},
  \bibinfo{journal}{Phys. Rev. B} \textbf{\bibinfo{volume}{61}},
  \bibinfo{pages}{11922} (\bibinfo{year}{2000}),
  \urlprefix\url{https://link.aps.org/doi/10.1103/PhysRevB.61.11922}.

\bibitem[{\citenamefont{Pelc et~al.}(2021)\citenamefont{Pelc, Spieker,
  Anderson, Krogstad, Biniskos, Bielinski, Yu, Sasagawa, Chauviere, Dosanjh
  et~al.}}]{Pelc2021}
\bibinfo{author}{\bibfnamefont{D.}~\bibnamefont{Pelc}},
  \bibinfo{author}{\bibfnamefont{R.~J.} \bibnamefont{Spieker}},
  \bibinfo{author}{\bibfnamefont{Z.~W.} \bibnamefont{Anderson}},
  \bibinfo{author}{\bibfnamefont{M.~J.} \bibnamefont{Krogstad}},
  \bibinfo{author}{\bibfnamefont{N.}~\bibnamefont{Biniskos}},
  \bibinfo{author}{\bibfnamefont{N.~G.} \bibnamefont{Bielinski}},
  \bibinfo{author}{\bibfnamefont{B.}~\bibnamefont{Yu}},
  \bibinfo{author}{\bibfnamefont{T.}~\bibnamefont{Sasagawa}},
  \bibinfo{author}{\bibfnamefont{L.}~\bibnamefont{Chauviere}},
  \bibinfo{author}{\bibfnamefont{P.}~\bibnamefont{Dosanjh}},
  \bibnamefont{et~al.} (\bibinfo{year}{2021}),
  \urlprefix\url{https://arxiv.org/abs/2103.05482}.

\bibitem[{\citenamefont{Chou et~al.}(1995)\citenamefont{Chou, Belk, Kastner,
  Birgeneau, and Aharony}}]{Chou1995}
\bibinfo{author}{\bibfnamefont{F.~C.} \bibnamefont{Chou}},
  \bibinfo{author}{\bibfnamefont{N.~R.} \bibnamefont{Belk}},
  \bibinfo{author}{\bibfnamefont{M.~A.} \bibnamefont{Kastner}},
  \bibinfo{author}{\bibfnamefont{R.~J.} \bibnamefont{Birgeneau}},
  \bibnamefont{and} \bibinfo{author}{\bibfnamefont{A.}~\bibnamefont{Aharony}},
  \bibinfo{journal}{Phys. Rev. Lett.} \textbf{\bibinfo{volume}{75}},
  \bibinfo{pages}{2204} (\bibinfo{year}{1995}),
  \urlprefix\url{https://link.aps.org/doi/10.1103/PhysRevLett.75.2204}.

\bibitem[{\citenamefont{Wakimoto et~al.}(2000)\citenamefont{Wakimoto, Ueki,
  Endoh, and Yamada}}]{Wakimoto2000}
\bibinfo{author}{\bibfnamefont{S.}~\bibnamefont{Wakimoto}},
  \bibinfo{author}{\bibfnamefont{S.}~\bibnamefont{Ueki}},
  \bibinfo{author}{\bibfnamefont{Y.}~\bibnamefont{Endoh}}, \bibnamefont{and}
  \bibinfo{author}{\bibfnamefont{K.}~\bibnamefont{Yamada}},
  \bibinfo{journal}{Phys. Rev. B} \textbf{\bibinfo{volume}{62}},
  \bibinfo{pages}{3547} (\bibinfo{year}{2000}),
  \urlprefix\url{https://link.aps.org/doi/10.1103/PhysRevB.62.3547}.

\bibitem[{\citenamefont{Sasagawa et~al.}(2002)\citenamefont{Sasagawa, Mang,
  Vajk, Kapitulnik, and Greven}}]{Sasagawa2002}
\bibinfo{author}{\bibfnamefont{T.}~\bibnamefont{Sasagawa}},
  \bibinfo{author}{\bibfnamefont{P.~K.} \bibnamefont{Mang}},
  \bibinfo{author}{\bibfnamefont{O.~P.} \bibnamefont{Vajk}},
  \bibinfo{author}{\bibfnamefont{A.}~\bibnamefont{Kapitulnik}},
  \bibnamefont{and} \bibinfo{author}{\bibfnamefont{M.}~\bibnamefont{Greven}},
  \bibinfo{journal}{Phys. Rev. B} \textbf{\bibinfo{volume}{66}},
  \bibinfo{pages}{184512} (\bibinfo{year}{2002}),
  \urlprefix\url{https://link.aps.org/doi/10.1103/PhysRevB.66.184512}.

\bibitem[{\citenamefont{Mendels et~al.}(1994)\citenamefont{Mendels, Alloul,
  Brewer, Morris, Duty, Johnston, Ansaldo, Collin, Marucco, Niedermayer
  et~al.}}]{Mendels1994}
\bibinfo{author}{\bibfnamefont{P.}~\bibnamefont{Mendels}},
  \bibinfo{author}{\bibfnamefont{H.}~\bibnamefont{Alloul}},
  \bibinfo{author}{\bibfnamefont{J.~H.} \bibnamefont{Brewer}},
  \bibinfo{author}{\bibfnamefont{G.~D.} \bibnamefont{Morris}},
  \bibinfo{author}{\bibfnamefont{T.~L.} \bibnamefont{Duty}},
  \bibinfo{author}{\bibfnamefont{S.}~\bibnamefont{Johnston}},
  \bibinfo{author}{\bibfnamefont{E.~J.} \bibnamefont{Ansaldo}},
  \bibinfo{author}{\bibfnamefont{G.}~\bibnamefont{Collin}},
  \bibinfo{author}{\bibfnamefont{J.~F.} \bibnamefont{Marucco}},
  \bibinfo{author}{\bibfnamefont{C.}~\bibnamefont{Niedermayer}},
  \bibnamefont{et~al.}, \bibinfo{journal}{Phys. Rev. B}
  \textbf{\bibinfo{volume}{49}}, \bibinfo{pages}{10035} (\bibinfo{year}{1994}),
  \urlprefix\url{https://link.aps.org/doi/10.1103/PhysRevB.49.10035}.

\bibitem[{\citenamefont{Lake et~al.}(2005)\citenamefont{Lake, Lefmann,
  Christensen, Aeppli, McMorrow, Ronnow, Vorderwisch, Smeibidl, Mangkorntong,
  Sasagawa et~al.}}]{Lake2005}
\bibinfo{author}{\bibfnamefont{B.}~\bibnamefont{Lake}},
  \bibinfo{author}{\bibfnamefont{K.}~\bibnamefont{Lefmann}},
  \bibinfo{author}{\bibfnamefont{N.~B.} \bibnamefont{Christensen}},
  \bibinfo{author}{\bibfnamefont{G.}~\bibnamefont{Aeppli}},
  \bibinfo{author}{\bibfnamefont{D.~F.} \bibnamefont{McMorrow}},
  \bibinfo{author}{\bibfnamefont{H.~M.} \bibnamefont{Ronnow}},
  \bibinfo{author}{\bibfnamefont{P.}~\bibnamefont{Vorderwisch}},
  \bibinfo{author}{\bibfnamefont{P.}~\bibnamefont{Smeibidl}},
  \bibinfo{author}{\bibfnamefont{N.}~\bibnamefont{Mangkorntong}},
  \bibinfo{author}{\bibfnamefont{T.}~\bibnamefont{Sasagawa}},
  \bibnamefont{et~al.}, \bibinfo{journal}{Nature Materials}
  \textbf{\bibinfo{volume}{4}}, \bibinfo{pages}{658} (\bibinfo{year}{2005}),
  \urlprefix\url{https://doi.org/10.1038/nmat1452}.

\bibitem[{\citenamefont{R\o{}mer et~al.}(2015)\citenamefont{R\o{}mer, Ray,
  Jacobsen, Udby, Andersen, Bertelsen, Holm, Christensen, Toft-Petersen,
  Skoulatos et~al.}}]{Romer2015}
\bibinfo{author}{\bibfnamefont{A.~T.} \bibnamefont{R\o{}mer}},
  \bibinfo{author}{\bibfnamefont{P.~J.} \bibnamefont{Ray}},
  \bibinfo{author}{\bibfnamefont{H.}~\bibnamefont{Jacobsen}},
  \bibinfo{author}{\bibfnamefont{L.}~\bibnamefont{Udby}},
  \bibinfo{author}{\bibfnamefont{B.~M.} \bibnamefont{Andersen}},
  \bibinfo{author}{\bibfnamefont{M.}~\bibnamefont{Bertelsen}},
  \bibinfo{author}{\bibfnamefont{S.~L.} \bibnamefont{Holm}},
  \bibinfo{author}{\bibfnamefont{N.~B.} \bibnamefont{Christensen}},
  \bibinfo{author}{\bibfnamefont{R.}~\bibnamefont{Toft-Petersen}},
  \bibinfo{author}{\bibfnamefont{M.}~\bibnamefont{Skoulatos}},
  \bibnamefont{et~al.}, \bibinfo{journal}{Phys. Rev. B}
  \textbf{\bibinfo{volume}{91}}, \bibinfo{pages}{174507}
  (\bibinfo{year}{2015}),
  \urlprefix\url{https://link.aps.org/doi/10.1103/PhysRevB.91.174507}.

\bibitem[{\citenamefont{Chakravarty and Orbach}(1990)}]{Chakravarty1990}
\bibinfo{author}{\bibfnamefont{S.}~\bibnamefont{Chakravarty}} \bibnamefont{and}
  \bibinfo{author}{\bibfnamefont{R.}~\bibnamefont{Orbach}},
  \bibinfo{journal}{Phys. Rev. Lett.} \textbf{\bibinfo{volume}{64}},
  \bibinfo{pages}{224} (\bibinfo{year}{1990}),
  \urlprefix\url{https://link.aps.org/doi/10.1103/PhysRevLett.64.224}.

\bibitem[{\citenamefont{Birgeneau et~al.}(1995)\citenamefont{Birgeneau,
  Aharony, Belk, Chou, Endoh, Greven, Hosoya, Kastner, Lee, Lee
  et~al.}}]{Birgeneau1995}
\bibinfo{author}{\bibfnamefont{R.}~\bibnamefont{Birgeneau}},
  \bibinfo{author}{\bibfnamefont{A.}~\bibnamefont{Aharony}},
  \bibinfo{author}{\bibfnamefont{N.}~\bibnamefont{Belk}},
  \bibinfo{author}{\bibfnamefont{F.}~\bibnamefont{Chou}},
  \bibinfo{author}{\bibfnamefont{Y.}~\bibnamefont{Endoh}},
  \bibinfo{author}{\bibfnamefont{M.}~\bibnamefont{Greven}},
  \bibinfo{author}{\bibfnamefont{S.}~\bibnamefont{Hosoya}},
  \bibinfo{author}{\bibfnamefont{M.}~\bibnamefont{Kastner}},
  \bibinfo{author}{\bibfnamefont{C.}~\bibnamefont{Lee}},
  \bibinfo{author}{\bibfnamefont{Y.}~\bibnamefont{Lee}}, \bibnamefont{et~al.},
  \bibinfo{journal}{Journal of Physics and Chemistry of Solids}
  \textbf{\bibinfo{volume}{56}}, \bibinfo{pages}{1913} (\bibinfo{year}{1995}),
  \urlprefix\url{https://doi.org/10.1016/0022-3697(95)00234-0}.

\bibitem[{\citenamefont{Aeppli et~al.}(1997)\citenamefont{Aeppli, Mason,
  Hayden, Mook, and Kulda}}]{Aeppli1997}
\bibinfo{author}{\bibfnamefont{G.}~\bibnamefont{Aeppli}},
  \bibinfo{author}{\bibfnamefont{T.~E.} \bibnamefont{Mason}},
  \bibinfo{author}{\bibfnamefont{S.~M.} \bibnamefont{Hayden}},
  \bibinfo{author}{\bibfnamefont{H.~A.} \bibnamefont{Mook}}, \bibnamefont{and}
  \bibinfo{author}{\bibfnamefont{J.}~\bibnamefont{Kulda}},
  \bibinfo{journal}{Science} \textbf{\bibinfo{volume}{278}},
  \bibinfo{pages}{1432} (\bibinfo{year}{1997}),
  \urlprefix\url{https://www.science.org/doi/10.1126/science.278.5342.1432}.

\bibitem[{\citenamefont{Wen et~al.}(2019)\citenamefont{Wen, Huang, Lee, Jang,
  Knight, Lee, Fujita, Suzuki, Asano, Kivelson et~al.}}]{Wen2019}
\bibinfo{author}{\bibfnamefont{J.~J.} \bibnamefont{Wen}},
  \bibinfo{author}{\bibfnamefont{H.}~\bibnamefont{Huang}},
  \bibinfo{author}{\bibfnamefont{S.~J.} \bibnamefont{Lee}},
  \bibinfo{author}{\bibfnamefont{H.}~\bibnamefont{Jang}},
  \bibinfo{author}{\bibfnamefont{J.}~\bibnamefont{Knight}},
  \bibinfo{author}{\bibfnamefont{Y.~S.} \bibnamefont{Lee}},
  \bibinfo{author}{\bibfnamefont{M.}~\bibnamefont{Fujita}},
  \bibinfo{author}{\bibfnamefont{K.~M.} \bibnamefont{Suzuki}},
  \bibinfo{author}{\bibfnamefont{S.}~\bibnamefont{Asano}},
  \bibinfo{author}{\bibfnamefont{S.~A.} \bibnamefont{Kivelson}},
  \bibnamefont{et~al.}, \bibinfo{journal}{Nature Communications}
  \textbf{\bibinfo{volume}{10}}, \bibinfo{pages}{3269} (\bibinfo{year}{2019}),
  \urlprefix\url{https://doi.org/10.1038/s41467-019-11167-z}.

\bibitem[{\citenamefont{Miao et~al.}(2021)\citenamefont{Miao, Fabbris, Koch,
  Mazzone, Nelson, Acevedo-Esteves, Gu, Li, Yilimaz, Kaznatcheev
  et~al.}}]{Miao2021}
\bibinfo{author}{\bibfnamefont{H.}~\bibnamefont{Miao}},
  \bibinfo{author}{\bibfnamefont{G.}~\bibnamefont{Fabbris}},
  \bibinfo{author}{\bibfnamefont{R.~J.} \bibnamefont{Koch}},
  \bibinfo{author}{\bibfnamefont{D.~G.} \bibnamefont{Mazzone}},
  \bibinfo{author}{\bibfnamefont{C.~S.} \bibnamefont{Nelson}},
  \bibinfo{author}{\bibfnamefont{R.}~\bibnamefont{Acevedo-Esteves}},
  \bibinfo{author}{\bibfnamefont{G.~D.} \bibnamefont{Gu}},
  \bibinfo{author}{\bibfnamefont{Y.}~\bibnamefont{Li}},
  \bibinfo{author}{\bibfnamefont{T.}~\bibnamefont{Yilimaz}},
  \bibinfo{author}{\bibfnamefont{K.}~\bibnamefont{Kaznatcheev}},
  \bibnamefont{et~al.}, \bibinfo{journal}{npj Quantum Materials}
  \textbf{\bibinfo{volume}{6}}, \bibinfo{pages}{31} (\bibinfo{year}{2021}),
  \urlprefix\url{https://doi.org/10.1038/s41535-021-00327-4}.

\bibitem[{\citenamefont{Croft et~al.}(2014)\citenamefont{Croft, Lester, Senn,
  Bombardi, and Hayden}}]{Croft2014}
\bibinfo{author}{\bibfnamefont{T.~P.} \bibnamefont{Croft}},
  \bibinfo{author}{\bibfnamefont{C.}~\bibnamefont{Lester}},
  \bibinfo{author}{\bibfnamefont{M.~S.} \bibnamefont{Senn}},
  \bibinfo{author}{\bibfnamefont{A.}~\bibnamefont{Bombardi}}, \bibnamefont{and}
  \bibinfo{author}{\bibfnamefont{S.~M.} \bibnamefont{Hayden}},
  \bibinfo{journal}{Phys. Rev. B} \textbf{\bibinfo{volume}{89}},
  \bibinfo{pages}{224513} (\bibinfo{year}{2014}),
  \urlprefix\url{https://link.aps.org/doi/10.1103/PhysRevB.89.224513}.

\bibitem[{\citenamefont{von Arx et al.~preprint (2022)}()}]{vonArx2022}
\bibinfo{author}{\bibfnamefont{K.}~\bibnamefont{von Arx et al.~preprint
  (2022)}}.

\bibitem[{\citenamefont{Baek et~al.}(2012)\citenamefont{Baek, Erb, B\"uchner,
  and Grafe}}]{Baek2012}
\bibinfo{author}{\bibfnamefont{S.-H.} \bibnamefont{Baek}},
  \bibinfo{author}{\bibfnamefont{A.}~\bibnamefont{Erb}},
  \bibinfo{author}{\bibfnamefont{B.}~\bibnamefont{B\"uchner}},
  \bibnamefont{and} \bibinfo{author}{\bibfnamefont{H.-J.} \bibnamefont{Grafe}},
  \bibinfo{journal}{Phys. Rev. B} \textbf{\bibinfo{volume}{85}},
  \bibinfo{pages}{184508} (\bibinfo{year}{2012}),
  \urlprefix\url{https://link.aps.org/doi/10.1103/PhysRevB.85.184508}.

\bibitem[{\citenamefont{Hunt et~al.}(1999)\citenamefont{Hunt, Singer, Thurber,
  and Imai}}]{Hunt1999}
\bibinfo{author}{\bibfnamefont{A.~W.} \bibnamefont{Hunt}},
  \bibinfo{author}{\bibfnamefont{P.~M.} \bibnamefont{Singer}},
  \bibinfo{author}{\bibfnamefont{K.~R.} \bibnamefont{Thurber}},
  \bibnamefont{and} \bibinfo{author}{\bibfnamefont{T.}~\bibnamefont{Imai}},
  \bibinfo{journal}{Phys. Rev. Lett.} \textbf{\bibinfo{volume}{82}},
  \bibinfo{pages}{4300} (\bibinfo{year}{1999}),
  \urlprefix\url{https://link.aps.org/doi/10.1103/PhysRevLett.82.4300}.

\bibitem[{\citenamefont{Coneri et~al.}(2010)\citenamefont{Coneri, Sanna, Zheng,
  Lord, and De~Renzi}}]{Coneri2010}
\bibinfo{author}{\bibfnamefont{F.}~\bibnamefont{Coneri}},
  \bibinfo{author}{\bibfnamefont{S.}~\bibnamefont{Sanna}},
  \bibinfo{author}{\bibfnamefont{K.}~\bibnamefont{Zheng}},
  \bibinfo{author}{\bibfnamefont{J.}~\bibnamefont{Lord}}, \bibnamefont{and}
  \bibinfo{author}{\bibfnamefont{R.}~\bibnamefont{De~Renzi}},
  \bibinfo{journal}{Phys. Rev. B} \textbf{\bibinfo{volume}{81}},
  \bibinfo{pages}{104507} (\bibinfo{year}{2010}),
  \urlprefix\url{https://link.aps.org/doi/10.1103/PhysRevB.81.104507}.

\bibitem[{\citenamefont{Bourgeois-Hope
  et~al.}(2019)\citenamefont{Bourgeois-Hope, Li, Laliberté, Badoux, Hayden,
  Momono, Kurosawa, Yamada, Takagi, Doiron-Leyraud et~al.}}]{BourgeoisHope2019}
\bibinfo{author}{\bibfnamefont{P.}~\bibnamefont{Bourgeois-Hope}},
  \bibinfo{author}{\bibfnamefont{S.~Y.} \bibnamefont{Li}},
  \bibinfo{author}{\bibfnamefont{F.}~\bibnamefont{Laliberté}},
  \bibinfo{author}{\bibfnamefont{S.}~\bibnamefont{Badoux}},
  \bibinfo{author}{\bibfnamefont{S.~M.} \bibnamefont{Hayden}},
  \bibinfo{author}{\bibfnamefont{N.}~\bibnamefont{Momono}},
  \bibinfo{author}{\bibfnamefont{T.}~\bibnamefont{Kurosawa}},
  \bibinfo{author}{\bibfnamefont{K.}~\bibnamefont{Yamada}},
  \bibinfo{author}{\bibfnamefont{H.}~\bibnamefont{Takagi}},
  \bibinfo{author}{\bibfnamefont{N.}~\bibnamefont{Doiron-Leyraud}},
  \bibnamefont{et~al.} (\bibinfo{year}{2019}),
  \urlprefix\url{https://arxiv.org/abs/1910.08126}.

\bibitem[{\citenamefont{Zhou et~al.}(2017)\citenamefont{Zhou, Hirata, Wu,
  Vinograd, Mayaffre, Kr{\"a}mer, Reyes, Kuhns, Liang, Hardy
  et~al.}}]{Zhou2017}
\bibinfo{author}{\bibfnamefont{R.}~\bibnamefont{Zhou}},
  \bibinfo{author}{\bibfnamefont{M.}~\bibnamefont{Hirata}},
  \bibinfo{author}{\bibfnamefont{T.}~\bibnamefont{Wu}},
  \bibinfo{author}{\bibfnamefont{I.}~\bibnamefont{Vinograd}},
  \bibinfo{author}{\bibfnamefont{H.}~\bibnamefont{Mayaffre}},
  \bibinfo{author}{\bibfnamefont{S.}~\bibnamefont{Kr{\"a}mer}},
  \bibinfo{author}{\bibfnamefont{A.~P.} \bibnamefont{Reyes}},
  \bibinfo{author}{\bibfnamefont{P.~L.} \bibnamefont{Kuhns}},
  \bibinfo{author}{\bibfnamefont{R.}~\bibnamefont{Liang}},
  \bibinfo{author}{\bibfnamefont{W.~N.} \bibnamefont{Hardy}},
  \bibnamefont{et~al.}, \bibinfo{journal}{Proceedings of the National Academy
  of Sciences} \textbf{\bibinfo{volume}{114}}, \bibinfo{pages}{13148}
  (\bibinfo{year}{2017}), ISSN \bibinfo{issn}{0027-8424},
  \eprint{https://www.pnas.org/content/114/50/13148.full.pdf},
  \urlprefix\url{https://www.pnas.org/content/114/50/13148}.

\bibitem[{\citenamefont{Ka\ifmmode \check{c}\else
  \v{c}\fi{}mar\ifmmode~\check{c}\else \v{c}\fi{}\'{\i}k
  et~al.}(2018)\citenamefont{Ka\ifmmode \check{c}\else
  \v{c}\fi{}mar\ifmmode~\check{c}\else \v{c}\fi{}\'{\i}k, Vinograd, Michon,
  Rydh, Demuer, Zhou, Mayaffre, Liang, Hardy, Bonn et~al.}}]{Kacmarcik2018}
\bibinfo{author}{\bibfnamefont{J.}~\bibnamefont{Ka\ifmmode \check{c}\else
  \v{c}\fi{}mar\ifmmode~\check{c}\else \v{c}\fi{}\'{\i}k}},
  \bibinfo{author}{\bibfnamefont{I.}~\bibnamefont{Vinograd}},
  \bibinfo{author}{\bibfnamefont{B.}~\bibnamefont{Michon}},
  \bibinfo{author}{\bibfnamefont{A.}~\bibnamefont{Rydh}},
  \bibinfo{author}{\bibfnamefont{A.}~\bibnamefont{Demuer}},
  \bibinfo{author}{\bibfnamefont{R.}~\bibnamefont{Zhou}},
  \bibinfo{author}{\bibfnamefont{H.}~\bibnamefont{Mayaffre}},
  \bibinfo{author}{\bibfnamefont{R.}~\bibnamefont{Liang}},
  \bibinfo{author}{\bibfnamefont{W.~N.} \bibnamefont{Hardy}},
  \bibinfo{author}{\bibfnamefont{D.~A.} \bibnamefont{Bonn}},
  \bibnamefont{et~al.}, \bibinfo{journal}{Phys. Rev. Lett.}
  \textbf{\bibinfo{volume}{121}}, \bibinfo{pages}{167002}
  (\bibinfo{year}{2018}),
  \urlprefix\url{https://link.aps.org/doi/10.1103/PhysRevLett.121.167002}.

\bibitem[{\citenamefont{Katanin and Sushkov}(2011)}]{Katanin2011}
\bibinfo{author}{\bibfnamefont{A.}~\bibnamefont{Katanin}} \bibnamefont{and}
  \bibinfo{author}{\bibfnamefont{O.~P.} \bibnamefont{Sushkov}},
  \bibinfo{journal}{Phys. Rev. B} \textbf{\bibinfo{volume}{83}},
  \bibinfo{pages}{094426} (\bibinfo{year}{2011}),
  \urlprefix\url{https://link.aps.org/doi/10.1103/PhysRevB.83.094426}.

\bibitem[{\citenamefont{Girod et~al.}(2021)\citenamefont{Girod, LeBoeuf,
  Demuer, Seyfarth, Imajo, Kindo, Kohama, Lizaire, Legros, Gourgout
  et~al.}}]{Girod2021}
\bibinfo{author}{\bibfnamefont{C.}~\bibnamefont{Girod}},
  \bibinfo{author}{\bibfnamefont{D.}~\bibnamefont{LeBoeuf}},
  \bibinfo{author}{\bibfnamefont{A.}~\bibnamefont{Demuer}},
  \bibinfo{author}{\bibfnamefont{G.}~\bibnamefont{Seyfarth}},
  \bibinfo{author}{\bibfnamefont{S.}~\bibnamefont{Imajo}},
  \bibinfo{author}{\bibfnamefont{K.}~\bibnamefont{Kindo}},
  \bibinfo{author}{\bibfnamefont{Y.}~\bibnamefont{Kohama}},
  \bibinfo{author}{\bibfnamefont{M.}~\bibnamefont{Lizaire}},
  \bibinfo{author}{\bibfnamefont{A.}~\bibnamefont{Legros}},
  \bibinfo{author}{\bibfnamefont{A.}~\bibnamefont{Gourgout}},
  \bibnamefont{et~al.}, \bibinfo{journal}{Phys. Rev. B}
  \textbf{\bibinfo{volume}{103}}, \bibinfo{pages}{214506}
  (\bibinfo{year}{2021}),
  \urlprefix\url{https://link.aps.org/doi/10.1103/PhysRevB.103.214506}.

\bibitem[{\citenamefont{Wu et~al.}(2013{\natexlab{b}})\citenamefont{Wu,
  Mayaffre, Kr{\"a}mer, Horvati{\'c}, Berthier, Kuhns, Reyes, Liang, Hardy,
  Bonn et~al.}}]{Wu2013b}
\bibinfo{author}{\bibfnamefont{T.}~\bibnamefont{Wu}},
  \bibinfo{author}{\bibfnamefont{H.}~\bibnamefont{Mayaffre}},
  \bibinfo{author}{\bibfnamefont{S.}~\bibnamefont{Kr{\"a}mer}},
  \bibinfo{author}{\bibfnamefont{M.}~\bibnamefont{Horvati{\'c}}},
  \bibinfo{author}{\bibfnamefont{C.}~\bibnamefont{Berthier}},
  \bibinfo{author}{\bibfnamefont{P.~L.} \bibnamefont{Kuhns}},
  \bibinfo{author}{\bibfnamefont{A.~P.} \bibnamefont{Reyes}},
  \bibinfo{author}{\bibfnamefont{R.}~\bibnamefont{Liang}},
  \bibinfo{author}{\bibfnamefont{W.~N.} \bibnamefont{Hardy}},
  \bibinfo{author}{\bibfnamefont{D.~A.} \bibnamefont{Bonn}},
  \bibnamefont{et~al.}, \bibinfo{journal}{Nature Communications}
  \textbf{\bibinfo{volume}{4}}, \bibinfo{pages}{2113}
  (\bibinfo{year}{2013}{\natexlab{b}}),
  \urlprefix\url{https://doi.org/10.1038/ncomms3113}.

\bibitem[{\citenamefont{Khaykovich et~al.}(2005)\citenamefont{Khaykovich,
  Wakimoto, Birgeneau, Kastner, Lee, Smeibidl, Vorderwisch, and
  Yamada}}]{Khaykovich2005}
\bibinfo{author}{\bibfnamefont{B.}~\bibnamefont{Khaykovich}},
  \bibinfo{author}{\bibfnamefont{S.}~\bibnamefont{Wakimoto}},
  \bibinfo{author}{\bibfnamefont{R.~J.} \bibnamefont{Birgeneau}},
  \bibinfo{author}{\bibfnamefont{M.~A.} \bibnamefont{Kastner}},
  \bibinfo{author}{\bibfnamefont{Y.~S.} \bibnamefont{Lee}},
  \bibinfo{author}{\bibfnamefont{P.}~\bibnamefont{Smeibidl}},
  \bibinfo{author}{\bibfnamefont{P.}~\bibnamefont{Vorderwisch}},
  \bibnamefont{and} \bibinfo{author}{\bibfnamefont{K.}~\bibnamefont{Yamada}},
  \bibinfo{journal}{Phys. Rev. B} \textbf{\bibinfo{volume}{71}},
  \bibinfo{pages}{220508} (\bibinfo{year}{2005}),
  \urlprefix\url{https://link.aps.org/doi/10.1103/PhysRevB.71.220508}.

\bibitem[{\citenamefont{Vignolle et~al.}(2007)\citenamefont{Vignolle, Hayden,
  McMorrow, R{\o}nnow, Lake, Frost, and Perring}}]{Vignolle2007}
\bibinfo{author}{\bibfnamefont{B.}~\bibnamefont{Vignolle}},
  \bibinfo{author}{\bibfnamefont{S.~M.} \bibnamefont{Hayden}},
  \bibinfo{author}{\bibfnamefont{D.~F.} \bibnamefont{McMorrow}},
  \bibinfo{author}{\bibfnamefont{H.~M.} \bibnamefont{R{\o}nnow}},
  \bibinfo{author}{\bibfnamefont{B.}~\bibnamefont{Lake}},
  \bibinfo{author}{\bibfnamefont{C.~D.} \bibnamefont{Frost}}, \bibnamefont{and}
  \bibinfo{author}{\bibfnamefont{T.~G.} \bibnamefont{Perring}},
  \bibinfo{journal}{Nature Physics} \textbf{\bibinfo{volume}{3}},
  \bibinfo{pages}{163} (\bibinfo{year}{2007}),
  \urlprefix\url{https://doi.org/10.1038/nphys546}.

\bibitem[{\citenamefont{Lipscombe et~al.}(2007)\citenamefont{Lipscombe, Hayden,
  Vignolle, McMorrow, and Perring}}]{Lipscombe2007}
\bibinfo{author}{\bibfnamefont{O.~J.} \bibnamefont{Lipscombe}},
  \bibinfo{author}{\bibfnamefont{S.~M.} \bibnamefont{Hayden}},
  \bibinfo{author}{\bibfnamefont{B.}~\bibnamefont{Vignolle}},
  \bibinfo{author}{\bibfnamefont{D.~F.} \bibnamefont{McMorrow}},
  \bibnamefont{and} \bibinfo{author}{\bibfnamefont{T.~G.}
  \bibnamefont{Perring}}, \bibinfo{journal}{Phys. Rev. Lett.}
  \textbf{\bibinfo{volume}{99}}, \bibinfo{pages}{067002}
  (\bibinfo{year}{2007}),
  \urlprefix\url{https://link.aps.org/doi/10.1103/PhysRevLett.99.067002}.

\bibitem[{\citenamefont{Tranquada}(2020)}]{Tranquada2020}
\bibinfo{author}{\bibfnamefont{J.~M.} \bibnamefont{Tranquada}},
  \bibinfo{journal}{Advances in Physics} \textbf{\bibinfo{volume}{69}},
  \bibinfo{pages}{437} (\bibinfo{year}{2020}),
  \eprint{https://doi.org/10.1080/00018732.2021.1935698},
  \urlprefix\url{https://doi.org/10.1080/00018732.2021.1935698}.

\bibitem[{\citenamefont{{Chaboussant, G.}
  et~al.}(1998)\citenamefont{{Chaboussant, G.}, {Julien, M.-H.},
  {Fagot-Revurat, Y.}, {Hanson, M.}, {L\'evy, L. P.}, {Berthier, C.},
  {Horvati\'{}c, M.}, and {Piovesana, O.}}}]{Chaboussant1998}
\bibinfo{author}{\bibnamefont{{Chaboussant, G.}}},
  \bibinfo{author}{\bibnamefont{{Julien, M.-H.}}},
  \bibinfo{author}{\bibnamefont{{Fagot-Revurat, Y.}}},
  \bibinfo{author}{\bibnamefont{{Hanson, M.}}},
  \bibinfo{author}{\bibnamefont{{L\'evy, L. P.}}},
  \bibinfo{author}{\bibnamefont{{Berthier, C.}}},
  \bibinfo{author}{\bibnamefont{{Horvati\'{}c, M.}}}, \bibnamefont{and}
  \bibinfo{author}{\bibnamefont{{Piovesana, O.}}}, \bibinfo{journal}{Eur. Phys.
  J. B} \textbf{\bibinfo{volume}{6}}, \bibinfo{pages}{167}
  (\bibinfo{year}{1998}),
  \urlprefix\url{https://doi.org/10.1007/s100510050539}.

\bibitem[{\citenamefont{Giamarchi et~al.}(2008)\citenamefont{Giamarchi,
  R{\"u}egg, and Tchernyshyov}}]{Giamarchi2008}
\bibinfo{author}{\bibfnamefont{T.}~\bibnamefont{Giamarchi}},
  \bibinfo{author}{\bibfnamefont{C.}~\bibnamefont{R{\"u}egg}},
  \bibnamefont{and}
  \bibinfo{author}{\bibfnamefont{O.}~\bibnamefont{Tchernyshyov}},
  \bibinfo{journal}{Nature Physics} \textbf{\bibinfo{volume}{4}},
  \bibinfo{pages}{198} (\bibinfo{year}{2008}),
  \urlprefix\url{https://doi.org/10.1038/nphys893}.

\bibitem[{\citenamefont{Chang et~al.}(2009)\citenamefont{Chang, Christensen,
  Niedermayer, Lefmann, R\o{}nnow, McMorrow, Schneidewind, Link, Hiess, Boehm
  et~al.}}]{Chang2009}
\bibinfo{author}{\bibfnamefont{J.}~\bibnamefont{Chang}},
  \bibinfo{author}{\bibfnamefont{N.~B.} \bibnamefont{Christensen}},
  \bibinfo{author}{\bibfnamefont{C.}~\bibnamefont{Niedermayer}},
  \bibinfo{author}{\bibfnamefont{K.}~\bibnamefont{Lefmann}},
  \bibinfo{author}{\bibfnamefont{H.~M.} \bibnamefont{R\o{}nnow}},
  \bibinfo{author}{\bibfnamefont{D.~F.} \bibnamefont{McMorrow}},
  \bibinfo{author}{\bibfnamefont{A.}~\bibnamefont{Schneidewind}},
  \bibinfo{author}{\bibfnamefont{P.}~\bibnamefont{Link}},
  \bibinfo{author}{\bibfnamefont{A.}~\bibnamefont{Hiess}},
  \bibinfo{author}{\bibfnamefont{M.}~\bibnamefont{Boehm}},
  \bibnamefont{et~al.}, \bibinfo{journal}{Phys. Rev. Lett.}
  \textbf{\bibinfo{volume}{102}}, \bibinfo{pages}{177006}
  (\bibinfo{year}{2009}),
  \urlprefix\url{https://link.aps.org/doi/10.1103/PhysRevLett.102.177006}.

\bibitem[{\citenamefont{Li et~al.}(2018)\citenamefont{Li, Zhong, Stone,
  Kolesnikov, Gu, Zaliznyak, and Tranquada}}]{Li2018}
\bibinfo{author}{\bibfnamefont{Y.}~\bibnamefont{Li}},
  \bibinfo{author}{\bibfnamefont{R.}~\bibnamefont{Zhong}},
  \bibinfo{author}{\bibfnamefont{M.~B.} \bibnamefont{Stone}},
  \bibinfo{author}{\bibfnamefont{A.~I.} \bibnamefont{Kolesnikov}},
  \bibinfo{author}{\bibfnamefont{G.~D.} \bibnamefont{Gu}},
  \bibinfo{author}{\bibfnamefont{I.~A.} \bibnamefont{Zaliznyak}},
  \bibnamefont{and} \bibinfo{author}{\bibfnamefont{J.~M.}
  \bibnamefont{Tranquada}}, \bibinfo{journal}{Phys. Rev. B}
  \textbf{\bibinfo{volume}{98}}, \bibinfo{pages}{224508}
  (\bibinfo{year}{2018}),
  \urlprefix\url{https://link.aps.org/doi/10.1103/PhysRevB.98.224508}.

\bibitem[{\citenamefont{Panagopoulos et~al.}(2003)\citenamefont{Panagopoulos,
  Tallon, Rainford, Cooper, Scott, and Xiang}}]{Panagopoulos2003}
\bibinfo{author}{\bibfnamefont{C.}~\bibnamefont{Panagopoulos}},
  \bibinfo{author}{\bibfnamefont{J.}~\bibnamefont{Tallon}},
  \bibinfo{author}{\bibfnamefont{B.}~\bibnamefont{Rainford}},
  \bibinfo{author}{\bibfnamefont{J.}~\bibnamefont{Cooper}},
  \bibinfo{author}{\bibfnamefont{C.}~\bibnamefont{Scott}}, \bibnamefont{and}
  \bibinfo{author}{\bibfnamefont{T.}~\bibnamefont{Xiang}},
  \bibinfo{journal}{Solid State Communications} \textbf{\bibinfo{volume}{126}},
  \bibinfo{pages}{47} (\bibinfo{year}{2003}),
  \urlprefix\url{https://doi.org/10.1016/S0038-1098(02)00667-1}.

\bibitem[{\citenamefont{Kimura et~al.}(2003{\natexlab{a}})\citenamefont{Kimura,
  Kofu, Matsumoto, and Hirota}}]{Kimura2003}
\bibinfo{author}{\bibfnamefont{H.}~\bibnamefont{Kimura}},
  \bibinfo{author}{\bibfnamefont{M.}~\bibnamefont{Kofu}},
  \bibinfo{author}{\bibfnamefont{Y.}~\bibnamefont{Matsumoto}},
  \bibnamefont{and} \bibinfo{author}{\bibfnamefont{K.}~\bibnamefont{Hirota}},
  \bibinfo{journal}{Phys. Rev. Lett.} \textbf{\bibinfo{volume}{91}},
  \bibinfo{pages}{067002} (\bibinfo{year}{2003}{\natexlab{a}}),
  \urlprefix\url{https://link.aps.org/doi/10.1103/PhysRevLett.91.067002}.

\bibitem[{\citenamefont{Post et~al.}(2021)\citenamefont{Post, Legros, Rickel,
  Singleton, McDonald, He, Bo\ifmmode \check{z}\else
  \v{z}\fi{}ovi\ifmmode~\acute{c}\else \'{c}\fi{}, Xu, Shi, Armitage
  et~al.}}]{Post2021}
\bibinfo{author}{\bibfnamefont{K.~W.} \bibnamefont{Post}},
  \bibinfo{author}{\bibfnamefont{A.}~\bibnamefont{Legros}},
  \bibinfo{author}{\bibfnamefont{D.~G.} \bibnamefont{Rickel}},
  \bibinfo{author}{\bibfnamefont{J.}~\bibnamefont{Singleton}},
  \bibinfo{author}{\bibfnamefont{R.~D.} \bibnamefont{McDonald}},
  \bibinfo{author}{\bibfnamefont{X.}~\bibnamefont{He}},
  \bibinfo{author}{\bibfnamefont{I.}~\bibnamefont{Bo\ifmmode \check{z}\else
  \v{z}\fi{}ovi\ifmmode~\acute{c}\else \'{c}\fi{}}},
  \bibinfo{author}{\bibfnamefont{X.}~\bibnamefont{Xu}},
  \bibinfo{author}{\bibfnamefont{X.}~\bibnamefont{Shi}},
  \bibinfo{author}{\bibfnamefont{N.~P.} \bibnamefont{Armitage}},
  \bibnamefont{et~al.}, \bibinfo{journal}{Phys. Rev. B}
  \textbf{\bibinfo{volume}{103}}, \bibinfo{pages}{134515}
  (\bibinfo{year}{2021}),
  \urlprefix\url{https://link.aps.org/doi/10.1103/PhysRevB.103.134515}.

\bibitem[{\citenamefont{Boebinger et~al.}(1996)\citenamefont{Boebinger, Ando,
  Passner, Kimura, Okuya, Shimoyama, Kishio, Tamasaku, Ichikawa, and
  Uchida}}]{Boebinger1996}
\bibinfo{author}{\bibfnamefont{G.~S.} \bibnamefont{Boebinger}},
  \bibinfo{author}{\bibfnamefont{Y.}~\bibnamefont{Ando}},
  \bibinfo{author}{\bibfnamefont{A.}~\bibnamefont{Passner}},
  \bibinfo{author}{\bibfnamefont{T.}~\bibnamefont{Kimura}},
  \bibinfo{author}{\bibfnamefont{M.}~\bibnamefont{Okuya}},
  \bibinfo{author}{\bibfnamefont{J.}~\bibnamefont{Shimoyama}},
  \bibinfo{author}{\bibfnamefont{K.}~\bibnamefont{Kishio}},
  \bibinfo{author}{\bibfnamefont{K.}~\bibnamefont{Tamasaku}},
  \bibinfo{author}{\bibfnamefont{N.}~\bibnamefont{Ichikawa}}, \bibnamefont{and}
  \bibinfo{author}{\bibfnamefont{S.}~\bibnamefont{Uchida}},
  \bibinfo{journal}{Phys. Rev. Lett.} \textbf{\bibinfo{volume}{77}},
  \bibinfo{pages}{5417} (\bibinfo{year}{1996}),
  \urlprefix\url{https://link.aps.org/doi/10.1103/PhysRevLett.77.5417}.

\bibitem[{\citenamefont{Harshman et~al.}(1988)\citenamefont{Harshman, Aeppli,
  Espinosa, Cooper, Remeika, Ansaldo, Riseman, Williams, Noakes, Ellman
  et~al.}}]{Harshman1988}
\bibinfo{author}{\bibfnamefont{D.~R.} \bibnamefont{Harshman}},
  \bibinfo{author}{\bibfnamefont{G.}~\bibnamefont{Aeppli}},
  \bibinfo{author}{\bibfnamefont{G.~P.} \bibnamefont{Espinosa}},
  \bibinfo{author}{\bibfnamefont{A.~S.} \bibnamefont{Cooper}},
  \bibinfo{author}{\bibfnamefont{J.~P.} \bibnamefont{Remeika}},
  \bibinfo{author}{\bibfnamefont{E.~J.} \bibnamefont{Ansaldo}},
  \bibinfo{author}{\bibfnamefont{T.~M.} \bibnamefont{Riseman}},
  \bibinfo{author}{\bibfnamefont{D.~L.} \bibnamefont{Williams}},
  \bibinfo{author}{\bibfnamefont{D.~R.} \bibnamefont{Noakes}},
  \bibinfo{author}{\bibfnamefont{B.}~\bibnamefont{Ellman}},
  \bibnamefont{et~al.}, \bibinfo{journal}{Phys. Rev. B}
  \textbf{\bibinfo{volume}{38}}, \bibinfo{pages}{852} (\bibinfo{year}{1988}),
  \urlprefix\url{https://link.aps.org/doi/10.1103/PhysRevB.38.852}.

\bibitem[{\citenamefont{Hayden et~al.}(1991)\citenamefont{Hayden, Aeppli, Mook,
  Rytz, Hundley, and Fisk}}]{Hayden1991}
\bibinfo{author}{\bibfnamefont{S.~M.} \bibnamefont{Hayden}},
  \bibinfo{author}{\bibfnamefont{G.}~\bibnamefont{Aeppli}},
  \bibinfo{author}{\bibfnamefont{H.}~\bibnamefont{Mook}},
  \bibinfo{author}{\bibfnamefont{D.}~\bibnamefont{Rytz}},
  \bibinfo{author}{\bibfnamefont{M.~F.} \bibnamefont{Hundley}},
  \bibnamefont{and} \bibinfo{author}{\bibfnamefont{Z.}~\bibnamefont{Fisk}},
  \bibinfo{journal}{Phys. Rev. Lett.} \textbf{\bibinfo{volume}{66}},
  \bibinfo{pages}{821} (\bibinfo{year}{1991}),
  \urlprefix\url{https://link.aps.org/doi/10.1103/PhysRevLett.66.821}.

\bibitem[{\citenamefont{Panagopoulos and Dobrosavljevi\ifmmode~\acute{c}\else
  \'{c}\fi{}}(2005)}]{Panagopoulos2005}
\bibinfo{author}{\bibfnamefont{C.}~\bibnamefont{Panagopoulos}}
  \bibnamefont{and}
  \bibinfo{author}{\bibfnamefont{V.}~\bibnamefont{Dobrosavljevi\ifmmode~\acute{c}\else
  \'{c}\fi{}}}, \bibinfo{journal}{Phys. Rev. B} \textbf{\bibinfo{volume}{72}},
  \bibinfo{pages}{014536} (\bibinfo{year}{2005}),
  \urlprefix\url{https://link.aps.org/doi/10.1103/PhysRevB.72.014536}.

\bibitem[{\citenamefont{Frachet}(2019)}]{Frachet-thesis}
\bibinfo{author}{\bibfnamefont{M.}~\bibnamefont{Frachet}},
  \bibinfo{journal}{PhD thesis, Universit\'e Grenoble-Alpes}
  (\bibinfo{year}{2019}).

\bibitem[{\citenamefont{Chubukov et~al.}(1994)\citenamefont{Chubukov, Sachdev,
  and Ye}}]{Chubukov1994}
\bibinfo{author}{\bibfnamefont{A.~V.} \bibnamefont{Chubukov}},
  \bibinfo{author}{\bibfnamefont{S.}~\bibnamefont{Sachdev}}, \bibnamefont{and}
  \bibinfo{author}{\bibfnamefont{J.}~\bibnamefont{Ye}}, \bibinfo{journal}{Phys.
  Rev. B} \textbf{\bibinfo{volume}{49}}, \bibinfo{pages}{11919}
  (\bibinfo{year}{1994}),
  \urlprefix\url{https://link.aps.org/doi/10.1103/PhysRevB.49.11919}.

\bibitem[{\citenamefont{Sonier et~al.}(2007)\citenamefont{Sonier, Callaghan,
  Ando, Kiefl, Brewer, Kaiser, Pacradouni, Sabok-Sayr, Sun, Komiya
  et~al.}}]{Sonier2007}
\bibinfo{author}{\bibfnamefont{J.~E.} \bibnamefont{Sonier}},
  \bibinfo{author}{\bibfnamefont{F.~D.} \bibnamefont{Callaghan}},
  \bibinfo{author}{\bibfnamefont{Y.}~\bibnamefont{Ando}},
  \bibinfo{author}{\bibfnamefont{R.~F.} \bibnamefont{Kiefl}},
  \bibinfo{author}{\bibfnamefont{J.~H.} \bibnamefont{Brewer}},
  \bibinfo{author}{\bibfnamefont{C.~V.} \bibnamefont{Kaiser}},
  \bibinfo{author}{\bibfnamefont{V.}~\bibnamefont{Pacradouni}},
  \bibinfo{author}{\bibfnamefont{S.~A.} \bibnamefont{Sabok-Sayr}},
  \bibinfo{author}{\bibfnamefont{X.~F.} \bibnamefont{Sun}},
  \bibinfo{author}{\bibfnamefont{S.}~\bibnamefont{Komiya}},
  \bibnamefont{et~al.}, \bibinfo{journal}{Phys. Rev. B}
  \textbf{\bibinfo{volume}{76}}, \bibinfo{pages}{064522}
  (\bibinfo{year}{2007}),
  \urlprefix\url{https://link.aps.org/doi/10.1103/PhysRevB.76.064522}.

\bibitem[{\citenamefont{Fradkin et~al.}(2015)\citenamefont{Fradkin, Kivelson,
  and Tranquada}}]{Fradkin2015}
\bibinfo{author}{\bibfnamefont{E.}~\bibnamefont{Fradkin}},
  \bibinfo{author}{\bibfnamefont{S.~A.} \bibnamefont{Kivelson}},
  \bibnamefont{and} \bibinfo{author}{\bibfnamefont{J.~M.}
  \bibnamefont{Tranquada}}, \bibinfo{journal}{Rev. Mod. Phys.}
  \textbf{\bibinfo{volume}{87}}, \bibinfo{pages}{457} (\bibinfo{year}{2015}),
  \urlprefix\url{https://link.aps.org/doi/10.1103/RevModPhys.87.457}.

\bibitem[{\citenamefont{Yamase et~al.}(2016)\citenamefont{Yamase, Eberlein, and
  Metzner}}]{Yamase2016}
\bibinfo{author}{\bibfnamefont{H.}~\bibnamefont{Yamase}},
  \bibinfo{author}{\bibfnamefont{A.}~\bibnamefont{Eberlein}}, \bibnamefont{and}
  \bibinfo{author}{\bibfnamefont{W.}~\bibnamefont{Metzner}},
  \bibinfo{journal}{Phys. Rev. Lett.} \textbf{\bibinfo{volume}{116}},
  \bibinfo{pages}{096402} (\bibinfo{year}{2016}),
  \urlprefix\url{https://link.aps.org/doi/10.1103/PhysRevLett.116.096402}.

\bibitem[{\citenamefont{Cui et~al.}(2020)\citenamefont{Cui, Sun, Ray, Zheng,
  Sun, and Chan}}]{Cui2020}
\bibinfo{author}{\bibfnamefont{Z.-H.} \bibnamefont{Cui}},
  \bibinfo{author}{\bibfnamefont{C.}~\bibnamefont{Sun}},
  \bibinfo{author}{\bibfnamefont{U.}~\bibnamefont{Ray}},
  \bibinfo{author}{\bibfnamefont{B.-X.} \bibnamefont{Zheng}},
  \bibinfo{author}{\bibfnamefont{Q.}~\bibnamefont{Sun}}, \bibnamefont{and}
  \bibinfo{author}{\bibfnamefont{G.~K.-L.} \bibnamefont{Chan}},
  \bibinfo{journal}{Phys. Rev. Research} \textbf{\bibinfo{volume}{2}},
  \bibinfo{pages}{043259} (\bibinfo{year}{2020}),
  \urlprefix\url{https://link.aps.org/doi/10.1103/PhysRevResearch.2.043259}.

\bibitem[{\citenamefont{Tranquada}(2013)}]{Tranquada2013}
\bibinfo{author}{\bibfnamefont{J.~M.} \bibnamefont{Tranquada}},
  \bibinfo{journal}{AIP Conference Proceedings}
  \textbf{\bibinfo{volume}{1550}}, \bibinfo{pages}{114} (\bibinfo{year}{2013}),
  \urlprefix\url{https://aip.scitation.org/doi/abs/10.1063/1.4818402}.

\bibitem[{\citenamefont{Ohsugi et~al.}(1994)\citenamefont{Ohsugi, Kitaoka,
  Ishida, Zheng, and Asayama}}]{Ohsugi1994}
\bibinfo{author}{\bibfnamefont{S.}~\bibnamefont{Ohsugi}},
  \bibinfo{author}{\bibfnamefont{Y.}~\bibnamefont{Kitaoka}},
  \bibinfo{author}{\bibfnamefont{K.}~\bibnamefont{Ishida}},
  \bibinfo{author}{\bibfnamefont{G.-q.} \bibnamefont{Zheng}}, \bibnamefont{and}
  \bibinfo{author}{\bibfnamefont{K.}~\bibnamefont{Asayama}},
  \bibinfo{journal}{Journal of the Physical Society of Japan}
  \textbf{\bibinfo{volume}{63}}, \bibinfo{pages}{700} (\bibinfo{year}{1994}),
  \eprint{https://doi.org/10.1143/JPSJ.63.700},
  \urlprefix\url{https://doi.org/10.1143/JPSJ.63.700}.

\bibitem[{\citenamefont{Dean et~al.}(2013)\citenamefont{Dean, Dellea,
  Springell, Yakhou-Harris, Kummer, Brookes, Liu, Sun, Strle, Schmitt
  et~al.}}]{Dean2013}
\bibinfo{author}{\bibfnamefont{M.~P.~M.} \bibnamefont{Dean}},
  \bibinfo{author}{\bibfnamefont{G.}~\bibnamefont{Dellea}},
  \bibinfo{author}{\bibfnamefont{R.~S.} \bibnamefont{Springell}},
  \bibinfo{author}{\bibfnamefont{F.}~\bibnamefont{Yakhou-Harris}},
  \bibinfo{author}{\bibfnamefont{K.}~\bibnamefont{Kummer}},
  \bibinfo{author}{\bibfnamefont{N.~B.} \bibnamefont{Brookes}},
  \bibinfo{author}{\bibfnamefont{X.}~\bibnamefont{Liu}},
  \bibinfo{author}{\bibfnamefont{Y.-J.} \bibnamefont{Sun}},
  \bibinfo{author}{\bibfnamefont{J.}~\bibnamefont{Strle}},
  \bibinfo{author}{\bibfnamefont{T.}~\bibnamefont{Schmitt}},
  \bibnamefont{et~al.}, \bibinfo{journal}{Nature Materials}
  \textbf{\bibinfo{volume}{12}}, \bibinfo{pages}{1019} (\bibinfo{year}{2013}),
  \urlprefix\url{https://doi.org/10.1038/nmat3723}.

\bibitem[{\citenamefont{Wakimoto et~al.}(2015)\citenamefont{Wakimoto, Ishii,
  Kimura, Fujita, Dellea, Kummer, Braicovich, Ghiringhelli, Debeer-Schmitt, and
  Granroth}}]{Wakimoto2015}
\bibinfo{author}{\bibfnamefont{S.}~\bibnamefont{Wakimoto}},
  \bibinfo{author}{\bibfnamefont{K.}~\bibnamefont{Ishii}},
  \bibinfo{author}{\bibfnamefont{H.}~\bibnamefont{Kimura}},
  \bibinfo{author}{\bibfnamefont{M.}~\bibnamefont{Fujita}},
  \bibinfo{author}{\bibfnamefont{G.}~\bibnamefont{Dellea}},
  \bibinfo{author}{\bibfnamefont{K.}~\bibnamefont{Kummer}},
  \bibinfo{author}{\bibfnamefont{L.}~\bibnamefont{Braicovich}},
  \bibinfo{author}{\bibfnamefont{G.}~\bibnamefont{Ghiringhelli}},
  \bibinfo{author}{\bibfnamefont{L.~M.} \bibnamefont{Debeer-Schmitt}},
  \bibnamefont{and} \bibinfo{author}{\bibfnamefont{G.~E.}
  \bibnamefont{Granroth}}, \bibinfo{journal}{Phys. Rev. B}
  \textbf{\bibinfo{volume}{91}}, \bibinfo{pages}{184513}
  (\bibinfo{year}{2015}),
  \urlprefix\url{https://link.aps.org/doi/10.1103/PhysRevB.91.184513}.

\bibitem[{\citenamefont{Monney et~al.}(2016)\citenamefont{Monney, Schmitt,
  Matt, Mesot, Strocov, Lipscombe, Hayden, and Chang}}]{Monney2016}
\bibinfo{author}{\bibfnamefont{C.}~\bibnamefont{Monney}},
  \bibinfo{author}{\bibfnamefont{T.}~\bibnamefont{Schmitt}},
  \bibinfo{author}{\bibfnamefont{C.~E.} \bibnamefont{Matt}},
  \bibinfo{author}{\bibfnamefont{J.}~\bibnamefont{Mesot}},
  \bibinfo{author}{\bibfnamefont{V.~N.} \bibnamefont{Strocov}},
  \bibinfo{author}{\bibfnamefont{O.~J.} \bibnamefont{Lipscombe}},
  \bibinfo{author}{\bibfnamefont{S.~M.} \bibnamefont{Hayden}},
  \bibnamefont{and} \bibinfo{author}{\bibfnamefont{J.}~\bibnamefont{Chang}},
  \bibinfo{journal}{Phys. Rev. B} \textbf{\bibinfo{volume}{93}},
  \bibinfo{pages}{075103} (\bibinfo{year}{2016}),
  \urlprefix\url{https://link.aps.org/doi/10.1103/PhysRevB.93.075103}.

\bibitem[{\citenamefont{Sordi et~al.}(2011)\citenamefont{Sordi, Haule, and
  Tremblay}}]{Sordi2011}
\bibinfo{author}{\bibfnamefont{G.}~\bibnamefont{Sordi}},
  \bibinfo{author}{\bibfnamefont{K.}~\bibnamefont{Haule}}, \bibnamefont{and}
  \bibinfo{author}{\bibfnamefont{A.-M.~S.} \bibnamefont{Tremblay}},
  \bibinfo{journal}{Phys. Rev. B} \textbf{\bibinfo{volume}{84}},
  \bibinfo{pages}{075161} (\bibinfo{year}{2011}),
  \urlprefix\url{https://link.aps.org/doi/10.1103/PhysRevB.84.075161}.

\bibitem[{\citenamefont{Peli et~al.}(2017)\citenamefont{Peli, Conte, Comin,
  Nembrini, Ronchi, Abrami, Banfi, Ferrini, Brida, Lupi et~al.}}]{Peli2017}
\bibinfo{author}{\bibfnamefont{S.}~\bibnamefont{Peli}},
  \bibinfo{author}{\bibfnamefont{S.~D.} \bibnamefont{Conte}},
  \bibinfo{author}{\bibfnamefont{R.}~\bibnamefont{Comin}},
  \bibinfo{author}{\bibfnamefont{N.}~\bibnamefont{Nembrini}},
  \bibinfo{author}{\bibfnamefont{A.}~\bibnamefont{Ronchi}},
  \bibinfo{author}{\bibfnamefont{P.}~\bibnamefont{Abrami}},
  \bibinfo{author}{\bibfnamefont{F.}~\bibnamefont{Banfi}},
  \bibinfo{author}{\bibfnamefont{G.}~\bibnamefont{Ferrini}},
  \bibinfo{author}{\bibfnamefont{D.}~\bibnamefont{Brida}},
  \bibinfo{author}{\bibfnamefont{S.}~\bibnamefont{Lupi}}, \bibnamefont{et~al.},
  \bibinfo{journal}{Nature Physics} \textbf{\bibinfo{volume}{13}},
  \bibinfo{pages}{806} (\bibinfo{year}{2017}),
  \urlprefix\url{https://doi.org/10.1038/nphys4112}.

\bibitem[{\citenamefont{Minola et~al.}(2017)\citenamefont{Minola, Lu, Peng,
  Dellea, Gretarsson, Haverkort, Ding, Sun, Zhou, Peets et~al.}}]{Minola2017}
\bibinfo{author}{\bibfnamefont{M.}~\bibnamefont{Minola}},
  \bibinfo{author}{\bibfnamefont{Y.}~\bibnamefont{Lu}},
  \bibinfo{author}{\bibfnamefont{Y.~Y.} \bibnamefont{Peng}},
  \bibinfo{author}{\bibfnamefont{G.}~\bibnamefont{Dellea}},
  \bibinfo{author}{\bibfnamefont{H.}~\bibnamefont{Gretarsson}},
  \bibinfo{author}{\bibfnamefont{M.~W.} \bibnamefont{Haverkort}},
  \bibinfo{author}{\bibfnamefont{Y.}~\bibnamefont{Ding}},
  \bibinfo{author}{\bibfnamefont{X.}~\bibnamefont{Sun}},
  \bibinfo{author}{\bibfnamefont{X.~J.} \bibnamefont{Zhou}},
  \bibinfo{author}{\bibfnamefont{D.~C.} \bibnamefont{Peets}},
  \bibnamefont{et~al.}, \bibinfo{journal}{Phys. Rev. Lett.}
  \textbf{\bibinfo{volume}{119}}, \bibinfo{pages}{097001}
  (\bibinfo{year}{2017}),
  \urlprefix\url{https://link.aps.org/doi/10.1103/PhysRevLett.119.097001}.

\bibitem[{\citenamefont{Kawasaki et~al.}(2010)\citenamefont{Kawasaki, Lin,
  Kuhns, Reyes, and Zheng}}]{Kawasaki2010}
\bibinfo{author}{\bibfnamefont{S.}~\bibnamefont{Kawasaki}},
  \bibinfo{author}{\bibfnamefont{C.}~\bibnamefont{Lin}},
  \bibinfo{author}{\bibfnamefont{P.~L.} \bibnamefont{Kuhns}},
  \bibinfo{author}{\bibfnamefont{A.~P.} \bibnamefont{Reyes}}, \bibnamefont{and}
  \bibinfo{author}{\bibfnamefont{G.-q.} \bibnamefont{Zheng}},
  \bibinfo{journal}{Phys. Rev. Lett.} \textbf{\bibinfo{volume}{105}},
  \bibinfo{pages}{137002} (\bibinfo{year}{2010}),
  \urlprefix\url{https://link.aps.org/doi/10.1103/PhysRevLett.105.137002}.

\bibitem[{\citenamefont{Joshi et~al.}(2020)\citenamefont{Joshi, Li,
  Tarnopolsky, Georges, and Sachdev}}]{Joshi2020}
\bibinfo{author}{\bibfnamefont{D.~G.} \bibnamefont{Joshi}},
  \bibinfo{author}{\bibfnamefont{C.}~\bibnamefont{Li}},
  \bibinfo{author}{\bibfnamefont{G.}~\bibnamefont{Tarnopolsky}},
  \bibinfo{author}{\bibfnamefont{A.}~\bibnamefont{Georges}}, \bibnamefont{and}
  \bibinfo{author}{\bibfnamefont{S.}~\bibnamefont{Sachdev}},
  \bibinfo{journal}{Phys. Rev. X} \textbf{\bibinfo{volume}{10}},
  \bibinfo{pages}{021033} (\bibinfo{year}{2020}),
  \urlprefix\url{https://link.aps.org/doi/10.1103/PhysRevX.10.021033}.

\bibitem[{\citenamefont{Shackleton et~al.}(2021)\citenamefont{Shackleton,
  Wietek, Georges, and Sachdev}}]{Shackleton2021}
\bibinfo{author}{\bibfnamefont{H.}~\bibnamefont{Shackleton}},
  \bibinfo{author}{\bibfnamefont{A.}~\bibnamefont{Wietek}},
  \bibinfo{author}{\bibfnamefont{A.}~\bibnamefont{Georges}}, \bibnamefont{and}
  \bibinfo{author}{\bibfnamefont{S.}~\bibnamefont{Sachdev}},
  \bibinfo{journal}{Phys. Rev. Lett.} \textbf{\bibinfo{volume}{126}},
  \bibinfo{pages}{136602} (\bibinfo{year}{2021}),
  \urlprefix\url{https://link.aps.org/doi/10.1103/PhysRevLett.126.136602}.

\bibitem[{\citenamefont{Badoux et~al.}(2016)\citenamefont{Badoux, Tabis,
  Lalibert{\'e}, Grissonnanche, Vignolle, Vignolles, B{\'e}ard, Bonn, Hardy,
  Liang et~al.}}]{Badoux2016}
\bibinfo{author}{\bibfnamefont{S.}~\bibnamefont{Badoux}},
  \bibinfo{author}{\bibfnamefont{W.}~\bibnamefont{Tabis}},
  \bibinfo{author}{\bibfnamefont{F.}~\bibnamefont{Lalibert{\'e}}},
  \bibinfo{author}{\bibfnamefont{G.}~\bibnamefont{Grissonnanche}},
  \bibinfo{author}{\bibfnamefont{B.}~\bibnamefont{Vignolle}},
  \bibinfo{author}{\bibfnamefont{D.}~\bibnamefont{Vignolles}},
  \bibinfo{author}{\bibfnamefont{J.}~\bibnamefont{B{\'e}ard}},
  \bibinfo{author}{\bibfnamefont{D.~A.} \bibnamefont{Bonn}},
  \bibinfo{author}{\bibfnamefont{W.~N.} \bibnamefont{Hardy}},
  \bibinfo{author}{\bibfnamefont{R.}~\bibnamefont{Liang}},
  \bibnamefont{et~al.}, \bibinfo{journal}{Nature}
  \textbf{\bibinfo{volume}{531}}, \bibinfo{pages}{210} (\bibinfo{year}{2016}),
  \urlprefix\url{https://doi.org/10.1038/nature16983}.

\bibitem[{\citenamefont{Putzke et~al.}(2021)\citenamefont{Putzke, Benhabib,
  Tabis, Ayres, Wang, Malone, Licciardello, Lu, Kondo, Takeuchi
  et~al.}}]{Putzke2021}
\bibinfo{author}{\bibfnamefont{C.}~\bibnamefont{Putzke}},
  \bibinfo{author}{\bibfnamefont{S.}~\bibnamefont{Benhabib}},
  \bibinfo{author}{\bibfnamefont{W.}~\bibnamefont{Tabis}},
  \bibinfo{author}{\bibfnamefont{J.}~\bibnamefont{Ayres}},
  \bibinfo{author}{\bibfnamefont{Z.}~\bibnamefont{Wang}},
  \bibinfo{author}{\bibfnamefont{L.}~\bibnamefont{Malone}},
  \bibinfo{author}{\bibfnamefont{S.}~\bibnamefont{Licciardello}},
  \bibinfo{author}{\bibfnamefont{J.}~\bibnamefont{Lu}},
  \bibinfo{author}{\bibfnamefont{T.}~\bibnamefont{Kondo}},
  \bibinfo{author}{\bibfnamefont{T.}~\bibnamefont{Takeuchi}},
  \bibnamefont{et~al.}, \bibinfo{journal}{Nature Physics}
  \textbf{\bibinfo{volume}{17}}, \bibinfo{pages}{826} (\bibinfo{year}{2021}),
  \urlprefix\url{https://doi.org/10.1038/s41567-021-01197-0}.

\bibitem[{\citenamefont{Kimura et~al.}(2003{\natexlab{b}})\citenamefont{Kimura,
  Kofu, Matsumoto, and Hirota}}]{Hirota2003}
\bibinfo{author}{\bibfnamefont{H.}~\bibnamefont{Kimura}},
  \bibinfo{author}{\bibfnamefont{M.}~\bibnamefont{Kofu}},
  \bibinfo{author}{\bibfnamefont{Y.}~\bibnamefont{Matsumoto}},
  \bibnamefont{and} \bibinfo{author}{\bibfnamefont{K.}~\bibnamefont{Hirota}},
  \bibinfo{journal}{Phys. Rev. Lett.} \textbf{\bibinfo{volume}{91}},
  \bibinfo{pages}{067002} (\bibinfo{year}{2003}{\natexlab{b}}),
  \urlprefix\url{https://link.aps.org/doi/10.1103/PhysRevLett.91.067002}.

\bibitem[{\citenamefont{H{\"u}cker et~al.}(2011)\citenamefont{H{\"u}cker,
  Zimmermann, Xu, Wen, Gu, Tian, Zarestky, and Tranquada}}]{Hucker2011}
\bibinfo{author}{\bibfnamefont{M.}~\bibnamefont{H{\"u}cker}},
  \bibinfo{author}{\bibfnamefont{M.~v.} \bibnamefont{Zimmermann}},
  \bibinfo{author}{\bibfnamefont{Z.~J.} \bibnamefont{Xu}},
  \bibinfo{author}{\bibfnamefont{J.~S.} \bibnamefont{Wen}},
  \bibinfo{author}{\bibfnamefont{G.~D.} \bibnamefont{Gu}},
  \bibinfo{author}{\bibfnamefont{W.}~\bibnamefont{Tian}},
  \bibinfo{author}{\bibfnamefont{J.}~\bibnamefont{Zarestky}}, \bibnamefont{and}
  \bibinfo{author}{\bibfnamefont{J.~M.} \bibnamefont{Tranquada}},
  \bibinfo{journal}{Journal of Superconductivity and Novel Magnetism}
  \textbf{\bibinfo{volume}{24}}, \bibinfo{pages}{1229} (\bibinfo{year}{2011}),
  \urlprefix\url{https://doi.org/10.1007/s10948-010-1122-0}.

\bibitem[{\citenamefont{Schafgans et~al.}(2010)\citenamefont{Schafgans,
  LaForge, Dordevic, Qazilbash, Padilla, Burch, Li, Komiya, Ando, and
  Basov}}]{Schafgans2010}
\bibinfo{author}{\bibfnamefont{A.~A.} \bibnamefont{Schafgans}},
  \bibinfo{author}{\bibfnamefont{A.~D.} \bibnamefont{LaForge}},
  \bibinfo{author}{\bibfnamefont{S.~V.} \bibnamefont{Dordevic}},
  \bibinfo{author}{\bibfnamefont{M.~M.} \bibnamefont{Qazilbash}},
  \bibinfo{author}{\bibfnamefont{W.~J.} \bibnamefont{Padilla}},
  \bibinfo{author}{\bibfnamefont{K.~S.} \bibnamefont{Burch}},
  \bibinfo{author}{\bibfnamefont{Z.~Q.} \bibnamefont{Li}},
  \bibinfo{author}{\bibfnamefont{S.}~\bibnamefont{Komiya}},
  \bibinfo{author}{\bibfnamefont{Y.}~\bibnamefont{Ando}}, \bibnamefont{and}
  \bibinfo{author}{\bibfnamefont{D.~N.} \bibnamefont{Basov}},
  \bibinfo{journal}{Phys. Rev. Lett.} \textbf{\bibinfo{volume}{104}},
  \bibinfo{pages}{157002} (\bibinfo{year}{2010}),
  \urlprefix\url{https://link.aps.org/doi/10.1103/PhysRevLett.104.157002}.

\bibitem[{\citenamefont{Ma et~al.}(2021)\citenamefont{Ma, Rule, Cronkwright,
  Dragomir, Mitchell, Smith, Chi, Kolesnikov, Stone, and Gaulin}}]{Ma2021}
\bibinfo{author}{\bibfnamefont{Q.}~\bibnamefont{Ma}},
  \bibinfo{author}{\bibfnamefont{K.~C.} \bibnamefont{Rule}},
  \bibinfo{author}{\bibfnamefont{Z.~W.} \bibnamefont{Cronkwright}},
  \bibinfo{author}{\bibfnamefont{M.}~\bibnamefont{Dragomir}},
  \bibinfo{author}{\bibfnamefont{G.}~\bibnamefont{Mitchell}},
  \bibinfo{author}{\bibfnamefont{E.~M.} \bibnamefont{Smith}},
  \bibinfo{author}{\bibfnamefont{S.}~\bibnamefont{Chi}},
  \bibinfo{author}{\bibfnamefont{A.~I.} \bibnamefont{Kolesnikov}},
  \bibinfo{author}{\bibfnamefont{M.~B.} \bibnamefont{Stone}}, \bibnamefont{and}
  \bibinfo{author}{\bibfnamefont{B.~D.} \bibnamefont{Gaulin}},
  \bibinfo{journal}{Phys. Rev. Research} \textbf{\bibinfo{volume}{3}},
  \bibinfo{pages}{023151} (\bibinfo{year}{2021}),
  \urlprefix\url{https://link.aps.org/doi/10.1103/PhysRevResearch.3.023151}.

\bibitem[{\citenamefont{Mazurenko et~al.}(2017)\citenamefont{Mazurenko, Chiu,
  Ji, Parsons, Kan{\'a}sz-Nagy, Schmidt, Grusdt, Demler, Greif, and
  Greiner}}]{Mazurenko2017}
\bibinfo{author}{\bibfnamefont{A.}~\bibnamefont{Mazurenko}},
  \bibinfo{author}{\bibfnamefont{C.~S.} \bibnamefont{Chiu}},
  \bibinfo{author}{\bibfnamefont{G.}~\bibnamefont{Ji}},
  \bibinfo{author}{\bibfnamefont{M.~F.} \bibnamefont{Parsons}},
  \bibinfo{author}{\bibfnamefont{M.}~\bibnamefont{Kan{\'a}sz-Nagy}},
  \bibinfo{author}{\bibfnamefont{R.}~\bibnamefont{Schmidt}},
  \bibinfo{author}{\bibfnamefont{F.}~\bibnamefont{Grusdt}},
  \bibinfo{author}{\bibfnamefont{E.}~\bibnamefont{Demler}},
  \bibinfo{author}{\bibfnamefont{D.}~\bibnamefont{Greif}}, \bibnamefont{and}
  \bibinfo{author}{\bibfnamefont{M.}~\bibnamefont{Greiner}},
  \bibinfo{journal}{Nature} \textbf{\bibinfo{volume}{545}},
  \bibinfo{pages}{462} (\bibinfo{year}{2017}),
  \urlprefix\url{https://doi.org/10.1038/nature22362}.

\bibitem[{\citenamefont{Koepsell et~al.}(2021)\citenamefont{Koepsell, Bourgund,
  Sompet, Hirthe, Bohrdt, Wang, Grusdt, Demler, Salomon, Gross
  et~al.}}]{Koepsell2021}
\bibinfo{author}{\bibfnamefont{J.}~\bibnamefont{Koepsell}},
  \bibinfo{author}{\bibfnamefont{D.}~\bibnamefont{Bourgund}},
  \bibinfo{author}{\bibfnamefont{P.}~\bibnamefont{Sompet}},
  \bibinfo{author}{\bibfnamefont{S.}~\bibnamefont{Hirthe}},
  \bibinfo{author}{\bibfnamefont{A.}~\bibnamefont{Bohrdt}},
  \bibinfo{author}{\bibfnamefont{Y.}~\bibnamefont{Wang}},
  \bibinfo{author}{\bibfnamefont{F.}~\bibnamefont{Grusdt}},
  \bibinfo{author}{\bibfnamefont{E.}~\bibnamefont{Demler}},
  \bibinfo{author}{\bibfnamefont{G.}~\bibnamefont{Salomon}},
  \bibinfo{author}{\bibfnamefont{C.}~\bibnamefont{Gross}},
  \bibnamefont{et~al.}, \bibinfo{journal}{Science}
  \textbf{\bibinfo{volume}{374}}, \bibinfo{pages}{82} (\bibinfo{year}{2021}),
  \urlprefix\url{https://www.science.org/doi/10.1126/science.abe7165}.

\bibitem[{\citenamefont{Zaninetti}(2017)}]{Zaninetti2017}
\bibinfo{author}{\bibfnamefont{L.}~\bibnamefont{Zaninetti}},
  \bibinfo{journal}{Advances in Astrophysics}  (\bibinfo{year}{2017}), ISSN
  \bibinfo{issn}{2415-6469},
  \urlprefix\url{http://dx.doi.org/10.22606/adap.2017.23005}.

\end{thebibliography}
\end{document}